\documentclass[aps,prd,floatfix,nofootinbib,amssymb,preprintnumbers,showpacs,twocolumn]{revtex4}
\usepackage{graphicx}
\begin{document}
\newcommand{\tr}{\mbox{tr}\,}
\preprint{GUTPA/01/01/01}
\title
{Perturbative Wilson loops with massive sea quarks on the lattice}
\author{Gunnar S.\ Bali}
\email{g.bali@physics.gla.ac.uk}
\affiliation{Department of Physics \& Astronomy, The University of Glasgow, Glasgow G12 8QQ, Scotland}
\author{Peter Boyle}
\email{pab@phys.columbia.edu}
\affiliation{Physics Department, Columbia University,
New York, NY 10027, USA}
\altaffiliation{Permanent address: Department of Physics \& Astronomy,
The University of Edinburgh, Edinburgh
EH9 3JZ, Scotland}
\date{\today}
\begin{abstract}
We present ${\mathcal O}(g^4)$ calculations of both
planar and non-planar Wilson
loops for various actions in the presence of sea quarks.
In particular, the plaquette, the static potential and the static self energy
are calculated to this order for massive Wilson,
Sheikholeslami-Wohlert and Kogut-Susskind fermions, including
the mass and $n_f$ dependence. The results can be
used to obtain $\alpha_{\overline{MS}}$ and $\overline{m_b}(\overline{m_b})$
from lattice simulations. We compare our perturbative calculations
to simulation data of the
static potential and report excellent qualitative agreement with
boosted perturbation theory predictions for distances $r<1\,\mbox{GeV}^{-1}$.
We are also able to resolve differences in the running of the coupling
between $n_f=2$ and $n_f=0$ static potentials.
We compute
perturbative estimates of the ``$\beta$-shifts'' of QCD
with sea quarks, relative to the quenched theory, which we find to agree
within 10~\% with non-perturbative simulations.
This is done by matching the respective static potentials at large
distances. The prospects of determining the QCD running coupling from low
energy hadron phenomenology in the near future are assessed.
We obtain the result $\Lambda^{(2)}_{\overline{MS}}r_0=0.69(15)$
for the two flavour QCD $\Lambda$-parameter from
presently available lattice data where $r_0^{-1}\approx
400$~MeV and estimate $\alpha_{\overline{MS}}^{(5)}(m_Z)=0.1133(59)$.
\end{abstract}
\maketitle

\section{Introduction\label{SecIntro}}
The calculation of Wilson loops in lattice perturbation theory is useful
in a number of ways. An important application is the prediction
of a strong coupling constant
$\alpha_{\overline{MS}}^{(5)}(m_Z)$
from low energy hadronic phenomenology by means of non-perturbative
lattice simulations~\cite{El-Khadra:1992vn,davi1,davi2,Spitz:1999tu,Booth:2001qp,davi3}.
Perturbative calculations of Wilson loops
are also employed in the context of mean field
improvement programmes of the lattice
action and operators~\cite{Parisi:1980pe,lepage}.

In the limit of infinite Euclidean time
separation, $T\rightarrow\infty$, Wilson loops
give access to the perturbative lattice potential.
Our results improve the understanding
of the violations of rotational symmetry at short distances of
this quantity,
as well as of the short distance effects of including various flavours
of sea quarks.
Furthermore, in the limit of large distances, $R\rightarrow\infty$,
the self-energy of static sources can be obtained from the potential,
enabling the calculation of
$\overline{m_b}(\overline{m_b})$ from non-perturbative
simulations of heavy-light mesons in the static
limit~\cite{Martinelli:1999vt}.

Perturbative computations of Wilson loops on the lattice now reach back as
far as two decades.
For this reason, much of the groundwork for the calculation covered in this
article is well known, and has been reproduced by us. In addition
to improving the precision of many previously existing results and presenting
them in a more comprehensive way,
we extend the calculations by incorporating the actions, sea quark masses
and observables that are relevant for present day lattice simulations. 
We also perform the first calculation
of non-planar Wilson loops. Because of the degree of replication we
present a brief survey of the existing literature here.

In 1981 M\"uller and R\"uhl~\cite{MullerRuhl} performed
a highly rigorous calculation of
the ${\mathcal O}(g^4)$ contribution to the Wilson loop for pure $U(1)$ and $SU(2)$
gauge theories and presented their result in terms of
lattice integrals. In particular, they obtained a closed form for
the $T\rightarrow\infty$ limit and hence the static potential.
The plaquette in $SU(N)$ pure gauge theory
was first calculated to this order by Di Giacomo and Rossi
\cite{DiGiacomo:1981wt}, again in 1981.
In the same year Hattori and Kawai~\cite{Hattori:1981ac}
estimated ${\mathcal O}(g^4)$ contributions
to larger Wilson loops, by modelling some of the relevant lattice
contributions by their continuum counterparts.
Rigorous results to this order were subsequently independently obtained
by three groups of authors: Weisz, Wohlert and Wetzel~\cite{Weisz:1984bn},
Curci, Paffuti and Tripiccione~\cite{Curci:1984wh} and
Heller and Karsch~\cite{Heller:1985hx}.
The first reference is the most general and also applies to Symanzik-improved
pure Yang-Mills actions. However, numerical results on Wilson loops
are only available for the Wilson action.
The  ${\mathcal O}(g^4)$ fermionic contribution to the plaquette
was obtained by Hamber and Wu~\cite{Hamber:1983ft}
for Wilson quarks and for Kogut-Susskind (KS) quarks
by Heller and Karsch~\cite{Heller:1985eq}.
More recently, Panagopoulos et al.\
calculated the ${\mathcal O}(g^6)$ $SU(N)$ gluonic contribution~\cite{Alles:1994dn}
as well as the ${\mathcal O}(g^4)$ and ${\mathcal O}(g^6)$
corrections~\cite{Alles:1998is} in presence of massive Wilson fermions.
The only numerical
calculation to-date of small Wilson loops for Symanzik improved 
gluonic actions was performed by Iso and Sakai \cite{Iso:1987dj}.

More recently, the Heller and Karsch calculations~\cite{Heller:1985hx}
have been extended by Martinelli and Sachrajda~\cite{Martinelli:1999vt}
in a manner similar to that of M\"uller and R\"uhl~\cite{MullerRuhl}: the
$R,T\rightarrow\infty$ limit of the perimeter term of
Wilson loops has been taken to obtain the static self energy
for $SU(3)$ gauge theory, with the addition of
massless Wilson and Sheikholeslami-Wohlert (SW) sea quarks.
 
The structure of this paper is as follows.
In Sec.~\ref{sec_maths} we describe the procedure 
used to calculate Wilson loops in perturbation theory
and present the perturbative expansions
of small Wilson loops, both planar and non-planar,
to ${\mathcal O}(g^4)$ using a variety of actions.
Although partially addressed by previous authors
\cite{Weisz:1984bn,Curci:1984wh,Heller:1985hx,Altevogt:1995cj}
the static potential
has not been well treated in literature, which we attempt to rectify
in Sec.~\ref{sec_potl}, where we also obtain the static
source self energy from the asymptotic $R\rightarrow\infty$ potential.
This sets the stage for
Sec.~\ref{sec:np} where these perturbative results are confronted
with state-of-the-art lattice data from
simulations with Wilson, SW and
KS fermions: we determine the
``$\beta$-shift'' and the running coupling
$\alpha_{\overline{MS}}$, before comparing perturbative and
non-perturbative potentials at short distances.
In Sec.~\ref{sec:conclude} we conclude and summarize.
Our article is augmented by two Appendices on the relations
between different lattice and continuum renormalization schemes
and on the perturbative $\beta$-shift.

\section{Wilson loops in lattice perturbation theory\label{sec_maths}}
In this section we explain our method and notations, before
displaying results on small Wilson loops with massive fermions
to ${\mathcal O}(g^4)$. We also include some findings
that apply to the Iwasaki and Symanzik-Weisz gauge actions.
\subsection{The Method}
We consider $SU(N)$ lattice gauge theory on an infinite four dimensional
isotropic hyper-cubic lattice
with lattice spacing $a$.
An oriented closed curve ${\mathcal C}$ of length $na$
touches a sequence of sites,
$\{x_i : x_i/a \in {\mathbb Z}^4\}$
such that $x_{i+1} - x_{i} \in \{\pm a\hat{\mu}:\mu =1, \ldots,4\}$,
$x_{n+1}=x_1$.
The Wilson loop $W({\mathcal C})$ around such a curve
is the expectation value of the
path ordered product of gauge links,
\begin{equation}
\label{eq:wloop1}
W({\mathcal C}) = \frac{1}{N}\left\langle
\tr\left[\prod_{x_i\in{\mathcal C}} U_{x_i,\mu_i} \right]\right\rangle,
\end{equation}
where $\mu_i\in\{\pm 1,\cdots,\pm 4\}$ denotes
the direction indicated by $x_{i+1} - x_{i}$
and $U_{x,-\mu} = U^\dagger_{x-\hat{\mu},\mu}$.

\begin{figure}[htb]
\includegraphics[width=7cm]{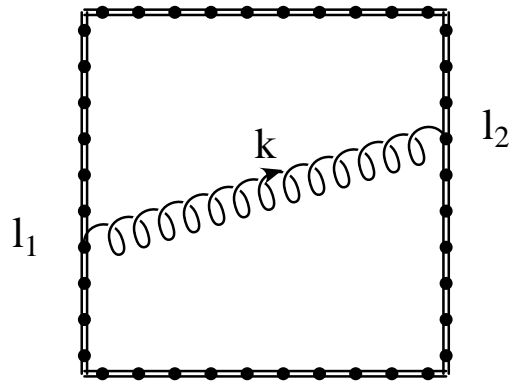}
\caption{
\label{WilsonTree} 
Tree-level [${\mathcal O}(g^2)$] contribution to the Wilson loop. The coordinate
space propagator is convoluted around the possible source and sink
links of the loop. This is done by constructing the corresponding
sum of phase factors in momentum space and Fourier transforming.}
\end{figure}

We write $W({\mathbf R},T)$ for a ``rectangular'' Wilson loop
where the curve ${\mathcal C}$ contains
two opposite lines with an extent of $T$ lattice units
pointing into the time ($\hat{4}$) direction,
that are separated by a spatial distance ${\mathbf R}a$.
The smallest such Wilson loop is an $a\times a$ square, the plaquette
$\Box=W(1,1)$ which, appropriately normalized, is the expectation value of the
Wilson gauge action,
\begin{equation}
S=-\beta\sum_{x,\mu>\nu}\frac{1}{N}\mbox{Re}\,\tr U_{x,\mu\nu},
\end{equation}
where $\beta=2N/g^2$ and $U_{x,\mu\nu}=U_{x,\mu}U_{x+a\hat{\mu},\nu}
U^{\dagger}_{x+a\hat{\nu},\mu}U^{\dagger}_{x,\nu}$.
Unless stated otherwise this is the gauge action which
we combine either with
the Wilson, SW or the KS fermionic actions.

Writing $U_l = e^{i ga A_l}$,
where $l=(x,\mu)$ denotes the link connecting $x$ with $x+a\hat{\mu}$
and $A_{l}$ is a short hand notation for $A_{\mu}(x+\frac{a}{2}\hat{\mu})$,
and expanding the exponentials within Eq.~(\ref{eq:wloop1})
one obtains,
\begin{equation}
\label{eq:wilexp}
W({\mathcal C}) = 1 - g^2 W_2 - g^3 W_3 - g^4 W_4-\cdots,
\end{equation}
with \cite{Weisz:1984bn}, 
\begin{equation}
W_2 = \frac{a^2}{2N} \tr T^aT^b \left\langle
\sum_{l_1,l_2} A_{l_1}^a A_{l_2}^b
\right\rangle,
\label{W2eqn}
\end{equation}
\begin{eqnarray}
\label{W3eqn}
W_3 &=& \frac{ia^3}{6N} \tr T^aT^bT^c \left\langle 
\sum_{l_1,l_2,l_3} A_{l_1}^a A_{l_2}^b A^c_{l_3}\right.\\\nonumber
&&+ \left.3 \sum_{l_1<l_2<l_3} \left( A_{l_1}^a A_{l_2}^b A^c_{l_3} - A_{l_3}^a A_{l_2}^b A^c_{l_1}\right)
\right\rangle,
\end{eqnarray}
\begin{eqnarray}
\label{W4eqn}
W_4 &=& -\frac{a^4}{24N} \tr T^aT^bT^cT^d \left\langle
24\sum_{l_1<l_2<l_3<l_4}
A_{l_1}^a A_{l_2}^b A_{l_3}^c A_{l_4}^d\right.\nonumber\\
&&+12 \sum_{l_1<l_2<l_3}\left( A_{l_1}^a A_{l_1}^b A_{l_2}^c A_{l_3}^d + A_{l_1}^aA_{l_2}^b A_{l_2}^c A_{l_3}^d\right.
\nonumber\\&&\qquad\qquad\qquad+\left.
A_{l_1}^a A_{l_2}^b A_{l_3}^c A_{l_3}^d \right)\\
&&+\sum_{l_1<l_2} \left( 6\, A_{l_1}^a A_{l_1}^b A_{l_2}^c A_{l_2}^d 
+ 4\, A_{l_1}^a A_{l_1}^b A_{l_1}^c A_{l_2}^d\right.\nonumber\\&&
\qquad\quad +\left.4\, A_{l_1}^a A_{l_2}^b A_{l_2}^c A_{l_2}^d \right)\nonumber\\
&& + \left.\sum_{l_1} A_{l_1}^a A_{l_1}^b A_{l_1}^c A_{l_1}^d 
\right\rangle.\nonumber
\end{eqnarray}

\begin{figure}[hbt]
\includegraphics[width=4cm]{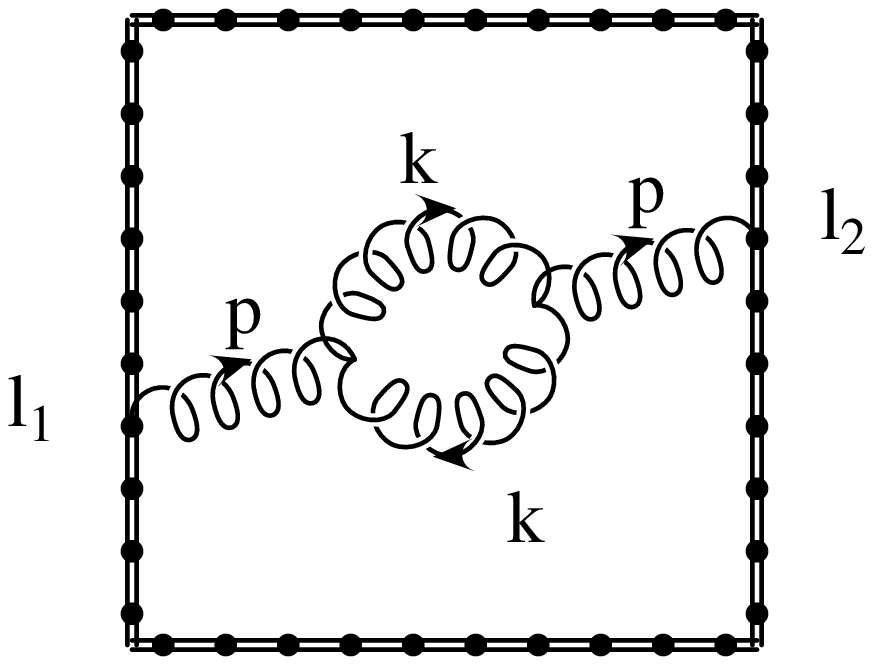}
\includegraphics[width=4cm]{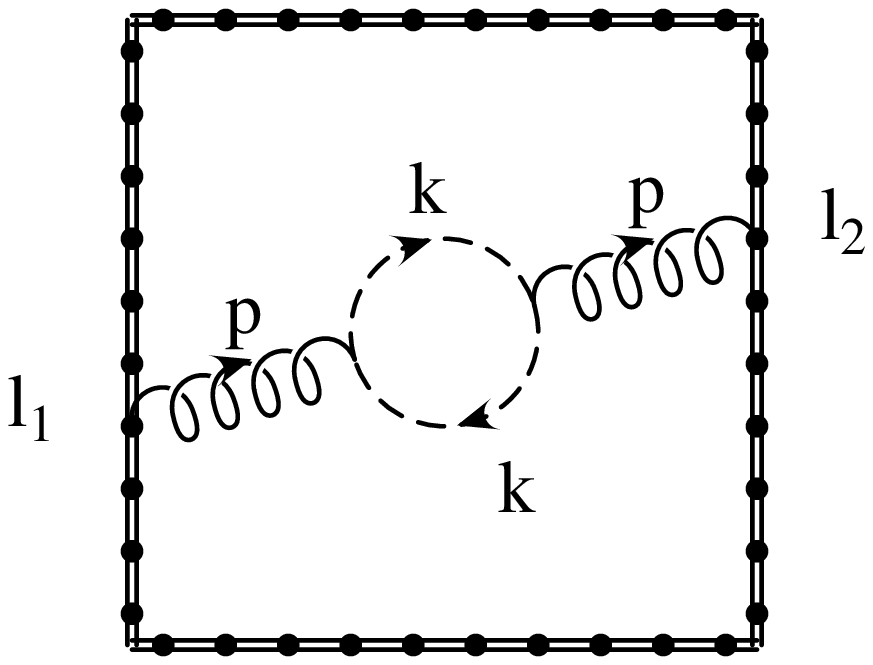}

\includegraphics[width=4cm]{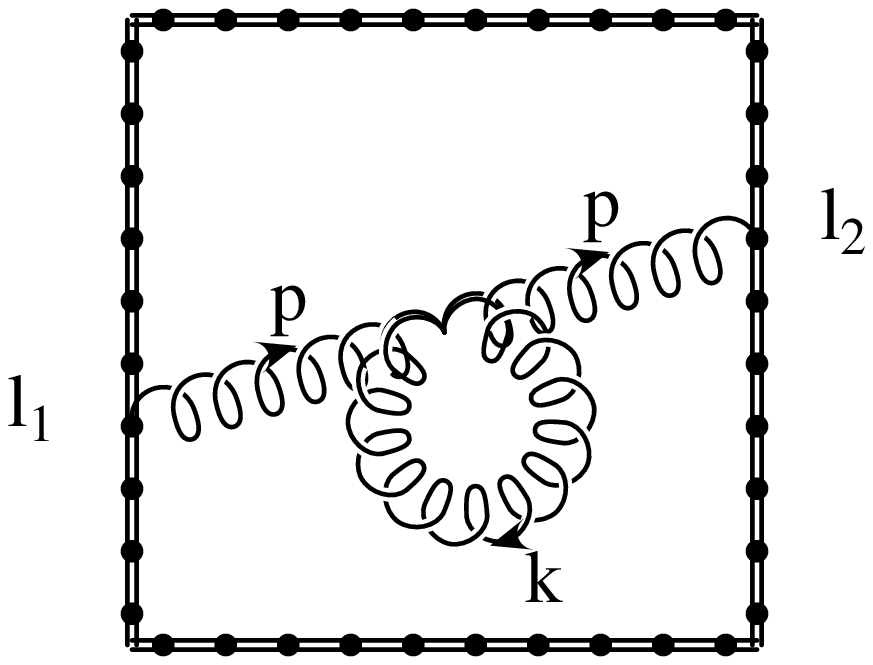}
\includegraphics[width=4cm]{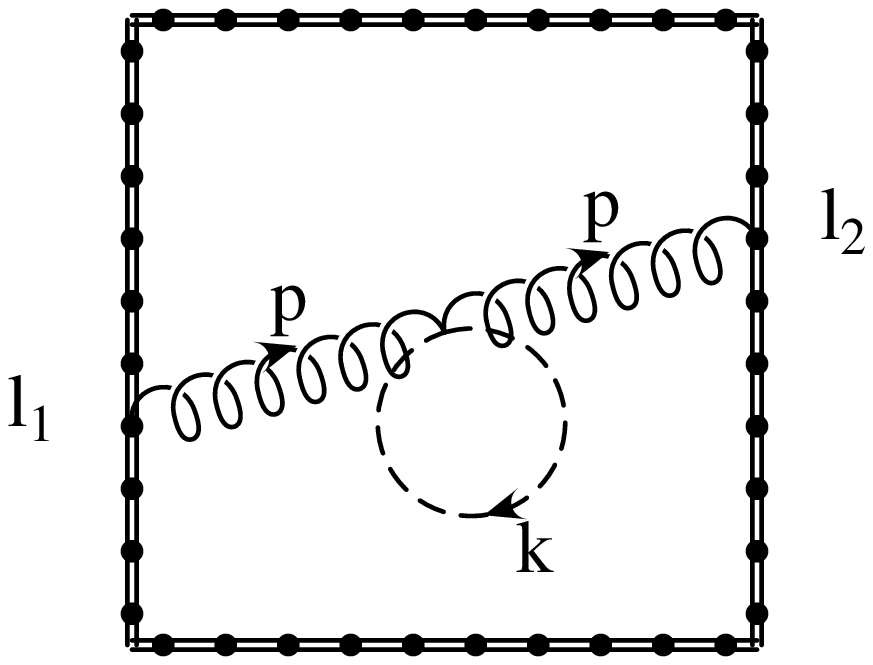}

\includegraphics[width=4cm]{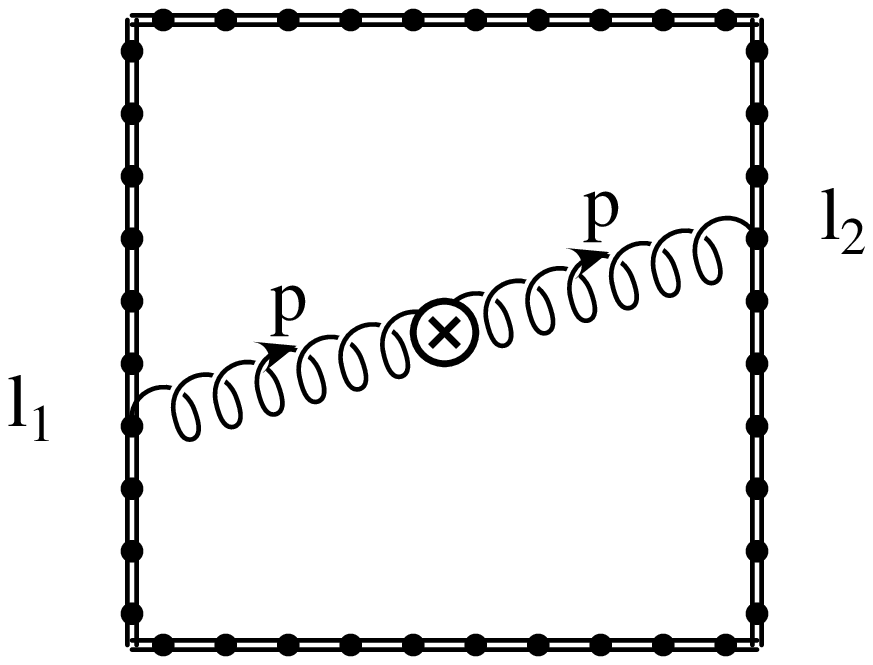}
\caption{
\label{WilsonVP_gluon} 
Pure gauge one-loop [${\mathcal O}(g^4)$] vacuum polarization contribution to
the Wilson loop. This is the sum of continuum like three-gluon and
ghost-ghost-gluon couplings, and both
a continuum like four-gluon tadpole contribution and lattice specific
four-gluon and two-ghost-two-gluon tadpole pieces.
The last measure diagram is due to the Jacobian involved in changing variables
in the path integral from $U_l$ to $A_l$.
}
\end{figure}

\begin{figure}[hbt]
\includegraphics[width=4cm]{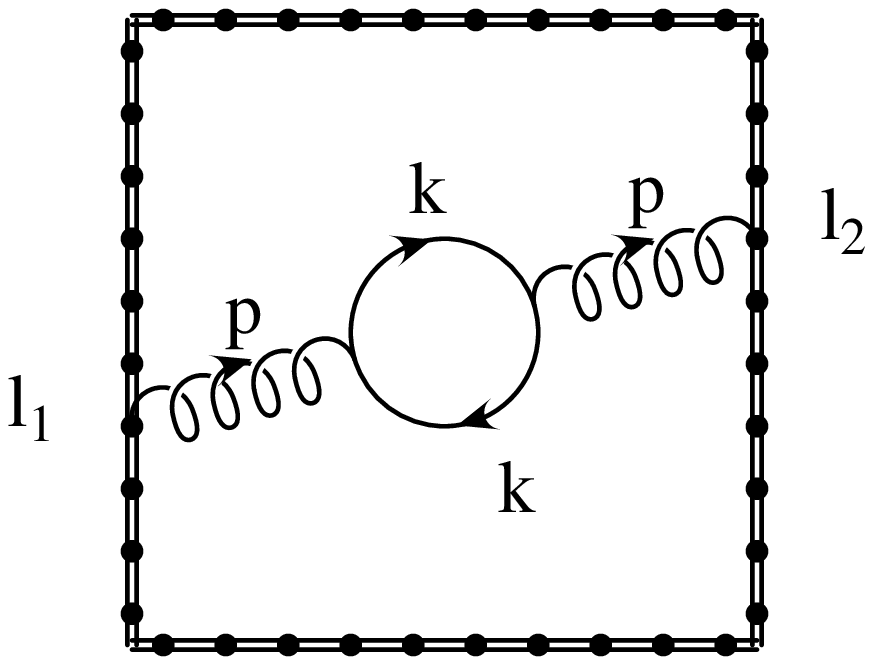}
\includegraphics[width=4cm]{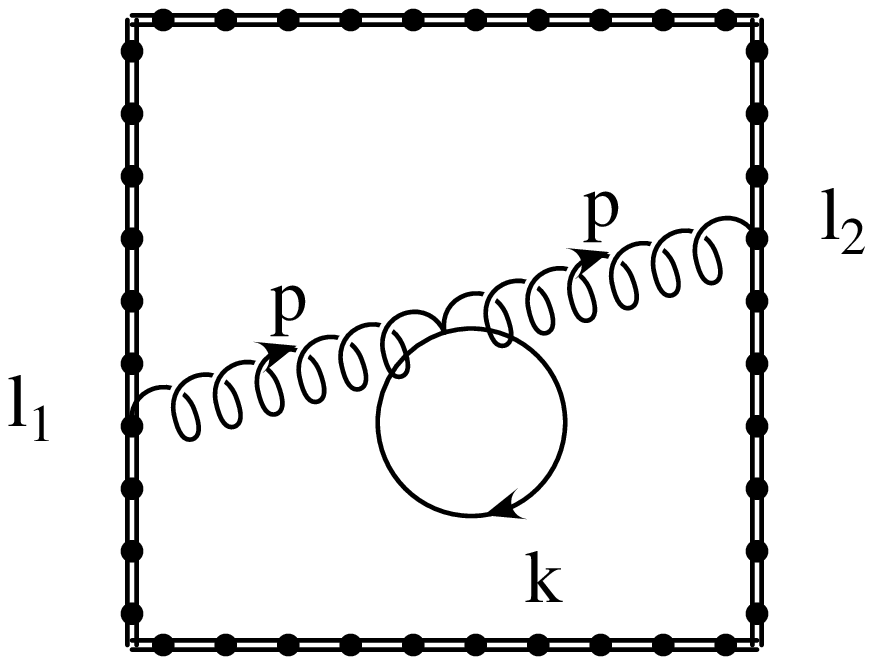}
\caption{
\label{WilsonFermion} 
The leading $n_f$ contribution arises at ${\mathcal O}(g^4)$ 
through the fermionic contribution to the vacuum polarization,
and the lattice specific fermionic tadpole contribution.
}
\end{figure}

\begin{figure}[hbt]
\includegraphics[width=6cm]{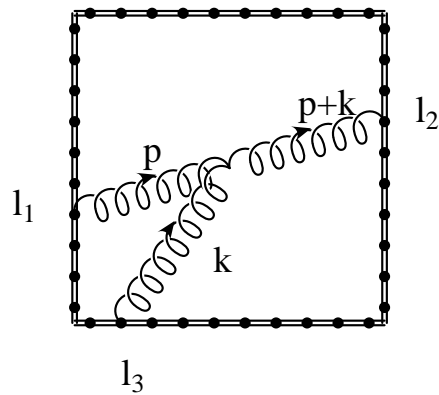}
\caption{
\label{WilsonSpider} 
The ${\mathcal O}(g^4)$ spider contribution to the Wilson loop. 
}
\end{figure}

\begin{figure}[hbt]
\includegraphics[width=6cm]{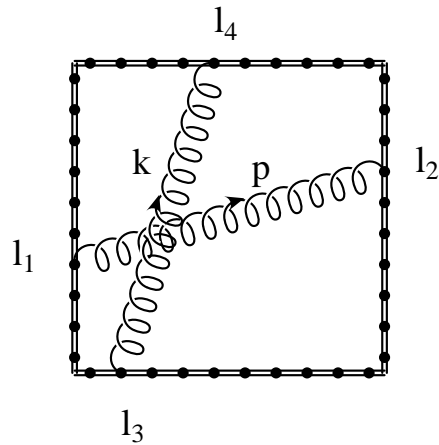}
\caption{
\label{WilsonNA} 
One-loop [${\mathcal O}(g^4)$] two-gluon exchange contribution to the Wilson loop. 
One part of this diagram is accompanied by a $C_F^2/2$ colour
factor, that arises when exponentiating the tree-level contribution.
The remainder is proportional to $C_FN/2$ and specific to non-Abelian
gauge theories.}
\end{figure}

The expectation values in the above formulae depend
on $g^2$ and are sums of gluonic position space $n$-point functions.
The leading order [${\mathcal O}(g^2)$] contribution
to the Wilson loop is the sum
of tree-level two-point functions depicted in 
Fig.~\ref{WilsonTree}.
This is easily expressed in momentum space as a sum of phase factors
multiplying the momentum space propagator:
\begin{eqnarray}
W_2&=&\sum_{l_1, l_2}\int_{-\pi/a}^{\pi/a}\frac{d^4k}{(2\pi)^4}
\frac{a^2}{2N}
\tr T^aT^b e^{i k [x_2-x_1+\frac{a}{2}(\hat{\mu}_2-\hat{\mu}_1)]}
\nonumber\\\nonumber
&&\times\langle A^a_{\mu_{1}} (-k) A^b_{\mu_{2}}(k) \rangle\\
&=&W_2^{(0)}+g^2W_2^{(2)}+ {\mathcal O}(g^4).\label{eq:w210}
\end{eqnarray}
For the Wilson gauge action the gluon propagator reads,
\begin{equation}
\label{eq:w22}
\langle A^a_{\mu_1}(-k)A^b_{\mu_2}(k)\rangle=
\frac{\delta^{ab}(\delta_{\mu_1,\mu_2}-\delta_{\mu_1,-\mu_2})}
{\hat{k}^2}+\Pi^{ab}_{\mu_1\mu_2}(k),
\end{equation}
where
$\hat{k}_{\mu}=2a^{-1}\sin\left(k_{\mu}a/2\right)$
and $\Pi$ is of order $g^2$.
The colour trace then results in the Casimir
factor of the fundamental representation,
$C_F=\frac{1}{N}\tr T^aT^a=(N^2-1)/(2N)$.
All formulae in this Section as well as in Sec.~\ref{sec_potl}
below apply to the general $SU(N)$ case unless stated otherwise.
In addition the constants $C_F$ are factorized out in such a way
that the results apply to external sources in any representation $D$
of the gauge group, under the replacement $C_F\mapsto C_D$.

In explicitly summing
the above phase factors one easily obtains the well known
tree-level expressions for planar Wilson loops.
The vacuum polarization tensor $\Pi$ has to be considered
to leading order
for an ${\mathcal O}(g^4)$ evaluation of Wilson loops which requires $W_2^{(2)}$.
The relevant graphs
for the gluonic sector are depicted in Fig.~\ref{WilsonVP_gluon}.
The fourth diagram involving a
two-gluon-two-ghost vertex (which is lattice specific)
as well as a lattice contribution
to the four-gluon vertex of the third diagram, result in a
lattice tadpole correction to $W_2^{(2)}$ that is
exactly $(2N^2-3)/(24N)$ times the tree-level value $W_2^{(0)}$.
The two additional diagrams of Fig.~\ref{WilsonFermion}
have to be considered in the case of dynamical fermions.

To the order at which we are working throughout this paper,
$W_3$ only contributes through its
contraction with the three gluon vertex, depicted in
Fig.~\ref{WilsonSpider}, and we write
\begin{equation}
\label{eq:w310}
W_3 = g W_3^{(1)} + {\mathcal O}(g^3).
\end{equation}
In momentum space the phase factor associated with a typical entry in
the sum Eq.~(\ref{W3eqn}) is 
\begin{eqnarray}
&&\int_{-\pi/a}^{\pi/a}\frac{d^4kd^4k'}{(2\pi)^8}
\frac{i}{N} \tr T^aT^bT^c
e^{i k [x_2-x_1+\frac{a}{2}(\hat{\mu}_2-\hat{\mu}_1)]}\\&&\times
e^{i k' [x_3-x_1+\frac{a}{2}(\hat{\mu}_3-\hat{\mu}_1)]}
\langle
A^a_{\mu_1} (-k-k') A^b_{\mu_2}(k) A^c_{\mu_3}(k')\rangle.
\nonumber
\end{eqnarray}
The term
\begin{equation}
W_4 = W_4^{(0)} +{\mathcal O}(g^2)
\label{eq:w410}
\end{equation}
only contributes to the Wilson loop at ${\mathcal O}(g^4)$
through its contraction with double gluon exchanges, as
depicted in Fig.~\ref{WilsonNA}.
The generic phase factor associated with this diagram is given
below, where due to the lack of an appropriately symmetrized contracting
vertex we perform the Wick contraction manually, preserving the ordering
of colour factors:
\begin{widetext}
\begin{equation}
\int_{-\pi/a}^{\pi/a}\frac{d^4k}{(2\pi)^4}\frac{d^4k'}{(2\pi)^4}
\frac{i}{N} \tr T^aT^bT^cT^d
\left\{
\begin{array}{c}
e^{i k [x_2-x_1+\frac{a}{2}(\hat{\mu}_2-\hat{\mu}_1)]}
e^{i k' [x_4-x_3+\frac{a}{2}(\hat{\mu}_4-\hat{\mu}_3)]}
\langle
A^a_{\mu_1} (-k) A^b_{\mu_2}(k) A^c_{\mu_3}(-k')A^d_{\mu_4}(k')\rangle\\
e^{i k [x_3-x_1+\frac{a}{2}(\hat{\mu}_3-\hat{\mu}_1)]}
e^{i k' [x_4-x_2+\frac{a}{2}(\hat{\mu}_4-\hat{\mu}_2)]}
\langle
A^a_{\mu_1} (-k) A^b_{\mu_2}(-k') A^c_{\mu_3}(k)A^d_{\mu_4}(k')\rangle\\
e^{i k [x_3-x_2+\frac{a}{2}(\hat{\mu}_3-\hat{\mu}_2)]}
e^{i k' [x_4-x_1+\frac{a}{2}(\hat{\mu}_4-\hat{\mu}_1)]}
\langle
A^a_{\mu_1} (-k') A^b_{\mu_2}(-k) A^c_{\mu_3}(k)A^d_{\mu_4}(k')\rangle
\end{array}
\right\}.
\end{equation}
\end{widetext}
After evaluating colour traces, this term can be split into 
an ``Abelian'' and a ``non-Abelian'' part. 
The Abelian part is proportional to $C_F^2/2$ and can be obtained by
expanding $\exp(-g^2W_2)$. The remainder is specific
to non-Abelian gauge theories and proportional to $C_FN/2$,
the same colour factor that also accompanies the vacuum polarization
diagrams. 

The integrands were coded with the help of a C++ package written by one of
the authors, and then integrated using VEGAS \cite{Lepage:1978sw}.
For finite Wilson loops to ${\mathcal O}(g^2)$
and ${\mathcal O}(g^4)$ we compute 4 and 8 dimensional momentum
integrals, respectively. In the case of the potential
one can analytically perform the limit $T\to\infty$. This allows us
to compute the momentum integrals in one less dimension, leaving
3 and 7 dimensional integrals.

The fermionic Wilson-SW and KS vacuum polarization graphs
have been computed after a further analytical integration
over the 4-component of the internal loop momentum, leaving
a 6 dimensional integration to be performed in the cases of the 
potential and the static self energy.

\begin{figure}[hbt]
\includegraphics[width=4cm]{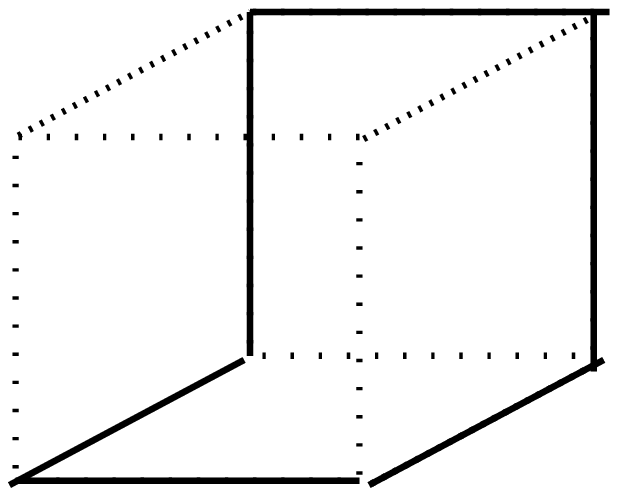}
\includegraphics[width=4cm]{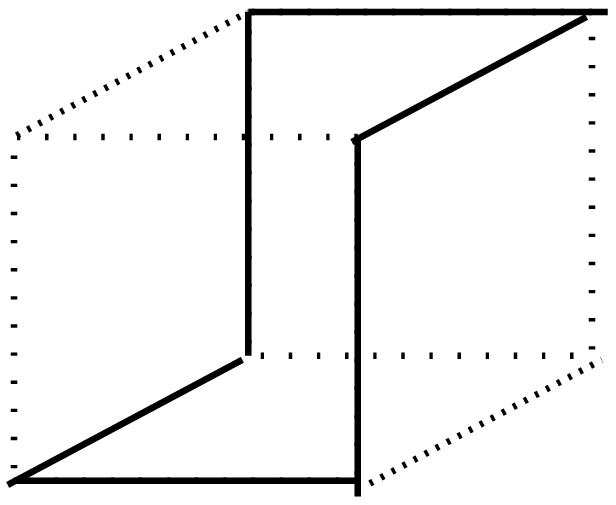}
\caption{
\label{ChairParallelogram} 
The non-planar Wilson loops in the unit cube which we refer to
as the chair and the ``parallelogram'', respectively.
}
\end{figure}

It is relatively easy to automate the
generation of the phase factors that are required for arbitrary
curves ${\mathcal C}$. We generate the perturbative
expansion for both small rectangular Wilson loops and for the 
non-planar loops shown in Fig.~\ref{ChairParallelogram}.
We refer to these non-planar loops as the chair and the (three-dimensional)
``parallelogram''; both are contained within the unit cube and are used
within certain improved gauge actions.

\subsection{Definitions}
We write the tree-level contribution to the expansion of the Wilson loop
with the Wilson gauge action as,
\begin{equation}
\label{eq:w20}
W_2^{(0)}  = C_F W_{T}.
\end{equation}
For Wilson and SW fermions with the Wilson
gauge action we write the vacuum polarization insertions as,
\begin{equation}\label{eq:wilex1}
W_2^{(2)}  = C_F\left(\frac{N}{2}W_{\Pi}+\frac{2N^2-3}{24N}W_T+n_f
X_{f}\right).
\end{equation}
For Wilson-SW fermions we define,
\begin{equation}
X_f=X_{f}^{(0)} + X_{f}^{(1)} c_{SW} +  X_{f}^{(2)} c_{SW}^2,
\end{equation}
where $c_{SW}=0$ in the Wilson case and $c_{SW}=1+{\mathcal O}(g^2)$ in the SW case.
The $X_f$ will depend on the quark mass in lattice units
$ma$. In the case of Wilson-SW
fermions and the order of perturbation theory we are working
at we have, $ma=m_0a=(\kappa^{-1}-\kappa^{-1}_c)/2$.
As anticipated above,
we have factorized out a lattice tadpole term that is proportional to $W_T$.
For KS fermions we define analogously,
\begin{equation}
\label{eq:wilex12}
X_f=X_f^{(KS)}.
\end{equation}
In this case $n_f$ has to be a multiple of four.
For each of the above quantities we may write a similar expansion in
terms of $\alpha_L=g^2/(4\pi)$,
rather than $g^2$, with lower case coefficients $w$ and $x$, where
$w_T = 4\pi W_T$, $w_{\Pi} = (4\pi)^2 W_{\Pi}$, $x = (4\pi)^2 X$ and
so forth.

We denote the spider  contribution by,
\begin{equation}
W_3^{(1)} = \frac{C_F N}{2} W_{S}. 
\end{equation}
The double gluon exchange can be written as the sum of Abelian
and non-Abelian pieces,
where the Abelian piece is the next term in the
exponentiation of the tree-level contribution,
\begin{equation} 
\label{eq:w40}
W_4^{(0)} = -\frac{C_F^2}{2} W_T^2 + \frac{C_F N}{2} W_{NA}.
\end{equation}

In combining the above Eqs.~(\ref{eq:w20}) -- (\ref{eq:w40}) with
Eqs.~(\ref{eq:wilexp}), (\ref{eq:w210}),
(\ref{eq:w310}) and (\ref{eq:w410}), we obtain the expansion of the Wilson loop,
\begin{equation}\label{eq:wilex2}
W=1-W_{\mbox{\scriptsize LO}}g^2-\left(W_{\mbox{\scriptsize NLO}}+C_Fn_fX_f\right)g^4,
\end{equation}
where
\begin{eqnarray}
W_{\mbox{\scriptsize LO}}&=&C_FW_T,\\
W_{\mbox{\scriptsize NLO}}&=&C_F\left(\frac{N}{2}W_{\Sigma}
+\frac{2N^2-3}{24N}W_T-\frac{C_F}{2}W_T^2\right),\label{eq:wnlo}\\
W_{\Sigma}&=&W_{\Pi}+W_{NA}+W_S\label{eq:wilex5}
\end{eqnarray}
The $W_T^2$ term within Eq.~(\ref{eq:wnlo})
originates from the exponentiation of the Abelian part of the
one gluon exchange, Eq.~(\ref{eq:w40}).
It is cancelled in the Taylor expansion of $\ln W$, which turns out to
be proportional to $C_F$, at least to ${\mathcal O}(g^4)$. The latter
observation, which is referred to as the ``Casimir scaling''
hypothesis in the literature (see e.g.\
Refs.~\cite{Bali:2000un,Deldar:2000vi}),
implies that, for given $N$ and $n_f$, $\ln W$ only depends
on the representation of the static colour sources, $D$, through its
proportionality to the corresponding Casimir charge $C_D=(\tr T^a_DT^a_D)/
\mbox{dim}_D$. Note that in the case of the fundamental representation,
by using the relation $C_F=(N^2-1)/(2N)$,
Eq.~(\ref{eq:wnlo}) can be rearranged into the form,
\begin{equation}
\label{eq:wfac}
W_{\mbox{\scriptsize NLO}}=
(N^2-1)\left(W_a+\frac{1}{N^2}W_b\right),
\end{equation}
where
\begin{eqnarray}
W_a&=&
\frac{1}{24}\left(6W_{\Sigma}+W_T-3W_T^2\right),\\
W_b&=&\frac{1}{16}\left(-W_T+2W_T^2\right).
\end{eqnarray}
For improved gluonic actions Eq.~(\ref{eq:wnlo}) does not apply and
hence $W_a$ and $W_b$  can take somewhat different forms.
The factorization Eq.~(\ref{eq:wfac}) is frequently employed
throughout the literature,
e.g.\ in Refs.~\cite{DiGiacomo:1981wt,Iso:1987dj,Alles:1994dn}.

\subsection{Small Wilson loop results}
The pure gauge results on small Wilson loops for the Wilson gluonic
action are displayed in Tab.~\ref{TableSmallLoops}.
We reproduce the known plaquette results of 
Di Giacomo and Rossi~\cite{DiGiacomo:1981wt}
and All\'es et al.~\cite{Alles:1998is}\footnote{In fact the result of
this reference is more precise than ours: they obtain
$W_{\mbox{\scriptsize NLO}}=0.033910993(1)$ while we quote
$W_{\mbox{\scriptsize NLO}}=0.033911(1)$ in the table.} and the small planar
Wilson loop results of Wohlert, Weisz and Wetzel~\cite{Weisz:1984bn}.
Boldface values have been calculated by us
for the first time. We also calculated tree-level Wilson loops for
the Symanzik-Weisz (SyW)~\cite{Weisz:1983zw} and
Iwasaki (I)~\cite{Iwasaki:1984cj} actions. These results are displayed in
Tabs.~\ref{TableSmallLoops2} and \ref{TableSmallLoops3}, respectively,
together with the
${\mathcal O}(g^4)$ values of Iso and Sakai~\cite{Iso:1987dj} (italicized)
which we have not validated ourselves.

\begin{table}[hbt]
\caption{\label{TableSmallLoops}
Pure gauge Wilson loops with the Wilson gauge action.
The different contributions 
are defined in Eqs.\ (\ref{eq:wilex2}) -- (\ref{eq:wilex5}).
The values $W_{\mbox{\scriptsize NLO}}$
apply to Wilson loops in the fundamental representation of
$SU(3)$ gauge theory. Boldface values are calculated here
for the first time.}
\begin{ruledtabular}
\begin{tabular}{cccc}
loop  & $W_T$ &  $W_{\Sigma}$ & $W_{\mbox{\scriptsize NLO}}$ \\
\hline
$1\times 1$    & 0.25  &0.0100109(4)&0.033911(1)\\
chair&\textbf{0.3922(1)}&\textbf{0.02204(2)}&\textbf{0.01629(1)}\\
parall.& \textbf{0.4267(1)} & 
	 \textbf{0.02730(2)} & \textbf{0.01128(1)} \\
$1\times 2$  & 0.43110(6) & 0.02463(2) & 0.00382(1)\\
$2\times 2$  & 0.68466(8) & 0.05303(2) &-0.12043(9)\\
\end{tabular} 
\end{ruledtabular}
\end{table} 

\begin{table}[hbt]
\caption{\label{TableSmallLoops2}
Pure gauge Wilson loop results with the Symanzik-Weisz action.
Italicized values have been obtained in Ref.~\cite{Iso:1987dj}.
$W_a$ and $W_b$ are defined in Eq.~(\ref{eq:wfac}) and
$W_{\mbox{\scriptsize NLO}}$ applies to $N=3$.}
\begin{ruledtabular}
\begin{tabular}{ccccc}
loop  &  $W_T^{SyW}$& $W_a^{SyW}$ & $W_b^{SyW}$&$W_{\mbox{\scriptsize NLO}}^{SyW}$ 
\\
\hline
$1\times 1$    & 0.18313(1) & \textit{-0.001133} & \textit{0.001333 }&
\textit{-0.00788}\\
chair & \textbf{0.28850(8)} & --- & --- & --- \\
parall. &  \textbf{0.3162(1)} & --- & --- & --- \\
$1\times 2$  & 0.33130(6) & \textit{-0.00678} & \textit{0.00830}&
\textit{-0.0469}
\end{tabular}
\end{ruledtabular}
\end{table} 

\begin{figure}[hbt]
\includegraphics[width=8cm]{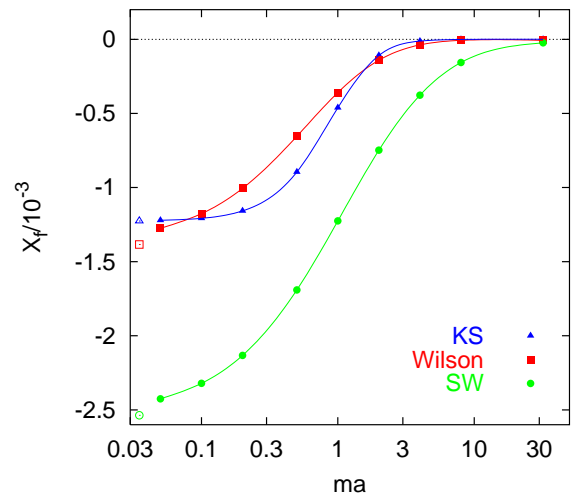}
\caption{
\label{PertPlaquette}
$n_f$ contributions to the one-loop plaquette for KS, Wilson,
and SW quarks.
All $X_f$ interpolate smoothly between the massless
values (open symbols) and zero
as the quark mass is increased.}
\end{figure}

We are able to reproduce the
known contributions to the plaquette for massless
KS~\cite{Heller:1985eq} and Wilson~\cite{Hamber:1983ft} quarks with increased
precision. The calculation of the SW contribution\footnote{Our result
has already been used in
Refs.~\cite{Marcantonio:2001fc,Booth:2001qp,Bali:2001yh}.},
as well as all results for massive quarks are new.
We also calculate the one-loop $n_f$ piece of the plaquette
with the Iwasaki gauge action which
has been used, for instance, 
in the dynamical fermion simulations of the CP-PACS group~\cite{Yoshie:2000wd}.
We display a selection of $m=0$ results
in Tab.~\ref{tab:fermiloop}.
In Fig.~\ref{PertPlaquette} we show the mass dependence of
$X_f$ for the case of the Wilson
gluonic action, combined with all three different fermionic actions.
The open symbols correspond to the respective $m=0$ limits.
The numbers that are plotted in the figure are also displayed
in Tab.~\ref{tab:fermiloop2}.

\begin{table}[hbt]
\caption{\label{TableSmallLoops3} The same as Tab.~\ref{TableSmallLoops2}
for the Iwasaki gauge action.}
\begin{ruledtabular}
\begin{tabular}{ccccc}
loop  & $W_T^I$ & $W_a^I$ & $W_b^I$ & $W_{\mbox{\scriptsize NLO}}^I$\\
\hline
$1\times 1$    & 0.10514(2) & \textit{-0.002269 } & \textit{0.003142 }&
\textit{-0.01536}\\
chair &  \textbf{ 0.16676(4)} & --- & --- & --- \\
parall. & \textbf{0.18494(4)} & --- & --- & --- \\
$1\times 2$  & 0.20166(4) & \textit{-0.00653} & \textit{0.00881}&
\textit{-0.04441}
\end{tabular} 
\end{ruledtabular}
\end{table} 

\begin{table}[hbt]
\caption{\label{tab:fermiloop}
Massless fermionic contribution to Wilson loops for the different
quark actions.
The plaquette is displayed
for both Wilson (W) and Iwasaki (I) gluonic actions. All other loops
are just shown for the Wilson gauge action.
$X_f$ are defined in Eqs.~(\ref{eq:wilex1}), (\ref{eq:wilex12})
and (\ref{eq:wilex2}).}
\begin{ruledtabular}
\begin{tabular}{ccccc}
loop &$X_{f}^{(KS)}/10^{-3}$&$X_{f}^{(0)}/10^{-3}$&$X_{f}^{(1)}/10^{-3}$&$X_{f}^{(2)}/10^{-3}$\\
\hline
W: $1\times 1$
           & -1.2258(7)
	   & -1.392(3)  
           & \textbf{0.0404(5)}   
           & \textbf{-1.1927(2)} \\
I: $1\times 1$
	   & \textbf{-0.2592(4)}
           & \textbf{-0.294(2)}
           & \textbf{0.00156(2)}
           & \textbf{-0.3396(2)} \\ 
chair
& \textbf{-2.674(2)}
& \textbf{-2.974(7)}
& \textbf{0.0724(9)} 
& \textbf{-1.9672(6)}\\
parall.
& \textbf{-2.890(2)}
& \textbf{-3.31(1)}
& \textbf{0.101(1)}
& \textbf{-2.3248(6)}\\
$1\times 2$
& -2.357(1)
& -2.652(6)
& \textbf{0.113(1)}
& \textbf{-2.9518(6)}\\
$2\times 2$
& -4.113(3)
& -4.76(2)
& \textbf{0.322(2)}
& \textbf{-6.684(1)}
\end{tabular}
\end{ruledtabular}
\end{table}

\begin{table}[hbt]
\caption{\label{tab:fermiloop2}
Mass dependence of the fermionic contribution
to the plaquette for the Wilson gluonic action.}
\begin{ruledtabular}
\begin{tabular}{ccccc}
$ma$ &$X_{f}^{(KS)}/10^{-3}$&$X_{f}^{(0)}/10^{-3}$&$X_{f}^{(1)}/10^{-3}$&$X_{f}^{(2)}/10^{-3}$\\
\hline
0.00&-1.2258(7) &-1.392(3)&0.0404(5)&-1.1927(2)    \\
0.05&-1.2211(7) &-1.275(2)&0.0267(5)&-1.1769(2)    \\
0.10&-1.2076(6) &-1.176(2)&0.0154(5)&-1.1603(2)    \\
0.20&-1.1576(6) &-1.006(2)&-0.0011(4)&-1.1256(2)   \\
0.50&-0.8954(5) &-0.655(2)&-0.0207(3)&-1.0158(2)   \\
1.00&-0.4617(4) &-0.359(2)&-0.0208(2)&-0.8455(1)   \\
2.00&-0.1091(2) &-0.141(1)&-0.0104(1)&-0.5967(1)   \\
4.00&-0.01212(6)&-0.038(1)&-0.00261(4)&-0.33661(4) \\
8.00&-0.00093(2)&-0.0070(6)&-0.00034(1)&-0.14920(2)
\end{tabular}
\end{ruledtabular}
\end{table}

In the region $ma<0.25$, which is most relevant for
lattice simulations, this dependence is relatively weak and can be
parametrized by,
\begin{equation}
X_f(ma)=X_f(0)+b\,ma+c(ma)^2,
\end{equation}
with $X_f(0)$ as in the first row of Tab.~\ref{tab:fermiloop2}
and $b\approx 0$, 0.0024 and 0.0024 and $c\approx 0.0017$, $-0.0023$
and $-0.0017$ for
KS, Wilson and SW fermions, respectively.
As expected the mass dependence remains weak for
any Wilson loop with linear dimensions that are small compared to the
inverse quark mass.

\section{The perturbative static potential\label{sec_potl}}
After a historical survey of the existing literature
we will briefly explain our method and the notations that we adopt,
before
presenting results on the ${\mathcal O}(g^4)$ potential.
We will then discuss limiting cases, including the
continuum limit and lattice artefacts as well as mass dependent
terms in the conversion between
lattice and $\overline{MS}$ schemes at finite lattice spacing. We conclude
with paragraphs on the perturbative $\beta$-shift, boosted lattice
perturbation theory and the static
source self energy.
\subsection{The method and definitions}
The one-loop potential has first been calculated for pure $SU(2)$ gauge
theory by M\"uller and R\"uhl~\cite{MullerRuhl}. Subsequently,
a closed formula for pure $SU(N)$ gauge theory with the standard Wilson
as well as Symanzik improved gauge
actions was derived by Weisz and Wohlert~\cite{Weisz:1984bn}.
The coefficients of the perturbative expansion have first been
evaluated numerically by Altevogt and Gutbrod~\cite{Altevogt:1995cj}
for $SU(2)$
pure gauge theory with Wilson glue 
on isotropic and anisotropic lattices.

Here we restrict our calculations to the case of the Wilson gauge action
on an isotropic lattice
but include massive Wilson, SW and KS fermions as well as $N>2$.
We also improve
the numerical precision with respect to earlier results and incorporate
off-axis lattice separations ${\mathbf R}$ of the colour sources.
We define the static potential,
\begin{eqnarray}
aV({\mathbf R}a)&=&-\lim_{T\rightarrow\infty}\frac{d\ln W({\mathbf R},T)}{dT}\\
&=&V_1({\mathbf R})g^2+V_2({\mathbf R})g^4+\cdots\\
&=&v_1({\mathbf R})\alpha_L+v_2({\mathbf R})\alpha_L^2,\label{eq:vis}
\end{eqnarray}
where $\alpha_L=g^2/(4\pi)$, $v_i({\mathbf R})=(4\pi)^i V_i({\mathbf R})$.
$W({\mathbf R},T)$ denotes a (generalized) rectangular Wilson loop
with the temporal extent of $T$ lattice units.
The expansion of the potential
in terms of $g^2$ is suitable for calculations
in lattice perturbation theory and for the comparison with data from lattice
simulations, while the expansion in terms of $\alpha_L$ turns out to
be more convenient to relate our results to those obtained in a
continuum scheme.

The only gluon exchanges that contribute to the $T$ dependence
of $W({\mathbf R},T)$ in the limit
$T\rightarrow\infty$ [at least up to ${\mathcal O}(g^6)$]
are those between temporal lines
of a Wilson loop. After exploiting translational invariance in time,
Eqs.~(\ref{eq:w210}) and (\ref{eq:w22}) result in,
\begin{equation}
\label{eq:v1per}
V_1({\mathbf R})=C_F\int \frac{d^3\!q}{(2\pi)^3}
\frac{2\sin^2\left(\frac{{\mathbf q}{\mathbf R}a}{2}\right)}
{\hat{\mathbf q}^2}.
\end{equation}
The ${\mathcal O}(g^4)$ coefficient $V_2$ consists of
two parts: one contribution from $W_2$, i.e.\ the vacuum polarization
tensor $\Pi$, inserted into Eq.~(\ref{eq:v1per}),
and a second
contribution that originates from the ``non-Abelian'' part $W_{NA}$
of $W_4$ [Eq.~(\ref{eq:w40})].
The spider diagrams, $W_3$, do not contribute to $V({\mathbf R})$
at this order.
A closed formula for $V_2$ can be found in Ref.~\cite{Weisz:1984bn},
and we extend this calculation by incorporating the fermionic
contributions to the vacuum polarization tensor.
In our calculation we produce $V_2$ automatically from the Feynman rules
and find agreement with the
published analytic form.

The potential can be factorized into an interaction part and
a part that is due to the static source self energy,
$V_S=2\delta m_{\mbox{\scriptsize stat}}$,
\begin{equation}\label{eq:edcba}
V({\mathbf R}a)=V_{\mbox{\scriptsize int}}({\mathbf R}a)+V_S.
\end{equation}
Since in perturbation theory $V_{\mbox{\scriptsize int}}\rightarrow 0$
as $r\rightarrow\infty$ we can identify the self energy $V_S$ with
$\lim_{r\rightarrow\infty}V(r)$ and write,
\begin{eqnarray}\label{eq:abcde}
aV_S&=&V_1(\infty)g^2+V_2(\infty)g^4+\cdots,\\
aV_{\mbox{\scriptsize int}}({\mathbf R}a)&=&V_{L,1}({\mathbf R})g^2+
V_{L,2}({\mathbf R})g^4+\cdots,
\end{eqnarray}
where $V_{L,i}({\mathbf R})=V_i({\mathbf R})-V_i(\infty)$.
The limit $R\rightarrow\infty$ can easily be realized by replacing
the factors $\sin^2[{\mathbf q}{\mathbf R}a/2]
=[1-\cos({\mathbf q}{\mathbf R}a)]/2$
within the external
3-dimensional Fourier transformations
in the expressions for $V_1$ [cf.\ Eq.~(\ref{eq:v1per})] and $V_2$ by
the constant value $1/2$.

\begin{table}
\caption{\label{tab:potpure}Gluonic contributions to the static potential,
[see Eqs.~(\ref{eq:defv1}) -- (\ref{eq:defv2})].
The last line ($R=\infty$) contains the respective
contributions to the static self energy $aV_S$.
The values for $V_2$ in the last column only apply to the case
$N=3, n_f=0$.}
\begin{ruledtabular}
\begin{tabular}{ccccc}
${\mathbf R}$&$V_T/10^{-2}$&$V_{\Pi}/10^{-2}$&$V_{NA}/10^{-2}$&$V_2/10^{-2}$\\\hline
(1,0,0)&16.6667&1.1920 (1)&0.1499 (1)& 7.3134 (3)\\
(2,0,0)&20.9842&1.6316 (2)&0.5470 (2)&10.1862 (6)\\
(3,0,0)&22.5186&1.8062 (3)&0.7435 (3)&11.3546 (8)\\
(4,0,0)&23.2442&1.8993 (5)&0.8526 (3)&11.9605(11)\\
(5,0,0)&23.6630&1.9608 (6)&0.9194 (4)&12.3335(14)\\
(6,0,0)&23.9366&2.0028 (7)&0.9663 (4)&12.5873(17)\\
(7,0,0)&24.1300&2.0345 (9)&1.0012 (5)&12.7740(19)\\
(8,0,0)&24.2742&2.0585(10)&1.0288 (5)&12.9176(22)\\\hline
(1,1,0)&19.7540&1.4532 (1)&0.4186 (1)& 9.2308 (4)\\
(1,1,1)&20.9152&1.5697 (2)&0.5443 (2)&10.0378 (5)\\
(2,1,0)&21.6800&1.6853 (2)&0.6337 (2)&10.6602 (7)\\
(2,1,1)&22.0766&1.7253 (3)&0.6858 (2)&10.9546 (7)\\
(2,2,0)&22.4676&1.7827 (3)&0.7384 (3)&11.2833 (8)\\
(2,2,1)&22.6572&1.8034 (4)&0.7647 (3)&11.4297 (9)\\
(2,2,2)&23.0080&1.8527 (4)&0.8163 (3)&11.7292(10)\\
(3,3,0)&23.4016&1.9153 (5)&0.8776 (4)&12.0863(12)\\
(4,2,0)&23.4889&1.9266(52)&0.8914(11)&12.1607(81)\\
(4,2,2)&23.6538&1.9530 (6)&0.9186 (4)&12.3137(15)\\
(3,3,3)&23.7512&1.9679 (7)&0.9345 (4)&12.4023(15)\\
(4,4,0)&23.8686&1.9884 (7)&0.9546 (4)&12.5162(16)\\
(4,4,2)&23.9512&2.0011 (8)&0.9695 (4)&12.5944(17)\\
(4,4,4)&24.1286&2.0310 (9)&1.0018 (5)&12.7675(20)\\\hline
$\infty$&25.2731&\multicolumn{2}{c}{$\sum:$ 3.5459(15)}&14.1122 (31)
\end{tabular}
\end{ruledtabular}
\end{table}

\begin{table}
\caption{\label{tab:potfermi0}
Fermionic contributions to the static potential
$V_f$ [see Eqs.~(\ref{eq:defv2}) and (\ref{eq:defvf})]
for massless KS [$V_f^{(KS)}$] and Wilson-SW [$V_f^{(i)}$, see
Eq.~(\ref{eq:defvf})] quarks.}
\begin{ruledtabular}
\begin{tabular}{ccccc}
${\mathbf R}$&$V_f^{(KS)}/10^{-3}$&$V_f^{(0)}/10^{-3}$&$V_f^{(1)}/10^{-3}$&$V_f^{(2)}/10^{-3}$\\\hline
(1,0,0)&-1.0830 (1)&-1.1318 (1)&0.0563 (2)&-1.5070 (1)\\
(2,0,0)&-1.4893 (3)&-1.6698 (1)&0.1469 (3)&-2.6859(16)\\
(3,0,0)&-1.7004 (5)&-1.9272 (2)&0.1978 (3)&-3.0033 (3)\\
(4,0,0)&-1.8123 (7)&-2.0723 (2)&0.2331(34)&-3.1578 (3)\\
(5,0,0)&-1.8900 (9)&-2.1659 (2)&0.2479 (4)&-3.2589 (3)\\
(6,0,0)&-1.9427(11)&-2.2328 (3)&0.2611 (4)&-3.3294 (4)\\
(7,0,0)&-1.9851(13)&-2.2818 (3)&0.2706 (4)&-3.3840 (4)\\
(8,0,0)&-2.0155(16)&-2.3207 (3)&0.2779 (5)&-3.4234 (4)\\\hline
(1,1,0)&-1.3418 (2)&-1.4784 (1)&0.0970 (2)&-2.0409 (2)\\
(1,1,1)&-1.4609 (3)&-1.6415 (1)&0.1252 (3)&-2.3156 (2)\\
(2,1,0)&-1.5687 (4)&-1.7723 (1)&0.1637 (3)&-2.6703 (3)\\
(2,1,1)&-1.6169 (4)&-1.8383 (2)&0.1755 (4)&-2.7298 (3)\\
(2,2,0)&-1.6774 (5)&-1.9121 (2)&0.1944 (4)&-2.8773 (3)\\
(2,2,1)&-1.7043 (5)&-1.9474 (2)&0.2010 (4)&-2.9099 (3)\\
(2,2,2)&-1.7615 (6)&-2.0182 (2)&0.2171 (4)&-3.0166 (4)\\
(3,3,0)&-1.8355 (8)&-2.1045 (2)&0.2366 (5)&-3.1557 (4)\\
(4,2,0)&-1.8499 (9)&-2.1233(11)&0.2389(10)&-3.1917 (8)\\
(4,2,2)&-1.8824 (9)&-2.1624 (3)&0.2476 (5)&-3.2294 (4)\\
(3,3,3)&-1.9005(10)&-2.1852 (3)&0.2526 (5)&-3.2536 (5)\\
(4,4,0)&-1.9269(11)&-2.2144 (3)&0.2584 (5)&-3.2955 (5)\\
(4,4,2)&-1.9428(12)&-2.2351 (3)&0.2624 (5)&-3.3180 (5)\\
(4,4,4)&-1.9804(14)&-2.2810 (3)&0.2714 (5)&-3.3720 (5)\\\hline
$\infty$&-2.3359 (4)&-2.6808 (4)&0.3266(10)&-3.7174(12)
\end{tabular}
\end{ruledtabular}
\end{table}

\begin{table}
\caption{\label{tab:potfermi1}
Fermionic contributions to the static potential
for massive KS and Wilson-SW quarks.}
\begin{ruledtabular}
\begin{tabular}{cccccc}
${\mathbf R}$&$ma$&$V_f^{(KS)}/10^{-3}$&$V_f^{(0)}/10^{-3}$&$V_f^{(1)}/10^{-3}$&$V_f^{(2)}/10^{-3}$\\\hline
       & 0  &-1.0830(1)&-1.1318(1)&0.0563(2)&-1.5070(1)\\
       &0.01&-1.0828(1)&-1.1120(1)&0.0525(4)&-1.5034(2)\\
       &0.03&-1.0811(1)&-1.0733(1)&0.0437(4)&-1.4952(2)\\
(1,0,0)&0.04&-1.0789(2)&-1.0545(1)&0.0396(4)&-1.4911(2)\\
       &0.05&-1.0777(1)&-1.0360(1)&0.0357(4)&-1.4869(2)\\
       &0.10&-1.0627(1)&-0.9491(1)&0.0177(2)&-1.4650(1)\\
       &0.25&-0.9720(1)&-0.7374(1)&-0.0142(1)&-1.3966(1)\\\hline
       & 0  &-1.3418(2)&-1.4784(1)&0.0970(2)&-2.0409(2)\\
       &0.01&-1.3416(2)&-1.4517(1)&0.0893(2)&-2.0360(2)\\
       &0.03&-1.3394(2)&-1.3997(1)&0.0754(2)&-2.0254(2)\\
(1,1,0)&0.04&-1.3367(4)&-1.3743(1)&0.0685(2)&-2.0202(1)\\
       &0.05&-1.3349(2)&-1.3495(1)&0.0623(3)&-2.0147(2)\\
       &0.10&-1.3151(2)&-1.2324(1)&0.0347(2)&-1.9871(1)\\
       &0.25&-1.1980(2)&-0.9489(1)&-0.0168(2)&-1.8976(1)\\\hline
       & 0  &-1.4609(3)&-1.6415(1)&0.1252(3)&-2.3156(2)\\
       &0.01&-1.4607(3)&-1.6114(1)&0.1157(3)&-2.3099(2)\\
       &0.03&-1.4580(3)&-1.5525(1)&0.0977(3)&-2.2985(2)\\
(1,1,1)&0.04&-1.4555(5)&-1.5238(2)&0.0894(2)&-2.2926(1)\\
       &0.05&-1.4530(3)&-1.4958(1)&0.0812(3)&-2.2868(2)\\
       &0.10&-1.4306(3)&-1.3635(1)&0.0462(3)&-2.2563(2)\\
       &0.25&-1.2990(3)&-1.0441(1)&-0.0182(3)&-2.1567(2)\\\hline
       & 0  &-1.4893(3)&-1.6698(1)&0.1469(3)&-2.6859(16)\\
       &0.01&-1.4889(3)&-1.6391(1)&0.1368(7)&-2.6778 (5)\\
       &0.03&-1.4863(3)&-1.5789(1)&0.1155(7)&-2.6644(5)\\
(2,0,0)&0.04&-1.4831(5)&-1.5496(1)&0.1057(7)&-2.6576(5)\\
       &0.05&-1.4813(3)&-1.5209(1)&0.0961(7)&-2.6508(5)\\
       &0.10&-1.4582(3)&-1.3855(1)&0.0539(3)&-2.6146(2)\\
       &0.25&-1.3242(3)&-1.0586(1)&-0.0214(3)&-2.4994(2)\\\hline
       & 0  &-1.5687(4)&-1.7723(1)&0.1637(3)&-2.6703(3)\\
       &0.01&-1.5684(4)&-1.7393(2)&0.1512(3)&-2.6641(2)\\
       &0.03&-1.5654(4)&-1.6748(1)&0.1283(3)&-2.6511(3)\\
(2,1,0)&0.04&-1.5615(6)&-1.6433(2)&0.1172(3)&-2.6460(5)\\
       &0.05&-1.5596(4)&-1.6125(1)&0.1071(3)&-2.6380(3)\\
       &0.10&-1.5345(4)&-1.4671(1)&0.0621(3)&-2.6038(3)\\
       &0.25&-1.3883(4)&-1.1170(1)&-0.0198(3)&-2.4912(2)
\end{tabular}
\end{ruledtabular}
\end{table}
\begin{table}
\caption{\label{tab:potfermi2}
Tab.~\ref{tab:potfermi1} continued.}
\begin{ruledtabular}
\begin{tabular}{cccccc}
${\mathbf R}$&$ma$&$V_f^{(KS)}/10^{-3}$&$V_f^{(0)}/10^{-3}$&$V_f^{(1)}/10^{-3}$&$V_f^{(2)}/10^{-3}$\\\hline
       & 0  &-1.6169(4)&-1.8383(2)&0.1755(4)&-2.7298(3)\\
       &0.01&-1.6164(4)&-1.8039(2)&0.1623(3)&-2.7233(2)\\
       &0.03&-1.6132(4)&-1.7364(1)&0.1374(4)&-2.7104(3)\\
(2,1,1)&0.04&-1.6099(6)&-1.7033(2)&0.1257(3)&-2.7037(2)\\
       &0.05&-1.6072(4)&-1.6712(2)&0.1145(4)&-2.6971(3)\\
       &0.10&-1.5801(4)&-1.5192(1)&0.0664(4)&-2.6624(3)\\
       &0.25&-1.4269(4)&-1.1536(1)&-0.0208(3)&-2.5479(3)\\\hline
       & 0  &-1.6774(5)&-1.9121(2)&0.1944(4)&-2.8773(3)\\
       &0.01&-1.6768(5)&-1.8760(2)&0.1794(4)&-2.8706(3)\\
       &0.03&-1.6735(5)&-1.8052(2)&0.1520(4)&-2.8569(3)\\
(2,2,0)&0.04&-1.6694(7)&-1.7705(2)&0.1388(4)&-2.8500(3)\\
       &0.05&-1.6668(5)&-1.7366(2)&0.1266(4)&-2.8431(3)\\
       &0.10&-1.6380(5)&-1.5768(1)&0.0729(4)&-2.8066(3)\\
       &0.25&-1.4754(5)&-1.1931(1)&-0.0228(3)&-2.6864(3)\\\hline
       & 0  &-1.7004(5)&-1.9272(2)&0.1978(3)&-3.0033(3)\\
       &0.01&-1.7001(5)&-1.8907(2)&0.1837(8)&-2.9979(6)\\
       &0.03&-1.6963(5)&-1.8192(2)&0.1550(8)&-2.9832(6)\\
(3,0,0)&0.04&-1.6923(6)&-1.7842(2)&0.1416(9)&-2.9758(6)\\
       &0.05&-1.6893(5)&-1.7499(2)&0.1289(9)&-2.9682(6)\\
       &0.10&-1.6595(5)&-1.5883(1)&0.0718(3)&-2.9282(2)\\
       &0.25&-1.4922(5)&-1.2006(1)&-0.0269(3)&-2.8014(2)\\\hline
       & 0  &-1.7043(5)&-1.9474(2)&0.2010(4)&-2.9099(3)\\
       &0.01&-1.7036(5)&-1.9104(2)&0.1858(3)&-2.9033(2)\\
       &0.03&-1.7003(5)&-1.8379(2)&0.1567(4)&-2.8894(3)\\
(2,2,1)&0.04&-1.6975(7)&-1.8024(2)&0.1435(3)&-2.8826(2)\\
       &0.05&-1.6932(5)&-1.7678(2)&0.1306(4)&-2.8754(3)\\
       &0.10&-1.6629(5)&-1.6041(1)&0.0751(4)&-2.8387(3)\\
       &0.25&-1.4954(5)&-1.2118(1)&-0.0236(3)&-2.7174(3)\\\hline
       & 0  &-2.3333(4)&-2.6808(4)&0.3266(10)&-3.7174(12)\\
       &0.01&-2.300 (6)&-2.6123(3)&0.2978 (2)&-3.7080(20)\\
       &0.03&-2.284 (6)&-2.4805(3)&0.2432 (4)&-3.6886 (4)\\
$\infty$&0.04&-2.256 (5)&-2.4180(3)&0.2166(12)&-3.6794(10)\\
       &0.05&-2.241 (6)&-2.3582(2)&0.1934(10)&-3.6728 (2)\\
       &0.10&-2.1603(3)&-2.0893(2)&0.0956 (2)&-3.6268 (2)\\
       &0.25&-1.845 (4)&-1.5082(2)&-0.0536(2)&-3.4728 (2)
\end{tabular}
\end{ruledtabular}
\end{table}

We write,
\begin{eqnarray}\label{eq:defv1}
V_1({\mathbf R})&=&C_FV_T({\mathbf R})\\\label{eq:defv2}
V_2({\mathbf R})&=&C_F\left\{\frac{N}{2}[V_{\Pi}({\mathbf R})
+V_{NA}({\mathbf R})]\right.\\\nonumber&&+\left.\frac{2N^2-3}{24N}
V_T({\mathbf R})+n_fV_f({\mathbf R})\right\},
\end{eqnarray}
where
$V_{\Pi}$ denotes the gluonic vacuum polarization and
$V_{NA}$ is a contribution, specific to non-Abelian gauge theories,
where the ordering of gluon vertices along the Wilson loop
has to be considered. In the case of KS quarks we denote the
fermionic vacuum polarization contribution to the potential by
$V_f^{(KS)}$ while for Wilson-SW quarks we split this contribution
into three parts,
\begin{equation}\label{eq:defvf}
V_f({\mathbf R})=V_f^{(0)}({\mathbf R})+
c_{SW}V_f^{(1)}({\mathbf R})+c_{SW}^2V_f^{(2)}({\mathbf R}).
\end{equation}

\subsection{Results on the potential and violations of rotational symmetry}
In Tab.~\ref{tab:potpure} we display $V_T$, $V_{\Pi}$
and $V_{NA}$ as well as $V_2$ for $N=3, n_f=0$
for small on- and off-axis distances ${\mathbf R}$.
At the origin all $V$ are zero
and $V_T(1,0,0)=1/6$.
In Tab.~\ref{tab:potfermi0} we display
the corresponding results on
the different $V_f$'s for massless quarks while in
Tabs.~\ref{tab:potfermi1} -- \ref{tab:potfermi2}
the results for distances $R\leq 3$ as well as for $R\rightarrow\infty$
are shown for quarks of various masses.

\begin{figure}[hbt]
\includegraphics[width=8cm]{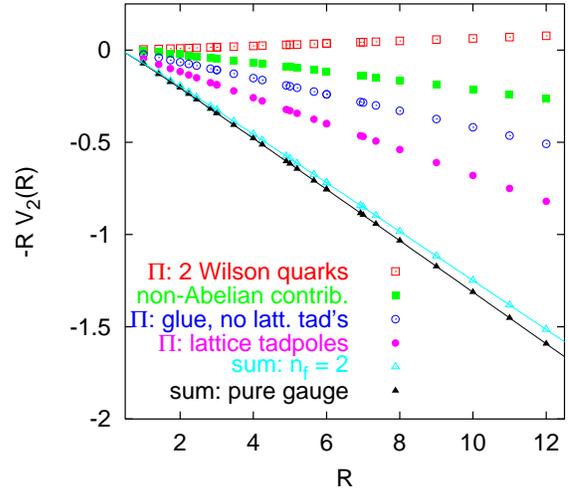}
\caption{
\label{PertGaugePotl} 
${\mathcal O}(g^4)$ contribution to the static potential, multiplied by $-R$
for $N=3$.
The solid triangles are the sum of the pure gauge contributions:
non-Abelian, vacuum polarization and the lattice tadpole contribution to
the vacuum polarization. The open triangles incorporate 2 flavours of Wilson
fermions.
}
\end{figure}

\begin{figure}[hbt]
\includegraphics[width=8cm]{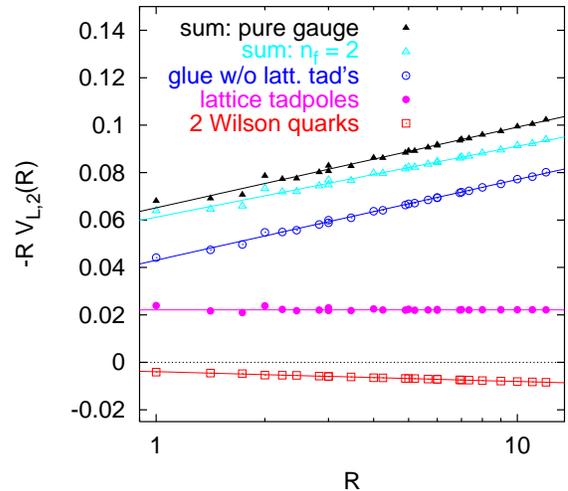}
\caption{
\label{PertGaugePotl2} 
The same as Fig.~\protect\ref{PertGaugePotl} but with the self energy
contribution subtracted, $V_{2,L}(R)=V_2(R)-V_2(\infty)$, and with
a logarithmic scale.}
\end{figure}

In Fig.~\ref{PertGaugePotl} we separately
display the different ${\mathcal O}(g^4)$ contributions to the
quantity $-R V_2({\mathbf R})$ as a function of $R$.
The factor $-R$ results in a cancellation of the
leading order Coulomb behaviour. Note that the lattice tadpole
contribution to the vacuum polarization (solid circles) is numerically
equally important as the sum of the remaining vacuum polarization diagrams
and the non-Abelian contribution. As one would na\"{\i}vely
expect, the effect of
fermions (open squares) goes into the opposite direction, relative to
the pure gauge contributions. 
In a continuum calculation using dimensional regularization
$-R V_2({\mathbf R})$ contains only a logarithmic dependence on R, but
in the lattice calculation the $R$ dependence of the points is 
dominated by a linear piece arising from the static source self energy
contribution to $V_2({\mathbf R})$, $V_{S,2}=V_2(\infty)$.
In Fig.~\ref{PertGaugePotl2}
we subtract this term, before multiplying the result
with $-R$. Note the logarithmic scale.
The static source self-energy was previously known to 
order $\alpha^2$ with massless
Wilson and SW quarks~\cite{Martinelli:1999vt} and
to ${\mathcal O}(\alpha^3)$ in the pure gauge
theory~\cite{DiRenzo:2001nd,Trottier:2001vj}. The
mass-dependence, as well as the results for KS sea quarks, are new.

In the limit
$r\gg a$, i.e.\ $R=r/a\gg 1$ rotational symmetry
should be restored such that we are able to compare
our result to calculations performed in a continuum scheme
(continuous curves in Figs.~\ref{PertGaugePotl} and \ref{PertGaugePotl2}).
The potential has been calculated
in the $\overline{MS}$ scheme to
order $\alpha^3$~\cite{Schroder:1999vy,Melles:2000dq}.
The conversion between the
$\overline{MS}$ and the lattice scheme has been worked
out to this order too for Wilson-SW fermions~\cite{Christou:1998ws,
Bode:2001uz}
(for KS fermions only to ${\mathcal O}(\alpha^2)$~\cite{Weisz:1981pu,Luscher:1995zz}),
and $V_S$
vanishes by definition to all orders in dimensional regularization.
The form of the large $R$ expectation is worked out
in Appendix~\ref{sec:larger} [Eqs.~(\ref{eq:resvc2}) -- (\ref{eq:respvc2})].
Here we display the parametrization of the curves in the limit of massless 
sea quarks:
\begin{eqnarray}
-rV_{L,2}({\mathbf r}/a)
&\stackrel{r/a\rightarrow\infty}{\longrightarrow}&
-rV_{c,2}(r)\label{eq:vvrru}\\\nonumber
&=&\frac{C_F}{4\pi}\left[b_1+a_1^R+2\beta_0\ln (r/a)\right].
\end{eqnarray}
The constants $b_1$
and $a_1^R=a_1+2\gamma\beta_0$ are defined
in Eqs.~(\ref{eq:b1}) and (\ref{eq:a1}), respectively.
$b_1$ originates from the conversion between the $\overline{MS}$
and the lattice scheme, Eq.~(\ref{eq:cmsbarl}).
The logarithmic running is proportional to the $\beta$-function coefficient
$\beta_0=(11N-2n_f)/(12\pi)$.

We can define an effective Coulomb coupling $C_F\alpha_R$
from the potential:
\begin{eqnarray}
-r\left[V(r)-V_S\right]&=&C_F\alpha_R\left(r^{-1}\right)\\
&=&C_F\alpha_L-RV_{L,2}(R)g^4.
\end{eqnarray}
Note that in this limit,
$C_F\alpha_L=-RV_{L,1}(R)$.
In Fig.~\ref{PertGaugePotl2} the enhancement of this effective coupling,
$C_F\alpha_R$, is demonstrated,
relative to the tree-level value as a function of $R$.
The points are indeed consistent with the logarithmic running
proportional to $2\beta_0C_F\ln R/(4\pi)$ that is expected from
the QCD $\beta$-function, Eq.~(\ref{eq:beta})
[as well as from Eq.~(\ref{eq:vvrru}) above].
The lattice tadpole terms do not contribute to this running
but renormalize the overall value of the coupling.
Lattice simulations are usually performed around $g^2\approx 1$,
where a fit to quenched lattice data~\cite{Bali:2000vr,Bali:2000gf}
yields $e\approx C_F\alpha_R[1/(3a)]\approx 0.3$.
The tree-level expectation in this case, however, is substantially smaller:
$C_F\alpha_R=C_F\alpha_L\approx 0.106$. As one can read off from the
figure, the one-loop value at $R=3$ adds about 0.08 to this, but still
the non-perturbative
result is underestimated by perturbation theory in terms of the bare coupling
$g^2$ by more than 30~\%. We will discuss the improvement resulting from
the use of so-called boosted perturbation theory in Sec.~\ref{sec_data} below.

\begin{figure}[hbt]
\includegraphics[width=8cm]{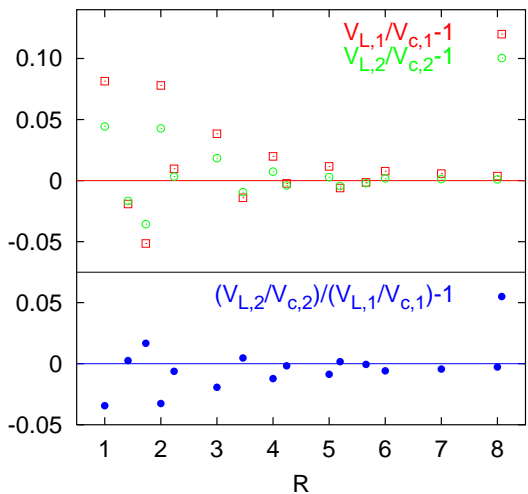}
\caption{
\label{fig:rotsym} 
Visualization of lattice artefacts. $V_{L,1}$ and $V_{L,2}$ denote the
tree-level and ${\mathcal O}(g^4)$ lattice potentials, where
the self energy pieces $V_i(\infty)$ have been subtracted.
$V_{c,i}$ are the respective large $R$ expectations.
}
\end{figure}

At small $R$ values the lattice results scatter around the continuous curves.
We shall investigate these violations of rotational symmetry
in more detail: in the top half of Fig.~\ref{fig:rotsym} we plot ratios of 
lattice and continuum ${\mathcal O}(g^2)$ (open squares, $V_{L,1}/V_{c,1}-1$)
and ${\mathcal O}(g^4)$
(open circles, $V_{L,2}/V_{c,2}-1$) contributions for on-axis as
well as for some off-axis
distances ${\mathbf R}$ for $n_f=0$; the relative violations of
rotational symmetry become smaller with bigger $R$ and the
one-loop violations are smaller than the tree-level ones.
In the lower part of the figure the ratio of the two ratios is displayed,
which amounts to replacing the continuum $1/R$ term that multiplies
the logarithmic running of $V_{L,2}$
by a lattice ``$[1/{\mathbf R}]$'' function.
In doing so we isolate those lattice artefacts that
appear specifically at order $g^4$; for instance,
the lattice tadpole contributions cancel from this combination.
The sign is opposite to
that of the tree-level differences,
which explains the weaker relative violations at ${\mathcal O}(g^4)$
in the comparison with ${\mathcal O}(g^2)$
in the upper part of the figure.

\begin{figure}[hbt]
\includegraphics[width=8cm]{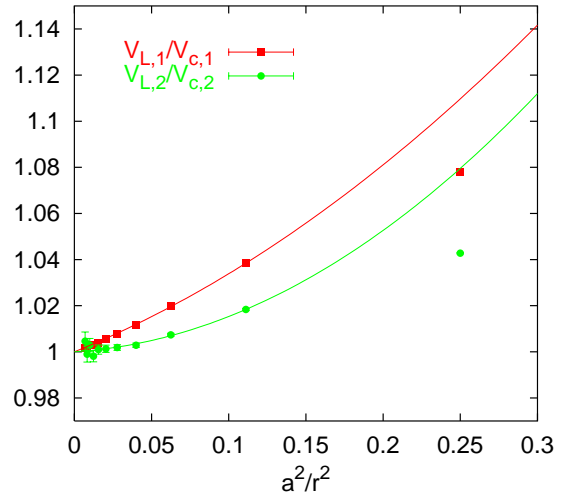}
\caption{
\label{fig:lattarte} 
The ratio of lattice and continuum order $\alpha$ and $\alpha^2$
pure gauge $SU(3)$ potentials
as a function of $a^2/r^2$ for on-axis separations.}
\end{figure}

\subsection{The continuum limit}
\label{sec:contu}
The limit of large $R=r/a$ has two interpretations: taken at a fixed
lattice spacing $a$ (sufficiently small for perturbation theory still to
be reliable at a distance $r$)
continuum perturbative predictions should be met.
We have already discussed this scenario above and indeed demonstrated
this agreement at large $R$
in Figs.~\ref{PertGaugePotl2} and~\ref{fig:rotsym}.
On the other hand taking the limit of large $R$ at fixed physical $r$
corresponds to the continuum limit $a\rightarrow 0$. In
Fig.~\ref{fig:lattarte} we investigate the
approach to the continuum limit by
displaying ratios of lattice and continuum
perturbation theory results $V_{L,i}/V_{c,i}$ for $i=1,2$ versus
$1/R^2=a^2/r^2$ for pure $SU(3)$ gauge theory
and on-axis separations $R\geq 2$.
For the Wilson gauge action we expect
lattice artefacts to be a polynomial in $a^2$ and indeed no
linear term is found. The solid curves represent
fits that are quadratic plus quartic
in the lattice spacing $a$. For off-axis separations
such as ${\mathbf R}\parallel(1,1,0)$ we find a similar picture, but with
different $a^2$ and $a^4$ coefficients.

\begin{figure}[hbt]
\includegraphics[width=8cm]{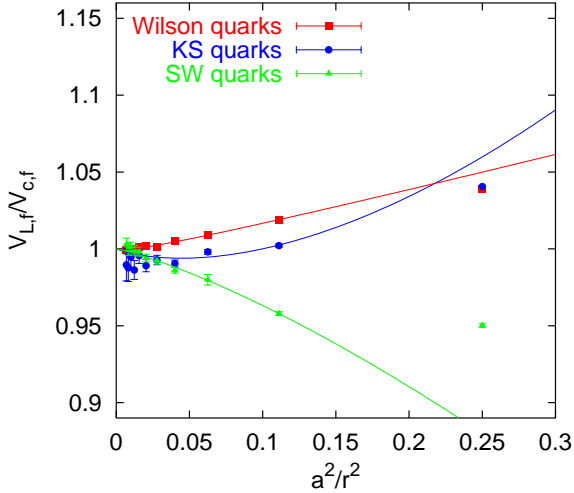}
\caption{
\label{fig:lattartef}
Lattice artefacts on the fermionic contribution to the on-axis potential
as a function of $a^2/r^2$ for different actions of massless
quarks. Note that the leading
order contribution is proportional to $a/r$ for Wilson fermions.
}
\end{figure}

\begin{figure}[hbt]
\includegraphics[width=8cm]{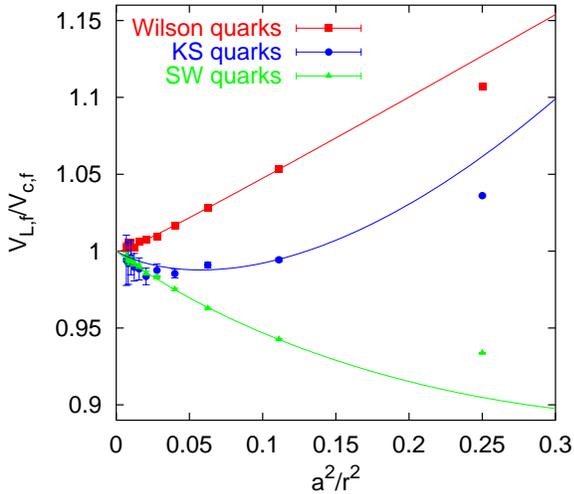}
\caption{
\label{fig:lattartef2}
Lattice artefacts on the fermionic contribution to the on-axis potential
as a function of $a^2/r^2$ for $ma=0.1$.
}
\end{figure}

The same
comparison is displayed for the fermionic contributions
$V_{L,f}({\mathbf R})=V_f({\mathbf R})-V_f(\infty)$
and $V_{c,f}(r)=\lim_{a\rightarrow 0}V_{L,f}(r/a)$
alone for massless quarks in Fig.~\ref{fig:lattartef}.
The fit curves are quadratic plus cubic for SW 
($c_{SW}=1$ to this order in perturbation theory) and KS fermions
while for Wilson fermions we indeed require the expected linear term and
attempt a linear plus quadratic fit. The coefficient of the linear
term turns out to be so small that the quadratic term already
dominates for
distances $a/r>0.1$. We see that in spite of the more favourable
functional dependence on $a$
the perturbative lattice artefacts of SW fermions
are numerically bigger over the whole displayed
$R$ range than those of Wilson
fermions. However, the KS action ``outperforms'' both Wilson and SW
for this observable.
As the quark mass is increased the violations of
the continuum symmetry become more pronounced in the Wilson case while
for the other two actions the change is only small, as
a comparison of 
Fig.~\ref{fig:lattartef} with Fig.~\ref{fig:lattartef2},
that corresponds to a quark mass $ma=0.1$, reveals:
``improvement'' enables one to simulate on coarser lattices
at the same physical quark mass.

\begin{figure}[hbt]
\includegraphics[width=8cm]{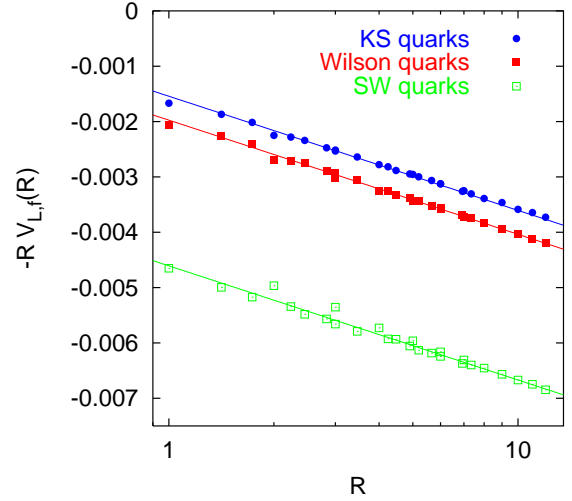}
\caption{
\label{fig:massless} 
The same as Fig.~\protect\ref{PertGaugePotl2} but for the fermionic
contribution alone.}
\end{figure}

\begin{figure}[hbt]
\includegraphics[width=8cm]{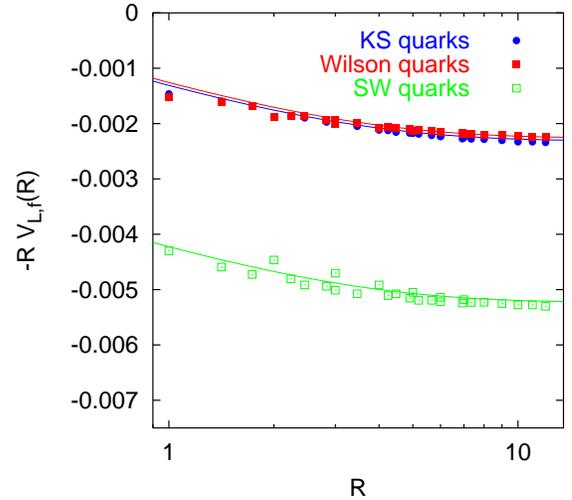}
\caption{
The same as Fig.~\protect\ref{fig:massless} but for $ma=0.1$.}
\label{fig:mass0.1} 
\end{figure}

{}From
Fig.~\ref{PertGaugePotl2} it is obvious
that for realistic values of $n_f$ the potential is
dominated by the gluonic contributions and
the additional violations of rotational symmetry due to sea quarks
that we have just discussed
are an interesting but numerically subleading effect.

We shall now return to the limit where continuum perturbation
theory is met, i.e.\ we investigate the behaviour at fixed
(small) $a$ and large $r=Ra$. In the massless case this limit and
the limit of large $R=r/a$ at fixed $r$ discussed above are equivalent,
up to non-perturbative effects.
However, as soon as a second scale
$ma>0$ is introduced,
the situation $r>m^{-1}$ becomes distinguishable from
$r<m^{-1}$.
In Fig.~\ref{fig:massless} we again display the self energy subtracted
fermionic contribution
$V_{L,f}({\mathbf R})$, this time
multiplied by $-R$, in analogy to Fig.~\ref{PertGaugePotl2}.
This combination, multiplied by $C_Fn_f$ isolates
the shift that is induced onto
an effective coupling $C_F\alpha_R$, due to sea quarks.
The logarithmic slope of the three curves that are displayed in
the figure is determined by
the fermionic contribution to $\beta_0$ and is therefore universal.
The inclusion of sea quarks not only reduces $\beta_0$, relative to
the pure gauge case,
and hence slows down the running of $\alpha_R$ but
it also decreases
the absolute value of the ${\mathcal O}(g^4)$ coefficient.
In the case of massive quarks, however,
this effect is (over)compensated by an increase in the coupling $g^2$,
relative to the pure gauge case, if we require the same physics
at scales $r\gg m^{-1}$ where the sea quarks decouple. This will be
discussed in Sec.~\ref{sec:nf2c} below.

\begin{figure}[hbt]
\includegraphics[width=8cm]{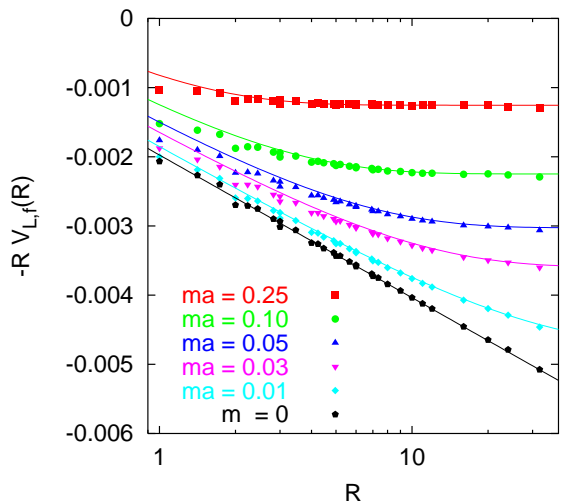}
\caption{
\label{fig:massive} 
The same as Fig.~\protect\ref{fig:massless} for
Wilson fermions of various masses.}
\end{figure}

While for massless flavours the coupling runs logarithmically,
for massive quarks
the running gradually
switches off
around\footnote{The constant $C_0\approx 5.2$~\cite{Brodsky:1999fr}
is introduced in
Eq.~(\ref{eq:a1run}).}
$R\approx 0.3/(ma)\approx 1/(\sqrt{C_0}ma)$: physics at length scales
$Ra\gg m^{-1}$ is insensitive
to the presence of massive sea quarks, at least in perturbation theory.
In Fig.~\ref{fig:mass0.1} we again compare the contributions to the
effective Coulomb couplings resulting {}from the three
different fermionic actions but for $ma=0.1$.
At this mass value Wilson and
KS results accidentally happen to be very close to each other.
In Fig.~\ref{fig:massive} we investigate the mass dependence
of the fermionic term somewhat more systematically by studying the example of
Wilson quarks alone. In order to relate
the lattice to a continuum scheme at large $R$,
we not only have to subtract the (sea quark mass-dependent) static source
self energy contribution to eliminate terms proportional to $R$ from the 
figure but also the vertical offset, $K_1(ma)$, at small $R$ changes due to
a mass dependent term that appears in
the conversion factor $b_1$ between the
$\overline{MS}$ and the lattice scheme at finite $a$.
We will discuss this effect in some detail in Sec.~\ref{sec:k1} below.

In the limit  $r\gg m^{-1}$ all curves indeed approach
a constant value: the running of the effective coupling is not affected
by the presence of massive sea quark flavours anymore.
The resulting potential is the same
as that in the pure gauge case, at least in perturbation theory, albeit
with a different overall
normalization of the effective Coulomb coupling.
We shall discuss this shift of the QCD coupling constant
at a given scale
and the possibility of matching to the quenched theory
in Sec.~\ref{sec:beta} below.

While at large $R$ the lattice data and the continuum curve
nicely coincide with each other [once the offset $K_1(ma)$
has been determined
from the large $R$ data and subtracted], at small $R$ we observe
large ${\mathcal O}(am)$ lattice artefacts, in addition to
mass independent ${\mathcal O}(a)$ effects: even at the relatively
small value $ma=0.03$
the lattice points lie systematically below the large $R$ expectation.
Fig.~\ref{fig:mass0.1} confirms this qualitative pattern for the
other two fermionic actions.

\subsection{``$\Delta K_1$''}
\label{sec:k1}
As we have seen above the choice of lattice action not only
affects the quality of rotational symmetry at small $R$ but also
the overall normalization.
The $\overline{MS}$ scheme is related to the lattice scheme via
\begin{equation}
\alpha_{\overline{MS}}(a^{-1})=\alpha_L+b_1\alpha^2+\cdots,
\end{equation}
with the conversion factor
\begin{equation}
\label{eq:b111}
b_1=-\pi/(2N)+k_1N+K_1(ma)n_f.
\end{equation}
The numerical constant $k_1$ is known for a variety
of gluonic actions and $K_1(0)$ is known for Wilson, SW and KS quarks and
independent of the gluonic action, cf.\ Eqs.~(\ref{eq:K1}) -- (\ref{eq:K12}).

The coefficients of the $\beta$-function in the $\overline{MS}$-scheme
as well as in the lattice scheme in the continuum limit do not depend
on the quark mass: both are ``mass-independent'' renormalization
schemes. In lattice simulations it is often
worthwhile to analyse quantities prior to an extrapolation to 
the continuum limit. One such example is determinations of the QCD
running coupling from small Wilson loops~\cite{El-Khadra:1992vn,davi1,davi2,Spitz:1999tu,Booth:2001qp,davi3}
that are ill-defined in this limit. We will discuss this technique
in Sec.~\ref{sec_alpha_s} below. Another example is simulations
within the framework of an effective field theory like
NRQCD~\cite{Thacker:1991bm}
that requires a finite momentum cut-off.

At finite $a$ and $m>0$ the lattice scheme becomes
mass-dependent, as indicated by the argument of the function,
\begin{equation}
K_1(ma)=K_1(0)+\Delta K_1(ma),
\end{equation}
within Eq.~(\ref{eq:b111}).
In this case universality is lost
and the coefficients of the perturbative $\beta$-function
acquire additional contributions that, like $\Delta K_1$,
will in general depend on the
dimensionful observable that is studied:
\begin{equation}
\beta_0^{\overline{MS}}-\beta_0^L=
n_f\frac{d\Delta K_1(ma)}{d\ln a^2}.
\end{equation}
By definition, $\Delta K_1(0)=0$.
However, the behaviour of $\Delta K_1$ is also constrained
in the limit $ma\rightarrow\infty$ where
the sea quarks decouple and therefore
$\beta_0^{\overline{MS}}=(11N)/(12\pi)$ while
$\beta_0^L=\beta_0^{\overline{MS}}-n_f/(6\pi)$, i.e.,
\begin{equation}
\label{eq:k1shiftt}
\Delta K_1\stackrel{ma\rightarrow\infty}{\longrightarrow}
-\frac{1}{3\pi}\ln(Dma),
\end{equation}
with a dimensionless constant $D$ that we calculate
in Sec.~\ref{sec:beta} below.

We have seen above that in the
limit $r\ll m^{-1}$ the behaviour of the massless theory is emulated
while the running of the potential for $r\gg m^{-1}$ is effectively
quenched:
the only scale that is relevant in perturbation theory
in these limits is the distance
$r$ (or, equivalently, momentum $q\approx r^{-1}$) and therefore
lattice artefacts disappear like $(a/r)^{\nu}$ [or $(aq)^{\nu}$]
with some positive integer power $\nu$, whose value depends on the
lattice action,
as $r\rightarrow\infty$.
However, in the intermediate range of quark masses
lattice corrections $(am)^{\nu}$ become relevant and the universality of the
$\beta$-function is lost.

The function $K_1(ma)$ of Eq.~(\ref{eq:b1})
can in principle
be read off from figures such as Figs.~\ref{fig:mass0.1}
and \ref{fig:massive} up to
short distance lattice artefacts:
$K_1(ma)=-C_F/(4\pi)\,V_{L,2}(1)-\sqrt{C_0}ma+{\mathcal O}[(ma)^2,a^{\nu}]$
[cf.\ Eqs.~(\ref{eq:vvrru}), (\ref{eq:a1rm}) and (\ref{eq:a1rm2})].
For the (improved)
SW action not only the lattice artefacts are much more pronounced
and qualitatively different from the Wilson and KS cases but also this
overall offset $K_1$ is enhanced (although not its mass dependence).

\begin{figure}[hbt]
\includegraphics[width=8cm]{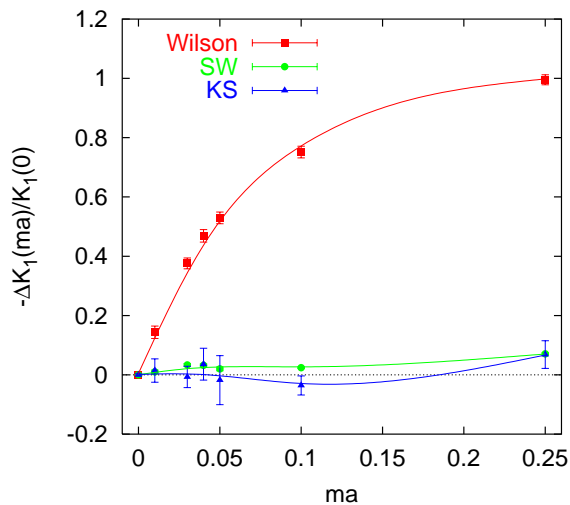}
\caption{
\label{fig:shift}
$\Delta K_1$ determined from the static position space potential, relative
to $K_1(0)$.
}
\end{figure}

\begin{table}
\caption{\label{tab:shift} $\Delta K_1$ as a function of $ma$ for
Wilson, SW and KS quarks. In the last row we display $-K_1(0)$
for comparison.}
\begin{ruledtabular}
\begin{tabular}{cccc}
$ma$&$\Delta K_1^{(KS)}$&$\Delta K_1^{(W)}$&$\Delta K_1^{(SW)}$\\\hline
0.01     &0.0005(13) &0.0121(18)&0.0038(20)\\
0.03     &-0.0002(12)&0.0316(15)&0.0131(19)\\
0.04     &0.0012(18) &0.0394(18)&0.0127(20)\\
0.05     &-0.0006(27)&0.0445(17)&0.0079(26)\\
0.10     &-0.0012(11)&0.0632(17)&0.0095(10)\\
0.25     &0.0023(15) &0.0837(14)&0.0280(14)\\\hline
$-K_1(0)$&0.032983419&0.0841444 &0.3957496 
\end{tabular}
\end{ruledtabular}
\end{table}

We determine
$K_1(ma)$ from two parameter fits,
\begin{eqnarray}
-\frac{4\pi R}{C_F}V_{L,2}&=&A_0+A_1\ln R+\frac{n_f}{3\pi}\mbox{Ein}
\left(C_0^{1/2}maR\right)\nonumber\\
&+&n_f\Delta K_1(ma)+\frac{c}{R^\nu},
\end{eqnarray}
to the $R>9$ on-axis
data points, with one redundant parameter $c$
whose role is to parametrize
any residual lattice artefacts.
The above functional form with the constants $A_0$ and $A_1$
of Eqs.~(\ref{eq:respvc21}) -- (\ref{eq:respvc22}) is motivated
by Eq.~(\ref{eq:resvc2}) of Appendix~\ref{sec:larger}.
The special function $\mbox{Ein}(x)$ denotes the normalized
exponential integral
defined in Eq.~(\ref{eq:defa1r}).

All fits turned out to be stable
within statistical errors under variations of the fit range as well
as consistent with the off-axis data.
The results on $\Delta K_1$ are displayed in Tab.~\ref{tab:shift}.
In Fig.~\ref{fig:shift} these shifts are
plotted relative to $-K_1(0)$, together with cubic splines.
Since $K_1(0)<0$, at large $ma$ the curves will
diverge towards the negative direction. Note that the slope of $\Delta K_1(ma)$
at $ma=0$ has also been determined in Ref.~\cite{Booth:2001qp}
for SW fermions, however, by matching the vacuum polarization
in momentum space, rather
than the position space potential and $\Delta K_1$ is not universal.

For the two ${\mathcal O}(a)$
improved actions $\Delta K_1(ma)$ is small (a few \%), relative
to $K_1(0)$, at least for $ma \leq 1/4$. This is very different in
the case of Wilson fermions, where even at $ma=0.05$ $\Delta K_1$
corrects $K_1(0)$ by more than 50~\%.
Using the static position space potential
as the prescription that defines $\Delta K_1$,
the $ma$ dependence is well parametrized
by $\Delta K_1(x)=ax+bx^2$ with $a = 1.20\pm 0.02$ and
$b=-5.7\pm 0.3$ for $x=ma\leq 0.1$ in the Wilson case.
For SW fermions the fit parameters read
$a=0.44\pm 0.08$ and $b=-3.5\pm 0.8$ while
the KS data are compatible with zero within our accuracy
over the whole range
$x\leq 0.25$ and can be fitted by a straight
line with $a=-0.04\pm 0.04$.

\subsection{The ``$\beta$-shift''}
\label{sec:beta}
We have seen above
that at distances $r\gg m^{-1}$ the running of the coupling is not
affected by the presence of sea quarks anymore: 
at large distances, at least in
perturbation theory, the effect of massive
quarks can be integrated out into a shift of the coupling constant of
the quenched theory. In contrast this is not possible
for a theory with massless quarks, which completely decouples from the
quenched case at all scales.

The theory with massive sea quarks and the quenched theory can
be matched in the infra red by use of
an intermediate mass-dependent scheme,
such as the $R$-scheme, defined through the inter-quark potential
in position space,
\begin{eqnarray}
\alpha_R\left(r^{-1}\right)&=&-\frac{r}{C_F}V_c(r)\\
&\stackrel{R=r/a\rightarrow\infty}{\longrightarrow}&
-\frac{Ra}{C_F}V_{\mbox{\scriptsize int}}(R)
\end{eqnarray}
Inserting the perturbative expansion of $V_{\mbox{\scriptsize int}}(R)$ 
in terms
of $\alpha_L$ one obtains,
\begin{eqnarray}
\alpha_R(\mu)&=&\alpha_L^{(n_f)}+e_1^{(n_f)}(R){\alpha_L^{(n_f)}}^2+\cdots\\
&=&\alpha_L^{(0)}+e_1^{(0)}(R){\alpha_L^{(0)}}^2+\cdots,
\end{eqnarray}
where $\mu=(Ra)^{-1}$, with known coefficients, $e_1^{(n_f)}(R)$.
Note that these coefficients will depend on the function $K_1(ma)$ discussed
above.
The requirement of a unique (static source self-energy subtracted) potential
$V_{\mbox{\scriptsize int}}(R)$
at $R\gg(ma)^{-1}$ then results in,
\begin{equation}
\alpha_L^{(0)}=\left\{1+\left[e_1^{(n_f)}(\infty)-e_1^{(0)}(\infty)\right]
\alpha\right\}\alpha_L^{(n_f)}+\cdots.
\end{equation}
One can now re-write the above equation in terms of
the lattice parameter $\beta=2N/g^2=N/(2\pi \alpha_L)$.
For $n_f$ degenerate quark flavours with mass $m$ and $N=3$ the result reads,
\begin{eqnarray}
\beta^{(n_f)}&=&\beta^{(0)}+\Delta\beta,\\
\label{eq:bshiftt}
\Delta\beta&=&\frac{n_f}{2\pi^2}\left[\ln(Dma)+3\pi\Delta K_1(ma)\right],
\end{eqnarray}
with the numerical values for the constant,
\begin{equation}
D=\sqrt{C_0}\exp\left[3\pi K_1(0)-\frac{5}{6}\right]:
\end{equation}
\begin{eqnarray}
D^W&=&0.448\pm 0.002,\\
D^{SW}&=&0.0238 \pm 0.0001,\\
D^{KS}&=&0.726\pm 0.002,
\end{eqnarray}
for Wilson, SW and KS
fermions, respectively.
Since $\Delta\beta\rightarrow 0$ as $ma\rightarrow\infty$,
the $\ln(Dma)$ term has to
be cancelled by $3\pi\Delta K_1(ma)$ at large $ma$ in Eq.~(\ref{eq:bshiftt}):
the constant $D$ is identical with that appearing within
Eq.~(\ref{eq:k1shiftt}).
We discuss the matching procedure in some more detail in
Appendix~\ref{sec:match}.

Na\"{\i}vely one would assume perturbation theory to
be applicable as long as the relative $\alpha$-shift is small.
To leading non-trivial order in perturbation theory the $\beta$-shift is linear
in $n_f$. At the next order, the situation will be complicated
by additional terms that are proportional to $n_f^2$ as can be seen
from Eq.~(\ref{eq:nextorder}) of Appendix~\ref{sec:match}.

\subsection{Boosted perturbation theory and $q^*$}
Lattice perturbation theory is well known for its bad convergence, partly
due to large contributions from lattice tadpole diagrams.
We recall that $V_2({\mathbf R})$ incorporates such a contribution:
$(2N^2-3)/(24N)V_1({\mathbf R})$ [cf.\ Eq.~(\ref{eq:defv2})].
Hence reordering
the series in terms of a better behaved expansion parameter
like $\alpha_V(q)$, the coupling defined by the
static QCD potential in momentum space,
\begin{equation}
\tilde{V}(q)=-4\pi C_F\frac{\alpha_V(q)}{q^2},
\end{equation}
is desirable in many cases.
In some respect this is similar to the situation in
continuum perturbation theory and resembles
an expansion in terms of a renormalized, rather than a ``bare'',
lattice coupling parameter.
We will refer to such techniques as ``boosted perturbation theory''.

\begin{figure}[hbt]
\includegraphics[width=8cm]{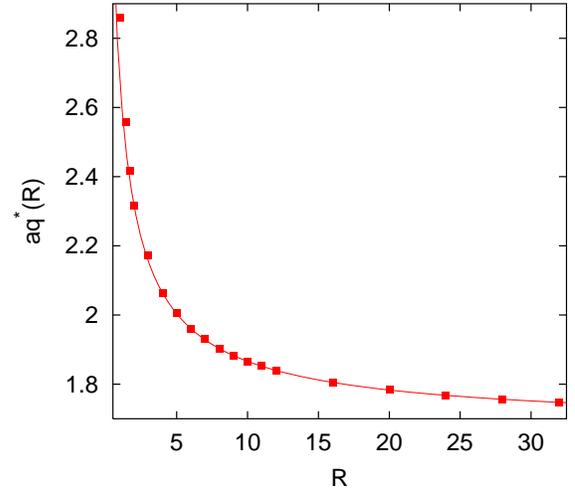}
\caption{
\label{fig:qstar}
$q^*a$ values for the static potential.}
\end{figure}

At ${\mathcal O}(\alpha^2)$ we can write,
\begin{equation}
\alpha_{V}(q^*)=
\alpha_L+\left[a_1+b_1-2\beta_0\ln(q^* a)\right]\alpha^2,
\end{equation}
with $a_1$, $b_1$ and $\beta_0$ as defined in
Eqs.~(\ref{eq:a1}), (\ref{eq:a1run}),
Eqs.~(\ref{eq:b1}) -- (\ref{eq:K1})
and Eq.~(\ref{eq:beta0}), respectively.
The ``optimal'' scale $q^*$ depends on the
underlying process.
It has been argued by Brodsky, Lepage and Mackenzie
(BLM)~\cite{Brodsky:1983gc,lepage}
that the logarithmic average of the momenta exchanged at tree-level
is a particularly good choice for the scale $q^*$ to be used
within a one-loop calculation.
The scale optimization procedure can also be generalized to
higher order perturbative calculations~\cite{Brodsky:1998mf,Hornbostel:2002af}.
We illustrate the original recipe for the case of the position space
potential. Here the tree-level calculation, Eq.~(\ref{eq:v1per}),
yields,
\begin{equation}
\ln\left[aq^*({\mathbf R})\right]=\frac{\int d^3\!q\ln q
\sin^2\left(\frac{{\mathbf q}{\mathbf R}a}{2}\right)/
\hat{\mathbf q}^2}{\int d^3\!q
\sin^2\left(\frac{{\mathbf q}{\mathbf R}a}{2}\right)/
\hat{\mathbf q}^2}.
\end{equation}
For $R\gg 1$ we find,
\begin{equation}\label{eq:qstarp}
aq^*(R)=
f_1+\frac{f_2}{R}+f_3\frac{\ln R}{R}+\cdots,
\end{equation}
with the numerical constants,
\begin{equation}
f_1\approx 1.672, f_2\approx 1.00, f_3\approx 0.41.
\end{equation}
The constants $f_2$ and $f_3$ have been obtained from a fit to $R\geq 4$
data while the $r\rightarrow\infty$ limit $f_1$ has been calculated directly.
The fit curve and $q^*$ values are displayed in Fig.~\ref{fig:qstar}.
Note that violations of rotational invariance in $q^*({\mathbf R})$ are
remarkably small.
Analogously we can obtain $q^*$ values for the inter-quark
force.
We display results for the plaquette $\Box$, the chair, the ``parallelogram'',
the ``rectangle'' $W(1,2)$, the static self energy
$\delta m_{\mbox{\scriptsize stat}}=V_S/2$,
the potential and the force $F_{R_1R_2}=a[V(R_2)-V(R_1)]/(R_2-R_1)$
at selected distances
in Tab.~\ref{tab:qstar}. In addition, the $q^*$ values for
the ``tadpole improved'' quantities
$m_{\mbox{\scriptsize stat}}+\ln\Box/(4a)$ as well as
$V(1)+\ln\Box/(2a)$ and $V(2)+\ln\Box/(2a)$
are included.

One can further convert between the $V$ and the $\overline{MS}$ scheme:
\begin{equation}
\alpha_{\overline{MS}}(\mu)=\alpha_V(q^*)-
\left[a_1+2\beta_0\ln(\mu/q^*)\right]\alpha^2+\cdots.
\end{equation}
It has been argued~\cite{Brodsky:1983gc} that the scale
\begin{equation}
\mu=e^{-5/6}q^*,
\end{equation}
at which the $n_f$ dependence of $a_1$ above
is cancelled (in the massless case)
was the optimal choice of scale for this conversion.
Note that the ratio $\mu/q^*$ above
is independent of the number of colours $N$ or
flavours $n_f$. The $\mu$ values are also included
into the table.

The average plaquette turns out to be
the most ultra violet quantity,
with $q^*\approx 3.402/a$, followed by the potential
$V(R=1)$. Due to the  self energy contribution
$V_S=V(\infty)=2\delta m_{\mbox{\scriptsize stat}}$,
$q^*$ is bounded from below by $1.672/a$ for the
potential at all distances. However, in the case of the force
$q^*_F\rightarrow 0$ as $R\rightarrow\infty$. As one
would expect $q^*$ also approaches
zero at large distances for the potential if $V_S$ is subtracted.
We find for instance $q^*a\approx 0.59$ for $V(1)-V_S$ as opposed to
$q^*a\approx 2.86$ for $V(1)$ alone.

\begin{table}[hbt]
\caption{\label{tab:qstar}BLM scales $q^*$ for some quantities.}
\begin{ruledtabular}
\begin{tabular}{ccc}
quantity&$q^*a$&$\mu a=e^{-5/6}q^*a$\\\hline
$\Box$       &3.402&1.478\\
chair        &3.300&1.434\\
``parallelogram''&3.128&1.360\\
$W(1,2)$     &3.066&1.332\\
$\delta m_{\mbox{\scriptsize stat}}$&1.672&0.727\\
$V(1)$       &2.860&1.243\\
$V(2)$       &2.317&1.007\\
$F_{12}$     &1.025&0.445\\
$F_{23}$     &0.904&0.393\\
$\delta m_{\mbox{\scriptsize stat}}+\ln\Box/(4a)$&0.835&0.363\\
$V(1)+\ln\Box/(2a)$&1.701&0.739\\
$V(2)+\ln\Box/(2a)$&1.316&0.572\\
$V(\sqrt{2})$&2.558&1.112 \\
$V(\sqrt{3})$&2.417&1.050\\
$V(3)$       &2.173&0.944\\
$V(4)$       &2.062&0.896\\
$V(5)$       &2.007&0.872\\
$V(6)$       &1.959&0.851\\
$V(7)$       &1.930&0.839\\
$V(8)$       &1.902&0.826\\
$V(16)$      &1.805&0.784\\
$V(24)$      &1.768&0.768\\
$V(28)$      &1.757&0.764\\
$V(32)$      &1.748&0.760
\end{tabular}
\end{ruledtabular}
\end{table}

\subsection{The static source self energy}
We define,
\begin{equation}
\label{eq:vsss}
V_S=\lim_{R\to \infty} V(R) = 2 \delta m_{\mbox{\scriptsize stat}},
\end{equation}
where $\delta m_{\mbox{\scriptsize stat}}$ is the lattice pole
mass of a fundamental static colour source, e.g.\ a quark in the
$m\rightarrow\infty$ limit. The above relation only holds in perturbation
theory where the interaction energy vanishes as $r\rightarrow\infty$.

The static (or residual) mass
$\delta m_{\mbox{\scriptsize stat}}$ has been calculated to
${\mathcal O}(\alpha^3)$
for pure gauge theory~\cite{DiRenzo:2001nd,Trottier:2001vj}
as well as to ${\mathcal O}(\alpha^2)$ for
massless Wilson and SW quarks~\cite{Martinelli:1999vt}.
This has enabled a number of authors to obtain the $b$
quark mass $\overline{m}_b(\overline{m_b})$ from
lattice simulations of 
static-light mesons. Since the sources are static the ${\mathcal O}(\alpha)$
result can be obtained from a tree-level calculation and we count this
as leading order (LO), discarding the trivial ${\mathcal O}(1)$
value $\delta m_{\mbox{\scriptsize stat}}=0$. The one-loop
${\mathcal O}(\alpha^2)$ result is then next to leading order (NLO)
and the two-loop ${\mathcal O}(\alpha^3)$ value NNLO. Note that 
the counting conventions employed in some of the literature
differ from the one defined above, in which
${\mathcal O}(\alpha^n)$
corresponds to an N${}^m$LO calculation with $m=n-1$ for which
diagrams involving $m$ loops have to be computed.
For instance in Ref.~\cite{Martinelli:1999vt} the value $m=n$ is used,
creating the (wrong) impression that a two-loop calculation is required
to obtain the ${\mathcal O}(\alpha^2)$ result.

\begin{table}[hbt]
\caption{\label{tab:mself}Fermionic contributions to the static
self energy of a fundamental source.
The labelling conventions are defined in Eqs.~(\ref{eq:vsss}) --
(\ref{eq:vsst}).}
\begin{ruledtabular}
\begin{tabular}{ccccc}
$ma$&$\delta M_f^{(KS)}/10^{-3}$&$\delta M_f^{(0)}/10^{-3}$&
$\delta M_f^{(1)}/10^{-3}$&$\delta M_f^{(2)}/10^{-3}$\\\hline
 0  &-1.1667 (2)&-1.3404 (2)&0.1633 (5)&-1.8587 (6)\\
0.01&-1.500  (3)&-1.3061 (2)&0.1489 (1)&-1.8540(11)\\
0.03&-1.142  (3)&-1.2402 (2)&0.1216 (2)&-1.8443 (2)\\
0.04&-1.128  (2)&-1.2090 (1)&0.1083 (6)&-1.8397 (5)\\
0.05&-1.120  (3)&-1.1791 (1)&0.0967 (5)&-1.8364 (1)\\
0.10&-1.0801 (1)&-1.0447 (1)&0.0478 (1)&-1.8134 (1)\\
0.25&-0.922  (2)&-0.7541 (1)&-0.0268(1)&-1.7364 (1)\\
1   &-0.2962 (1)&-0.2336 (2)& --- & --- \\
2   &-0.06131(3)&-0.08264(6)& --- & --- \\
4   & ---       &-0.02030(2)& --- & ---
\end{tabular}
\end{ruledtabular}
\end{table}

\begin{figure}[hbt]
\includegraphics[width=8cm]{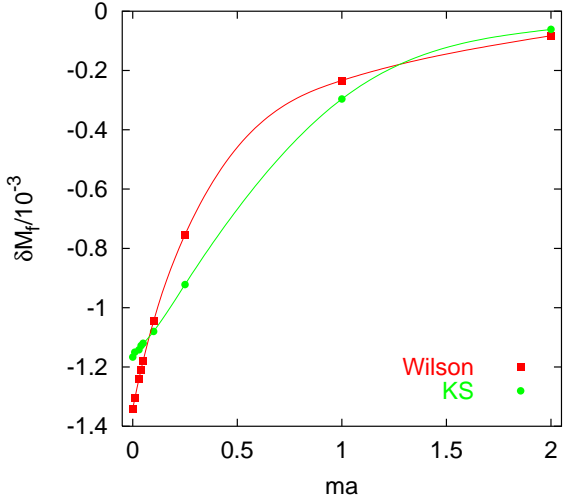}
\caption{
\label{fig:deltam}
The fermionic contribution to the static self energy
$\delta m_{\mbox{\tiny stat}}$ for Wilson and KS quarks as
a function of the quark mass.
}
\end{figure}

\begin{figure}[hbt]
\includegraphics[width=8cm]{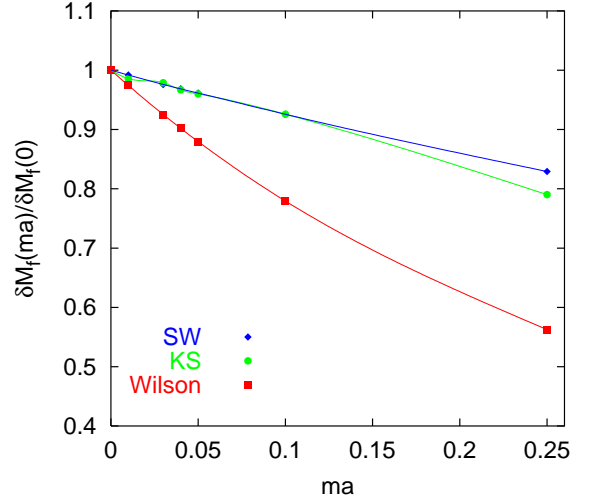}
\caption{
\label{fig:deltam2}
The fermionic contribution to the static self energy
for Wilson, SW and KS quarks,
normalized to the respective zero quark mass results.}
\end{figure}

We have computed the static quark mass shift as a function of the
mass of three species of sea quarks: KS, Wilson and SW.
The results, most of which can also be read off
{}from the last seven lines of Tab.~\ref{tab:potfermi2},
are displayed in Tab.~\ref{tab:mself}.
The labelling conventions are identical to those of
Eqs.~(\ref{eq:abcde}) -- (\ref{eq:defvf}):
\begin{eqnarray}
a\delta m_{\mbox{\scriptsize stat}}&=&\delta M_1 g^2 +\delta M_2 g^4,\\
\delta M_2&=&\delta M_{pg}+C_Fn_f\delta M_f,\label{eq:55}
\end{eqnarray}
where
\begin{eqnarray}
\delta M_1&=&\frac{C_F}{2}V_1(\infty),\\
\delta M_{pg}&=&\frac{C_F}{2}\left\{\frac{N}{2}\left[V_{\Pi}(\infty)+V_{NA}(\infty)
\right]\right.\nonumber\\
&+&\left.\frac{2N^2-3}{24N}V_T(\infty)\right\}
\end{eqnarray}
and $\delta M_f=\delta M_f^{(KS)}$ for KS quarks and
\begin{equation}
\label{eq:vsst}
\delta M_f=\delta M_f^{(0)}+c_{SW}\delta M_f^{(1)}+c_{SW}^2\delta
M_f^{(2)}
\end{equation}
for Wilson-SW fermions.
We find the numerical values,
\begin{eqnarray}
\delta M_1&\approx& 0.126365504\, C_F\\
\delta M_{pg}&\approx&0.05292\, C_F,
\end{eqnarray}
where the latter number applies to $SU(3)$ gauge theory
(cf.\ the last line of Tab.~\ref{tab:potpure}). Note that this result
differs in the least significant digit from the value
obtained in Ref.~\cite{Martinelli:1999vt} that can be translated into
$\delta M_{pg}\approx 0.05297C_F$.
The mass dependence of $\delta M_f$ [which in Eq.~(\ref{eq:55})
is accompanied by a
factor $C_Fn_f$] is visualized in Fig.~\ref{fig:deltam} for
Wilson and KS quarks.
$\delta M_f$ approaches zero like $1/(ma)$ as $ma\rightarrow \infty$.

Finally, in Fig.~\ref{fig:deltam2} we compare the results on
$\delta M_f(ma)$ for quark masses $ma\leq 0.25$,
normalized to $\delta M_f(0)$, from all
three quark actions with each other. While in absolute terms
$\delta M_f$ for SW quarks turns out to be
much bigger than for the other two actions
(cf.\ Tab.~\ref{tab:mself}) we find the relative variation
with the quark mass to be much weaker for both ${\mathcal O}(a)$ improved
actions than for Wilson fermions. This is consistent with
our observations for small Wilson loops and $\Delta K_1$ above.
For $ma\leq 0.1$ $\delta M_f(ma)$ is well parametrized by
a quadratic function,
\begin{equation}
\delta M_f(ma)=\delta M_f(0)+b\,ma+c(ma)^2,
\end{equation}
with $\delta M_f(0)=-0.0011667(3), -0.0013404(2)$ and $ -0.0030351(11)$,
$b=0.00101(8)$, 0.00350(1) and 0.00245(5) and $c=-0.0014(8)$, $-0.00543(5)$
and $-0.0020(4)$
for KS, Wilson and SW quarks, respectively.

\begin{table}[hbt]
\caption{\label{tab:const}
Fermionic contributions to the static
self energy as defined in Eqs.~(\ref{eq:selft}) --
(\ref{eq:selftend}). The first three rows refer to the expansion of the
tadpole improved observable in terms of $\alpha_L$, the last rows
to an expansion in terms of $\alpha_{\overline{MS}}(0.363/a)$.}
\begin{ruledtabular}
\begin{tabular}{ccccc}
&KS&Wilson&SW\\\hline
$a_f$&-0.1853&-0.2090&-0.5051\\
$b_f$&0.21  &0.61  &0.39  \\
$c_f$&-0.38  &-1.02  &-0.33  \\\hline
$\overline{a}_f$&-0.03499&-0.05860&-0.3548\\
$\overline{b}_f$&0.26&-0.67&-0.081\\
$\overline{c}_f$&-0.38&5.08&3.41
\end{tabular}
\end{ruledtabular}
\end{table}

Starting from the expansion of the plaquette,
\begin{equation}
\Box=1-c_1\alpha_L-c_2\alpha^2+\cdots,
\end{equation}
we can rearrange the above expansion of the self energy in the
following way,
\begin{eqnarray}
\label{eq:selft}
a\delta m_{\mbox{\scriptsize stat}}&=&-\frac{\ln\Box}{4}+m_1^{t}\alpha_L
+ m_2^{t}\alpha_L^2\\\label{eq:bbboo}
&=&-\frac{\ln\Box}{4}+m_1^{t}\alpha_{\overline{MS}}(\mu)
+\overline{m}_2^{t}\alpha^2,
\end{eqnarray}
where $\mu=q^*e^{-5/6}=0.363/a$,
$m_1^t=(4\pi)\delta M_1-c_1/4$, $m_2^t=(4\pi)^2\delta M_2-c_2/4-c_1^2/8$.
We will refer to practices such as adding the $\ln\Box$ term
non-perturbatively and
subtracting it perturbatively as ``tadpole improvement'' of the
observable while in the
second step, Eq.~(\ref{eq:bbboo})
we ``boost'' the perturbation theory. Here we choose to
convert everything into the $\overline{MS}$ scheme.
We prefer this to plaquette based schemes
or the $V$ scheme as now all mass dependence is made explicit in the
coefficient function $\overline{m}_2^{t}(ma)$.
A conversion of the result into
another scheme of choice including mass dependent schemes
is easily possible with the help of
Appendix~\ref{AppendixMSlat}.
We will obtain the $\overline{MS}$ coupling
from the average plaquette and other quantities
in Sec.~\ref{sec_alpha_s} below.

For $SU(3)$ gauge theory with $n_f$ quark flavours
of masses $ma\leq 0.1$ we find,
\begin{eqnarray}
m_1^t&=&1.0700768,\\
m_2^{t}&=&7.6104+n_f\left[a_f+b_f
ma+c_f(ma)^2\right],\\
\overline{m}_2^{t}&=&-0.5839+n_f\left[\overline{a}_f+\overline{b}_f
ma+\overline{c}_f(ma)^2\right],\label{eq:selftend}
\end{eqnarray}
with the constants $a_f$, $b_f$, $c_f, \ldots$ of Tab.~\ref{tab:const}.
Tadpole improvement reduces the pure gauge value
$\delta m_{pg}\approx 11.14$
to $m_{pg}^t\approx 7.61$ and boosting reduces this further down to
$\overline{m}_{pg}^t\approx -0.584$.
The fermionic coefficients $\delta m_f\mapsto a_f\mapsto \overline{a}_f$
also undergo a reduction in each step.
The NNLO coefficient
$\delta m_3\approx 86\mapsto m_3^t\approx 67.5\mapsto
\overline{m}_3^{t}\approx 2.85$
is only known for $n_f=0$.

When applying the above result
to extract say $m_b$ from lattice simulations in the static limit
one can readily ignore the
mass dependent terms for Wilson
fermions, which are ${\mathcal O}(a)$ effects. However, it
does not harm to include them either. In the cases of massive
KS and SW sea quarks which are ${\mathcal O}(a)$ improved
at least the $b_f$s have to be included.

\section{Comparison with non-perturbative data}
\label{sec:np}
We shall compare and apply our perturbative calculations to
non-perturbative
results obtained in lattice simulations. For this purpose we
will use several data sets that have been obtained by different collaborations.
All quenched reference data are from simulations of
one of the co-authors of this
article and collaborators
(Ref.~\cite{Bali:2000gf} and references therein) while $n_f=2$ data
have been provided by the SESAM/T$\chi$L
Collaboration~\cite{Bali:2000vr,Bali:2002} (Wilson fermions),
the UKQCD Collaboration~\cite{Allton:2001sk,Booth:2001qp} (SW fermions)
and the MILC Collaboration~\cite{Tamhankar:2000ce}
(KS fermions\footnote{While KS quarks are only defined
for multiples of four mass-degenerate fermion flavours, many
authors have attempted to emulate $n_f=2$ (or even $n_f=3$) by 
using the (positive)
square root of the $n_f=4$ fermionic determinant in their simulation.
The MILC Collaboration adopt this strategy.
However, it is not clear that the resulting lattice action
corresponds to a local field theory like QCD.
In our analysis
we shall set this problem aside. We remark
that the perturbation theory generated by the use of this action
indeed corresponds to replacing $n_f=4$ by
$n_f=2$ to all orders, at least as long as no external fermion
lines are encountered. The comparison with simulation results, however,
might be meaningless should no universal continuum limit exist.}).

We will apply our perturbative results to the ``$\beta$-shift''
encountered when including sea quarks --- relative to the quenched
approximation. Subsequently we determine the
QCD running coupling ``constant'' for $n_f=0$ and $n_f=2$ from
lattice data of the correlation length~\cite{Sommer:1994ce} $R_0=r_0/a$,
the average plaquette and the
short distance static potential.
In this context we will also make use of the CP-PACS~\cite{Yoshie:2000wd}
ensemble obtained with $n_f=2$ SW fermions and the Iwasaki gluonic
action~\cite{Iwasaki:1984cj}.
Finally we shall also compare
perturbative and non-perturbative lattice potentials with the
aim of parametrizing lattice artefacts and to resolve
the differences in the running of the Coulomb couplings
between quenched and un-quenched data sets.

\subsection{The non-perturbative ``$\beta$-shift''}
\label{sec_beta_shift}
We study the situation of $n_f$ mass-degenerate flavours of sea quarks.
Two parameters can be varied: the coupling $\beta=6/g^2$ and the
lattice quark mass $ma$ which in the case of Wilson-SW fermions
is related to the parameter $\kappa$. Each $\kappa$-$\beta$
(or $ma$-$\beta$)
combination can be translated into
a pion mass $m_{\pi}r_0$ and a lattice spacing $a/r_0$ where
$r_0$ is a correlation length defined below. As $\beta\rightarrow\infty$
we reach the continuum limit,
$a/r_0\rightarrow 0$. Sending the quark mass
$mr_0$ to zero corresponds to a vanishing pion mass,
$m_{\pi}r_0\rightarrow 0$ (at finite $\beta$: only
up to violations of chiral symmetry). However, in general
the two limits will ``mix'': varying $\beta$ at
fixed $ma$ will not only affect $a/r_0$ but also to some
extent the ratio $m_{\pi}r_0$
while a variation of $ma$, keeping the coupling fixed, will
result in a
change of $a/r_0$ as well.

In view of the computational cost of lattice simulations incorporating
sea quarks, predicting by what amount
the coupling has to be shifted in order to compensate for the
change in $a/r_0$ that is induced by varying the quark mass $ma$
is certainly desirable. We will explore this possibility
by comparing the expectation Eq.~(\ref{eq:bshiftt}) on
the perturbative $\beta$-shift against
results from lattice simulations with $n_f=2$ sea quarks
of masses
$0.0098\leq ma\leq 0.3$.

At large distances the potentials
will be dominated by non-perturbative effects: the quenched
potential will linearly rise {\em ad infinitum}. However, as
soon as sea quarks are introduced the $Z_3$ symmetry of the action
is broken and at some distance~\cite{Bali:2000vr}
$r_c\approx (2.10+1.56\,mr_0)r_0$
string breaking will set in; obviously the linear quenched
behaviour at large distances can never be emulated by
a theory with a completely flat potential at large $r>r_c$.
Sea quarks will also affect the running of the coupling
at short distances $r<0.3\,m^{-1}$.
Hence, the best we can hope for is that sea quarks decouple
within a window of distances $m^{-1}\ll r<r_c$.

\begin{table}[hbt]
\caption{\label{tab:bshiftwil}
Comparison between non-perturbative $\beta$-shifts
$\Delta\beta_{r_0}$ and the respective perturbative
predictions $\Delta\beta_{\mbox{\tiny 1-l}}$ for two flavours
of Wilson fermions~\cite{Bali:2000vr,Bali:2002}.}
\begin{ruledtabular}
\begin{tabular}{cccccc}
$\beta$&$\kappa$&$ma$&$r_0/a$&$-\Delta\beta_{r_0}$&
$-\Delta\beta_{\mbox{\tiny 1-l}}$\\\hline
5.5&0.158&0.0597&4.03(3)&0.352(5)&0.318(3)\\
5.5&0.159&0.0398&4.39(3)&0.395(5)&0.371(4)\\
5.5&0.1596&0.0280&4.68(3)&0.427(4)&0.416(4)\\
5.5&0.160&0.0202&4.89(3)&0.450(4)&0.456(4)\\
5.6&0.156&0.0504&5.10(3)&0.373(5)&0.342(3)\\
5.6&0.1565&0.0402&5.28(5)&0.399(5)&0.369(4)\\
5.6&0.157&0.0300&5.46(5)&0.410(5)&0.406(4)\\
5.6&0.1575&0.0199&5.89(3)&0.454(5)&0.457(4)\\
5.6&0.158&0.0098&6.23(6)&0.488(7)&0.538(5)
\end{tabular}
\end{ruledtabular}
\end{table}

\begin{table}[hbt]
\caption{\label{tab:bshiftsw}
Comparison between non-perturbative $\beta$-shifts
$\Delta\beta_{r_0}$ and the respective perturbative
predictions $\Delta\beta_{\mbox{\tiny 1-l}}$ for two flavours
of SW fermions~\cite{Allton:2001sk,Booth:2001qp}.}
\begin{ruledtabular}
\begin{tabular}{cccccc}
$\beta$&$\kappa$&$ma$&$r_0/a$&$-\Delta\beta_{r_0}$&
$-\Delta\beta_{\mbox{\tiny 1-l}}$\\\hline
5.2&0.135 &0.0459(2)&4.75(4)&0.736(5)&0.679(3)\\
5.2&0.1355&0.0236(2)&5.04(4)&0.766(5)&0.750(3)\\
5.2&0.13565&0.0179(9)&5.21(5)&0.784(6)&0.780(6)\\
5.25&0.1352&0.0427(2)&5.14(5)&0.723(6)&0.686(3)\\
5.26&0.1345&0.0720(2)&4.71(5)&0.671(5)&0.632(4)\\
5.29&0.134 &0.0927(3)&4.81(5)&0.641(5)&0.609(3)\\
5.29&0.1350&0.0535(2)&5.26(7)&0.699(7)&0.664(3)\\
5.29&0.1355&0.0350(1)&5.62(9)&0.736(9)&0.708(3)
\end{tabular}
\end{ruledtabular}
\end{table}

\begin{table}[hbt]
\caption{\label{tab:bshiftks}
Comparison between non-perturbative $\beta$-shifts
$\Delta\beta_{r_0}$ and the respective perturbative
predictions $\Delta\beta_{\mbox{\tiny 1-l}}$ for ``two'' flavours
of KS fermions~\cite{Tamhankar:2000ce}.
The $\beta=5.6$, $ma=0.025$ data point is from
Ref.~\cite{Heller:1994rz}.}
\begin{ruledtabular}
\begin{tabular}{ccccc}
$\beta$&$ma$&$r_0/a$&$-\Delta\beta_{r_0}$&
$-\Delta\beta_{\mbox{\tiny 1-l}}$\\\hline
5.3&0.3&1.65(3)&0.079(25)&0.149(3)\\
5.35&0.3&1.79(1)&0.084(20)&0.149(3)\\
5.415&0.3&1.97(5)&0.076(9)&0.149(3)\\
5.3&0.2&1.75(5)&0.119(25)&0.196(2)\\
5.35&0.2&1.87(1)&0.111(13)&0.196(2)\\
5.415&0.2&2.15(1)&0.125(10)&0.196(2)\\
5.3&0.15&1.78(1)&0.130(20)&0.226(3)\\
5.35&0.15&1.94(1)&0.132(12)&0.226(3)\\
5.415&0.15&2.27(1)&0.154(11)&0.226(3)\\
5.3&0.1&1.86(2)&0.157(15)&0.267(2)\\
5.35&0.1&2.04(1)&0.161(10)&0.267(2)\\
5.415&0.1&2.45(1)&0.194(11)&0.267(2)\\
5.5&0.1&3.10(4)&0.225(7)&0.267(2)\\
5.3&0.075&1.91(2)&0.172(15)&0.295(2)\\
5.35&0.075&2.15(1)&0.190(10)&0.295(2)\\
5.3&0.05&1.96(2)&0.188(13)&0.336(2)\\
5.35&0.05&2.35(3)&0.237(13)&0.336(2)\\
5.415&0.05&2.72(1)&0.246(9)&0.336(2)\\
5.5&0.05&3.42(3)&0.273(6)&0.336(2)\\
5.3&0.025&2.03(1)&0.209(11)&0.406(1)\\
5.35&0.025&2.36(1)&0.240(12)&0.406(1)\\
5.415&0.025&2.95(1)&0.286(8)&0.406(1)\\
5.5&0.025&3.80(2)&0.324(5)&0.406(1)\\
5.6&0.025&4.8(1)&0.341(11)&0.406(1)\\
5.415&0.0125&3.12(2)&0.306(7)&0.476(1)\\
5.5&0.0125&3.98(1)&0.346(4)&0.476(1)\\
5.6&0.08&4.08(1)&0.259(4)&0.289(4)\\
5.6&0.04&4.54(2)&0.312(4)&0.358(1)\\
5.6&0.02&4.84(1)&0.345(3)&0.429(1)\\
5.6&0.01&4.99(1)&0.361(3)&0.499(1)
\end{tabular}
\end{ruledtabular}
\end{table}

We shall define a
\begin{equation}
\label{eq:dbr0def}
\Delta\beta_{r_0}(n_f,r_0/a,ma)=\beta^{(n_f)}(r_0/a,ma)
-\beta^{(0)}(r_0/a),
\end{equation}
by non-perturbatively matching un-quenched and quenched $\beta$ values
that correspond to the same scale~\cite{Sommer:1994ce} $r_0\approx 0.5$~fm,
implicitly defined through the relation,
\begin{equation}
\label{eq:r0d}
\left.r_0^2\frac{dV(r)}{r}\right|_{r=r_0}=1.65.
\end{equation}
If non-perturbative
effects cancel from Eq.~(\ref{eq:dbr0def}) and leading order
perturbation theory applies then
$\Delta\beta_{r_0}$ will only depend on $ma$ (and $n_f$) but
not on the lattice spacing $a/r_0$. Perturbation theory will obviously
break down for $ma\ll 1$ in which case
the coefficients of the perturbative expansion explode
[cf.\ Eq.~(\ref{eq:bshiftt})].
This is a reflection of the fact that the running of the
coupling in the theory with
massless sea quarks differs from the $n_f=0$ case {\em at all scales}.

\begin{figure}[hbt]
\includegraphics[width=8cm]{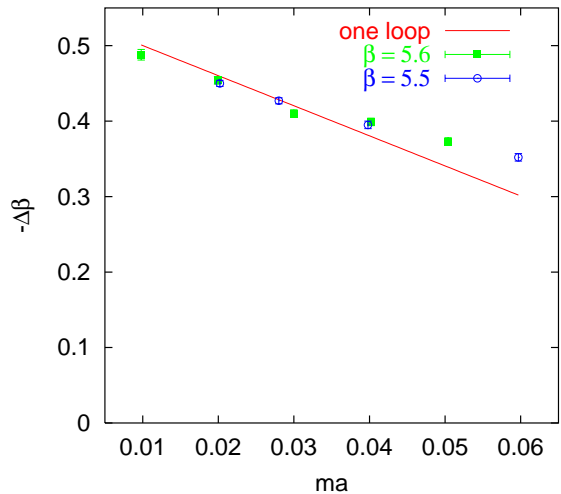}
\caption{
\label{fig:wilbshift}
The non-perturbative $\beta$-shift for two flavours of Wilson
fermions, in comparison with the perturbative prediction (curve).}
\end{figure}

\begin{figure}[hbt]
\includegraphics[width=8cm]{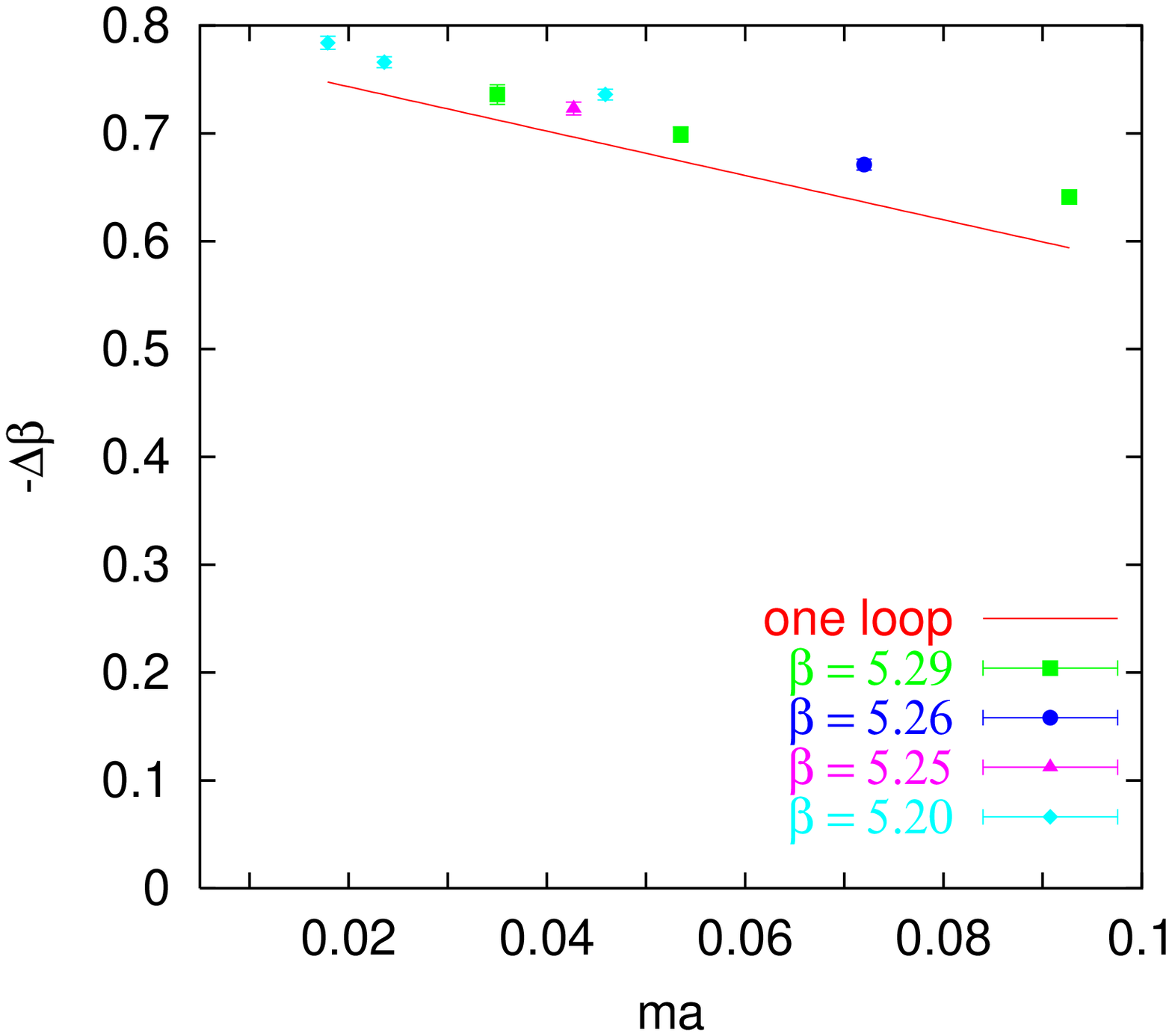}
\caption{
\label{fig:swbshift}
The non-perturbative $\beta$-shift for two flavours of SW
fermions.}
\end{figure}

\begin{figure}[hbt]
\includegraphics[width=8cm]{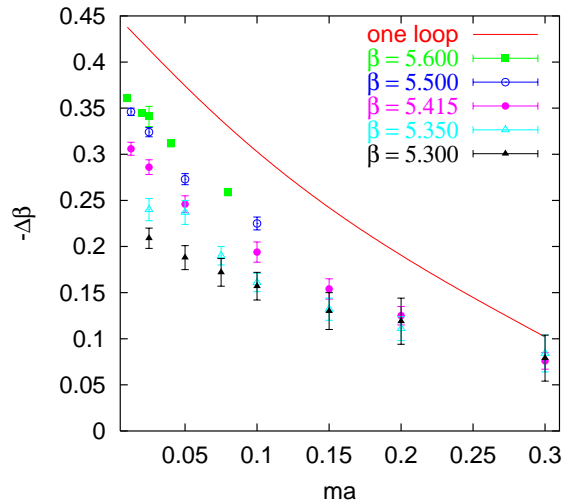}
\caption{
\label{fig:ksbshift}
The non-perturbative $\beta$-shift for ``two'' flavours of KS
fermions.}
\end{figure}

On the other hand, unless $ma\ll 1/D$
the relative importance of the non-universal term
$\Delta K_1(ma)$ will increase and in general the matching
of the running of the inter-quark force at a scale $r_0$ is not
equivalent to the matching of the running of the potential at
larger distances anymore. In principle one can work out the
matching condition for $r_0$ in perturbation theory. It turns
out, however, that at distances $r_0\approx 0.5$~fm the non-perturbative
contribution to the lattice potential is already substantial.
For instance a perturbative $n_f=0$ calculation yields
$r_0/a=5.33$ at $\beta\approx 1.49$ while non-perturbatively
this result is obtained at $\beta\approx 6.0$ and $\beta$
depends only logarithmically on $a$!

Given the above situation, it is difficult to identify a
sensible way of combining perturbative
and non-perturbative results at large distances.
Consequently, we refrain from attempting to do this but separately
employ purely perturbative and purely
non-perturbative matching strategies:
on the lattice simulation side of things we match $r_0/a$.
In perturbation theory we match a quantity that is suitable for
perturbative treatment, namely the inter-quark potential at large distances,
rather than a quantity that is inspired by (non-perturbative) phenomenology,
like $r_0$.
We then hope that some, ideally most,
non-perturbative effects cancel each other on the
right hand side of Eq.~(\ref{eq:dbr0def})
and leave the $\beta$-shift within the available window of
$r_0/a$ and $ma$ untouched. Clearly,
the perturbative matching will break down if either
$ma$ or $\beta$ are too small.
In addition the strategy that
we adopt requires $ma<1/D$ to limit violations of universality
as well as $mr_0> 0.3$ to ensure that the sea quarks do
not affect the running of the coupling at distances around $r_0$.

The quenched $\beta$ value that corresponds to a given $r_0/a$
is determined by use of the interpolation,
\begin{equation}
r_0/a=\exp\left(d_0+d_1x+d_2x^2+d_3x^3\right),\quad x=\beta-5.9,
\end{equation}
with $d_0\approx 1.489, d_1\approx 1.982, d_2\approx -0.630,
d_3\approx -1.522$,
obtained from a fit to lattice results within the window~\cite{Bali:2000gf}
$5.5\leq\beta\leq 6.2$.
In principle we can also
use the more recent precision
data of Refs.~\cite{Guagnelli:1998ud,Necco:2002xg}.
However, the difference is insignificant in view of
the present level of accuracy of un-quenched data
and the matching of some of the MILC data
requires an interpolation that extends to lattice spacings
coarser than those investigated in the latter two references.
In the un-quenched simulations, different groups used different
procedures to extract $r_0/a$ which partly explains
why different collaborations can obtain very different error
estimates with similar computational efforts. Aiming
only at a qualitative
comparison we shall not attempt to reanalyse all data
in one and the same way but prefer to cite the published values and
errors.

In the case of KS fermions~\cite{Tamhankar:2000ce,Heller:1994rz}
there is
no additive mass renormalization.
For Wilson fermions~\cite{Bali:2000vr,Bali:2002}
we define the lattice quark mass via,
\begin{equation}
\label{eq:made}
ma=\frac{1}{2}\left(\frac{1}{\kappa}-\frac{1}{\kappa_c}\right),
\end{equation}
where $\kappa_c$ has been obtained from a chiral extrapolation
of the non-perturbative $m_{\pi}^2$ at fixed\footnote{
In principle other extrapolations are possible like keeping
$r_0/a$ fixed~\cite{Allton:2001sk}.
However, the whole matching idea is based on a
semi-quenched philosophy: what value of the $n_f=0$
coupling will produce the same infra red physics, e.g.\ $r_0/a$?
Since at $\kappa_c$ the low energy physics will be very different
anyway, the ``natural'' choice in this case seems to be
keeping the coupling
constant fixed. Having said this, to the order of perturbation
theory at which we work both approaches are equivalent anyway.}
$\beta$\footnote{To all orders in perturbation theory
$ma$ is defined via Eq.~(\ref{eq:made}).
To the order that we work at,
$\kappa_{c,{\mbox{\tiny pert}}}=1/8$, which means that
the $\kappa$ values employed in the lattice simulation
all correspond to negative masses, $ma$, when directly plugged into
the perturbative expansion.
Obviously this is not a sensible choice,
and hence we formulate our perturbation theory in
terms of the mass $ma$ rather than $\kappa$. Subsequently,
we determine the $\kappa$ that corresponds to a given $ma$
value non-perturbatively. In the case of SW fermions we
proceed in an
analogous way: on the perturbation theory side we use
$ma$ and $c_{SW}=1$, which is the value consistent with
the order at which we
work. However, in the simulation we use the $\kappa$ and $c_{SW}$ values
respectively which result in the same $ma$ value and
that eliminate
${\mathcal O}(a)$ lattice artefacts non-perturbatively.}.
In the case of the SW data~\cite{Allton:2001sk,Booth:2001qp}
we do not know $\kappa_c$ but we can use the published values
of the PCAC quark mass~\cite{Allton:2001sk}, which is
equivalent to any other definition to leading order perturbation
theory.

We compile results from the three collaborations in
Tabs.~\ref{tab:bshiftwil}  -- \ref{tab:bshiftks}.
We have sorted Tab.~\ref{tab:bshiftks}
(with the exception of $\beta=5.6$)
with respect to the quark mass
and then $\beta$
while the other tables are sorted the other way around.
Unfortunately, no PCAC mass value is available for the most
critical UKQCD data set
[$(\beta,\kappa)=(5.2,0.13565)$], however, we arrived at
the estimate $ma=0.0179(9)$ by extrapolating $ma$ as
a linear function of
$(m_{\pi}r_0)^2$
to the value~\cite{Hepburn:2002fh} $(m_{\pi}r_0)^2=0.132(2)$.
The perturbative prediction Eq.~(\ref{eq:bshiftt}) is included
in the last column of the tables where the error is due to
the uncertainty in $\Delta K_1(ma)$.

In Figs.~\ref{fig:wilbshift} -- \ref{fig:ksbshift} we compare
the non-perturbative results with the expectations.
In all cases $ma$ is much smaller than $1/D$ as desired. However,
$mr_0$, varying from 0.06 to 0.24 in the Wilson case,
from 0.09 to 0.46 in the SW case and from 0.04 to 0.6
in the KS case, is not exactly big relative to $r_0=0.3\,m^{-1}$
(the distance below which the perturbative running of
the coupling changes and sea quark effects cannot completely
be compensated for by a scale redefinition alone).
Nonetheless,
the first two figures reveal an excellent agreement with the expectation
within 10~\% for Wilson-type quarks, even for such
light quark masses.
Moreover, no significant $\beta$-dependence is observed, again in
agreement with
Eq.~(\ref{eq:bshiftt}), indicating higher order corrections to be small.
This is very different for the KS fermions depicted in
Fig.~\ref{fig:ksbshift}: here the agreement with the prediction only
improves
as the lattice spacing is reduced.

\begin{figure}[hbt]
\includegraphics[width=8cm]{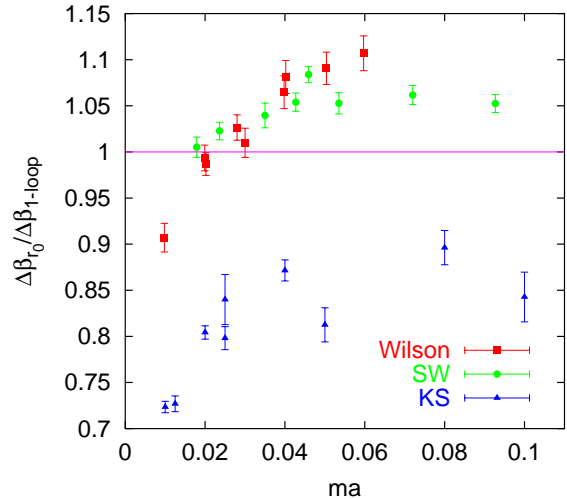}
\caption{
\label{fig:compbshift}
$\beta$-shift: the non-perturbative result, normalized to the
prediction.}
\end{figure}

In Fig.~\ref{fig:compbshift} we finally compare the three formulations
within the window $0.1\geq ma\geq 0.009$ for $4<r_0/a<6.5$ data.
The Wilson and SW results seem to fall onto almost universal
curves that differ by less than 10~\% from the prediction while
the KS results deviate by much more and some $\beta$-dependence is evident,
Whether this is due to a slower convergence of the
perturbative series, due to the inexact updating algorithm employed
or due to $n_f$ not being a multiple of four
is an open question. Around $m\approx 0.05\,r_0^{-1}\approx 20$~MeV
the reliability of the
matching for Wilson-SW quarks finally appears to break down
(left-most data point in Fig.~\ref{fig:compbshift}): the
behaviour becomes ``truly un-quenched''.

We remind the reader that $\beta$-shifts
are in general not independent of the quantity that is used to match
quenched and un-quenched theories. This ambiguity exists non-perturbatively
as well as in perturbation theory.
The qualitative agreement between prediction and simulation 
for Wilson-type quarks indicates that,
at least at present masses, physics at hadronic scales is not
yet strongly affected by quark loops, which is consistent with the
phenomenological success of the quenched approximation.

In simulations with non-perturbatively improved SW quarks
the lines of constant $a/r_0$ and constant $m_{\pi}r_0$ are
significantly tilted with respect to both axes of the $\beta$-$ma$
plane~\cite{Irving:2001hs}. This observation
is consistent with the small value,
$D^{SW}
\approx 0.0238$,
but causes practical problems as going to lighter quark masses at
sensible $a/r_0$ values requires simulations at small $\beta$s
for which the non-perturbative determination of the improvement coefficient
$c_{SW}$ causes problems~\cite{Jansen:1998mx}. In the worst case it might
even be conceivable that
the slope of the variation
of $a/r_0$ with $\kappa\rightarrow\kappa_c$ at fixed $\beta$ eventually
diverges and that the continuum limit $\beta\rightarrow\infty$
does not at all exist for light quarks~\cite{Aoki:1983qi,Aoki:2001xq}.
The variation of $r_0$ with the quark mass, however, is
reduced when actions with only moderately negative
values of $K_1(0)$ [and therefore
$D$ values that are of order one]
like the Wilson or KS action are employed.

Our result might be taken as an indication that the
perturbative matching of $n_f-1$ to $n_f$  QCD
couplings in the $\overline{MS}$ scheme via the intermediate
mass-dependent $V$-scheme~\cite{Brodsky:1998mf}
(or by calculating and matching a physical
amplitude~\cite{Bernreuther:1983zp})
is likely to be quite reliable,
even at masses as light as or lighter than
that of the charm quark. In this case the matching condition reads,
\begin{equation}
\frac{1}{\alpha^{(n_f)}_{\overline{MS}}(\mu)}=
\frac{1}{\alpha^{(n_f-1)}_{\overline{MS}}(\mu)}+\frac{1}{3\pi}\left[
\ln\left(\sqrt{C_0}\frac{m}{\mu}\right)-\frac{5}{6}\right]\alpha,
\end{equation}
which may be rewritten as,
\begin{equation}
\label{eq:ssss1}
\alpha^{(n_f)}_{\overline{MS}}(\mu)=\alpha^{(n_f-1)}_{\overline{MS}}(\mu),
\end{equation}
where,
\begin{equation}
\label{eq:ssss2}
\mu=\sqrt{C_0}\exp\left(-\frac{5}{6}\right)m=0.990(3)m,
\end{equation}
to this order in perturbation theory\footnote{
Eqs.~(\ref{eq:ssss1}) -- (\ref{eq:ssss2}) can also be cast into
$\alpha^{(n_f)}(m)=\alpha^{(n_f-1)}(m)+c\,\alpha^2$ where the
numerical value $c\approx 0.0033$ is of the same magnitude
as~\cite{Bernreuther:1983zp}
$c=7/(72\pi^2)\approx 0.0099$. Note that
the difference between pole and $\overline{MS}$ masses
is irrelevant to this order in perturbation theory.}.
Clearly, the applicability of the matching
method in this case
is ultimately limited by the reliability of the perturbative
running of the $\overline{MS}$ coupling at small scales.

\subsection{Determining $\alpha_{\overline{MS}}$\label{sec_alpha_s}}
Lattice simulations yield correlation lengths and masses
that are functions of a set of input parameters, namely
the inverse lattice coupling $\beta$ and lattice quark masses
$ma$ (or in the case of Wilson-type fermions $\kappa$-values that can
be related to the quark masses). One can then for instance
obtain the strong coupling
constant in the $\overline{MS}$ scheme,
\begin{equation}
\label{eq:ssss}
\alpha_{\overline{MS}}(a^{-1})=
\alpha_L+b_1\alpha_L^2+\cdots,
\end{equation}
once a physical scale has been assigned to the lattice spacing
$a$. One such possibility would be to ``measure'' $R_0=r_0/a$ on
the lattice and to equate $r_0\approx 0.5$~fm. Other input scales
with a more direct connection to experiment
are possible, for instance the proton mass $m_N$ or
the pion decay constant $f_{\pi}$. If QCD with
the right number of quark flavours and masses is simulated
the resulting $a$ should become independent of the choice of the
experimental input quantity. In this way the strong coupling constant
can be determined from low energy hadron phenomenology.

There are of course higher order perturbative as well as
non-perturbative corrections to Eq.~(\ref{eq:ssss})
which, however, will vanish as $a\rightarrow 0$.
In practice these corrections are big in the range of lattice spacings that
will ever be realistically accessible~\cite{Bali:1993ru,lepage}.
One way to improve the convergence is to convert between the
$\overline{MS}$ and the lattice schemes at an optimized
BLM scale~\cite{Brodsky:1983gc,lepage,Brodsky:1998mf}
$\mu=e^{-5/6}q^*$, rather than at $a^{-1}$. We refer
to this reorganization of the perturbative series as boosted
perturbation theory.
Another possibility (that can be combined with the BLM scheme)
is to ``measure'' the coupling on the
lattice. This can either be done from a quantity that depends on the
lattice spacing like the
average plaquette~\cite{Parisi:1980pe,lepage,El-Khadra:1992vn,davi1,davi2}
or at a scale $\mu\ll a^{-1}$
from quantities that have a well defined continuum
limit~\cite{Booth:1992bm,Bali:1993ru,Necco:2001gh,Garden:2000fg}.
Here we shall follow the former strategy.

\begin{table}[hbt]
\caption{\label{tab:plaqquench}Pure gauge $SU(3)$ results
on $r_0$ and the average plaquette $\Box$.}
\begin{ruledtabular}
\begin{tabular}{ccc}
$\beta$&$r_0/a$&$\Box$\\\hline
5.5  &2.01(3)&0.49680(2)\\
5.6  &2.44(6)&0.52451(3)\\
5.7  &2.86(5)&0.54919(3)\\
5.8  &3.64(5)&0.56765(2)\\
5.9  &4.60(9)&0.58184(2)\\
6.0  &5.33(3)&0.59368(1)\\
6.2  &7.29(4)&0.61363(1)\\
6.3  &8.39(7)&0.62243(1)\\
6.4 &9.89(16)&0.63064(1)\\
6.6&12.73(14)&0.64567(1)
\end{tabular}
\end{ruledtabular}
\end{table}

\begin{table}[hbt]
\caption{\label{tab:shortquench}Pure gauge $SU(3)$ results
on the short distance
potential and force $F_{R_1R_2}=a[V(R_2)-V(R_1)]$.}
\begin{ruledtabular}
\begin{tabular}{ccccc}
$\beta$&$aV(1)$&$aV(2)$&$F_{12}$&$F_{23}$\\\hline
5.5  &0.636 (2)&1.155(21)&0.519(22)&0.29 (17)\\
5.6  &0.566 (1)&0.963 (4)&0.398 (5)&0.325(16)\\
5.7  &0.506 (2)&0.804 (5)&0.297 (5)&0.233(15)\\
5.8  &0.464 (1)&0.706 (3)&0.242 (3)&0.160 (6)\\
5.9  &0.4330(9)&0.645 (2)&0.212 (2)&0.127 (4)\\
6.0  &0.4111(2)&0.5974(2)&0.1863(2)&0.1032(4)\\
6.2  &0.3778(1)&0.5337(1)&0.1559(1)&0.0783(2)\\
6.4  &0.3514(2)&0.4889(4)&0.1375(4)&0.0648(5)\\
6.6  &0.3293(2)&0.4538(3)&0.1244(3)&0.0546(5)
\end{tabular}
\end{ruledtabular}
\end{table}
\begin{table}[hbt]
\caption{\label{tab:shortwil}Average plaquette, short range potential
and force with
two flavours of Wilson quarks. The corresponding $r_0/a$ and $ma$ values
can be found in Tab.~\ref{tab:bshiftwil}.}
\begin{ruledtabular}
\begin{tabular}{cccccc}
$\beta$&$\kappa$&$\Box$&$aV(1)$&$aV(2)$&$F_{12}$\\\hline
5.5&0.158 &0.55547(4)&0.4756(3)&0.7156(10)&0.2400(10)\\
5.5&0.159 &0.55815(3)&0.4685(3)&0.6965 (9)&0.2280 (9)\\
5.5&0.1596&0.55967(3)&0.4644(2)&0.6853 (7)&0.2209 (7)\\
5.5&0.160 &0.56077(5)&0.4616(3)&0.6761 (8)&0.2145 (8)\\
5.6&0.156 &0.56989(1)&0.4460(2)&0.6511 (6)&0.2051 (6)\\
5.6&0.1565&0.57073(1)&0.4440(3)&0.6450 (9)&0.2010 (9)\\
5.6&0.157 &0.57160(1)&0.4422(3)&0.6387(11)&0.1965(10)\\
5.6&0.1575&0.57257(1)&0.4394(2)&0.6336 (6)&0.1942 (6)\\
5.6&0.158 &0.57337(1)&0.4373(3)&0.6303(13)&0.1930(12)
\end{tabular}
\end{ruledtabular}
\end{table}

\subsubsection{The method}
We collect
the average plaquette, the potential at $R=1$ and at $R=2$
obtained in quenched simulations as well as from simulations with
sea quarks in Tabs.~\ref{tab:plaqquench} -- \ref{tab:shortks}.
In addition we include the ``force'',
\begin{eqnarray}
F_{12}&=&a\left[V(2)-V(1)\right],\\
F_{23}&=&a\left[V(3)-V(2)\right].
\end{eqnarray}
We will restrict our discussion to a one-loop determination of the
running coupling. We shall also calculate two-loop corrections
for the pure gauge case. Two-loop results are also known
for the case of the plaquette with massive Wilson
quarks\footnote{Unfortunately, these results are of limited use since they
have been obtained at $\kappa^{-1}$ values that correspond to negative
quark masses, after subtracting $\kappa_c^{-1}$ to the same
order in perturbation
theory.}~\cite{Alles:1998is}.

\begin{table}[hbt]
\caption{\label{tab:shortsw}
The same as Tab.~\ref{tab:shortwil} but with
two flavours of SW quarks. The corresponding $r_0/a$ and $ma$ values
can be found in Tab.~\ref{tab:bshiftsw}.}
\begin{ruledtabular}
\begin{tabular}{cccccc}
$\beta$&$\kappa$&$\Box$&$aV(1)$&$aV(2)$&$F_{12}$\\\hline
5.2 &0.135  &0.53368(1)&0.4823(2)&0.6970 (8)&0.2147 (8)\\
5.2 &0.1355 &0.53629(1)&0.4762(2)&0.6832 (8)&0.2070 (8)\\
5.2 &0.13565& ---      &0.4749(2)&0.6794 (6)&0.2045 (6)\\
5.25&0.1352 &0.54113(2)& --- & --- & --- \\
5.26&0.1345 &0.53973(1)&0.4739(4)&0.6839(11)&0.2100(11)\\
5.29&0.134  &0.54241(1)&0.4707(3)&0.6782(10)&0.2075(10)\\
5.29&0.1350 &0.54552(3)& --- & --- & --- \\
5.29&0.1355 &0.54708(3)& --- & --- & --- 
\end{tabular}
\end{ruledtabular}
\end{table}
\begin{table}[hbt]
\caption{\label{tab:shortks}
The same as Tab.~\ref{tab:shortwil} but with
``two'' flavours of KS quarks. The corresponding $r_0/a$ values
can be found in Tab.~\ref{tab:bshiftks}.}
\begin{ruledtabular}
\begin{tabular}{ccccccc}
$\beta$&$ma$&$\Box$&$aV(1)$&$aV(2)$&$F_{12}$\\\hline
5.3  &0.3   &0.46980(6)&0.7055(35)&1.35  (14)&0.64  (14)\\
5.3  &0.2   &0.47554(8)&0.6848(34)&1.114  (9)&0.429  (9)\\
5.3  &0.15  &0.4792 (4)&0.6740(17)&1.299  (6)&0.625  (6)\\
5.3  &0.1   &0.48444(8)&0.6534(24)&1.291  (8)&0.638  (8)\\
5.3  &0.075 &0.48714(8)&0.6512 (4)&1.1580(18)&0.5068(18)\\
5.3  &0.05  &0.48918(9)&0.6419(14)&1.135 (35)&0.493 (35)\\
5.3  &0.025 &0.49238(8)&0.6301(10)&1.096 (20)&0.466 (20)\\
5.35 &0.3   &0.48243(5)&0.6724(14)&1.258  (6)&0.586  (6)\\
5.35 &0.2   &0.48912(5)&0.6482(13)&1.163 (36)&0.515 (36)\\
5.35 &0.15  &0.49373(5)& ---      & ---      & ---      \\		
5.35 &0.1   &0.49951(4)&0.6154(10)&1.024 (18)&0.409 (18)\\
5.35 &0.075 &0.50283(5)&0.6066(10)&1.032 (18)&0.425 (18)\\
5.35 &0.05  &0.50565(6)&0.5980 (9)&1.014 (13)&0.416 (13)\\
5.35 &0.025 &0.51097(6)&0.5822 (8)&0.976 (10)&0.394 (10)\\
5.415&0.3   &0.50065(6)&0.6223(15)&1.113 (37)&0.491 (37)\\
5.415&0.2   &0.50841(8)&0.5980(14)&1.009 (15)&0.411 (15)\\
5.415&0.15  &0.51366(6)&0.5834 (8)&0.977 (13)&0.394 (13)\\
5.415&0.1   &0.52007(7)&0.5627 (9)&0.948  (8)&0.385  (8)\\
5.415&0.05  &0.52692(5)&0.5422 (5)&0.8711(42)&0.3289(42)\\
5.415&0.025 &0.53066(4)&0.5295 (5)&0.8449(35)&0.3154(35)\\
5.415&0.0125&0.5329 (1)&0.5222 (5)&0.8231(35)&0.3009(35)\\
5.5  &0.1   &0.54328(2)&0.5070 (3)&0.8001(17)&0.2931(17)\\
5.5  &0.05  &0.54724(2)&0.4935 (1)&0.7592 (5)&0.2657 (5)\\
5.5  &0.025 &0.54932(2)&0.4862 (2)&0.7375(11)&0.2513(11)\\
5.5  &0.0125&0.55027(2)&0.4829 (2)&0.7289(11)&0.2460(11)\\
5.6  &0.08  &0.56325(2)&0.4616 (1)&0.6904 (7)&0.2288 (7)\\
5.6  &0.04  &0.56418(1)&0.4566 (1)&0.6763 (5)&0.2197 (5)\\
5.6  &0.02  &0.56479(1)&0.4537 (1)&0.6670 (7)&0.2133 (7)\\
5.6  &0.01  &0.56499(1)&0.4528 (1)&0.6619 (6)&0.2091 (6)
\end{tabular}
\end{ruledtabular}
\end{table}

Obviously many ways, some better than others,
to extract $\alpha_{\overline{MS}}$ exist that
are consistent with perturbation theory at a given order.
For convenience we adopt
the procedure detailed in Refs.~\cite{davi1,davi2} but point
out that many alternative ways are equally justified (e.g.\ that
of Ref.~\cite{Booth:2001qp}). We only differ from 
Refs.~\cite{davi1,davi2} in as far as, once the coupling has been converted
into the $\overline{MS}$ scheme, we evolve the scale and extract the
$\Lambda$-parameter by numerically integrating the four-loop $\beta$ function,
rather than using a perturbatively truncated formula.

We can write the plaquette as,
\begin{equation}
\Box=1-c_1\alpha_L-c_2\alpha_L^2-c_3\alpha_L^3,\label{eq:expand0}
\end{equation}
where $c_1$ and $c_2$ can be read off from Tab.~\ref{TableSmallLoops}
with the help of Eqs.~(\ref{eq:wilex2}) -- (\ref{eq:wilex5})
for the pure gauge case and the fermionic contributions
can be identified in Tabs.~\ref{tab:fermiloop} and \ref{tab:fermiloop2}.
In particular one finds $c_1=4\pi C_F/4$.
$c_3$ for the pure gauge case and sources in the fundamental
representation reads~\cite{Alles:1994dn,Alles:1998is}:
\begin{eqnarray}
c_3\approx&&\frac{(N^2-1)N}{8}\left[0.0063538
-\frac{0.0181239}{N^2}\right.\nonumber\\&&
\left.+\frac{0.0185221}{N^4}\right](4\pi)^3.
\end{eqnarray}
We now define the quantity,
\begin{eqnarray}
P&=&-\frac{\ln\Box}{c_1}\\
&=&\alpha_L+\tilde{c}_1\alpha_L^2+\tilde{c}_2\alpha_L^3,\label{eq:expand}\\
\tilde{c}_1&=&\frac{c_2}{c_1}+\frac{c_1}{2},\\
\tilde{c}_2&=&\frac{c_3}{c_1}+c_2+\frac{c_1^2}{3}.
\end{eqnarray}
The expansion of $P$ in terms of $\alpha_L$ suffers from
big coefficients. For instance one obtains $\tilde{c}_1\approx 3.37$,
$\tilde{c}_2\approx 17.69$ in $SU(3)$ pure gauge theory.
The origin of these big numbers can be traced to
the lattice tadpole diagrams~\cite{lepage}. Large coefficients
are also encountered
when $\alpha_L$ is converted into the more ``physical''
coupling $\alpha_V$ [or $\alpha_{\overline{MS}}$ which is
``close'' to $\alpha_V$, cf.~Eq.~(\ref{eq:cvms})].
By re-expressing the series Eq.~(\ref{eq:expand})
in terms of $\alpha_V$, taken at a suitable scale
$q^*$~\cite{Brodsky:1983gc}, one might hope that the two
effects cancel in part~\cite{lepage}:
\begin{equation}
\label{eq:pmsbar}
P=\alpha_{V}(q^*)+X_1(q^*a)\alpha_{V}^2(q^*)
+X_2(q^* a)\alpha_{V}^3+\cdots.
\end{equation}
We obtain,
\begin{eqnarray}
X_1(x)&=&\tilde{c}_1+B_1(x)\\
X_2(x)&=&\tilde{c}_2+B_2(x)+2\tilde{c}_1B_1(x),
\end{eqnarray}
where
\begin{eqnarray}
\label{eq:defB11}
B_1(x)&=&-b_1-a_1+2\beta_0\ln x\\
B_2(x)&=&-b_2-a_2+2(a_1b_1+b_1^2+a_1^2)\nonumber\\
&+&2\beta_1\ln x+B_1^2(x)-B_1^2(1).
\end{eqnarray}
The constant $b_1$ is defined in Eqs.~(\ref{eq:b1}) -- (\ref{eq:K12}),
$b_2$ in Eqs.~(\ref{eq:b2}) -- (\ref{eq:K3}) and $\beta_i$ in
Eqs.~(\ref{eq:beta0}) -- (\ref{eq:beta1}). The $a_i$ are defined
in Eqs.~(\ref{eq:a1}) and (\ref{eq:a2run}) and have to be taken at
the scale $q^*$ for massive sea quarks.
The numerical values for $n_f=0$ and
$q^*a\approx 3.402$ read: $X_1\approx -1.191$, $X_2\approx -1.688$.
We can now truncate Eq.~(\ref{eq:pmsbar}) at ${\mathcal O}(\alpha^2)$
and obtain,
\begin{eqnarray}
\alpha_{V}(q^*)&=&\alpha_P-X_2\alpha_P^3+\cdots,\\
\alpha_P&=&\frac{1}{2X_1}\left(\sqrt{4PX_1+1}-1\right).
\end{eqnarray}
Note that the coefficients $X_i$ inherit a mass dependence from
$a_i$, $b_i$ (through $\Delta K_i$) and $\tilde{c}_i$.
With
\begin{eqnarray}
\alpha_{\overline{MS}}(\mu)&=&
\alpha_V(q^*)+Y_1\alpha_V^2(q^*)+Z_2\alpha_V^3(q^*),\\
\mu&=&e^{-5/6}q^*\approx 1.478/a,\\
Y_1&=&-\left(a_1-\frac{5}{3}\beta_0\right),\\
Z_2&=&-\left[a_2-a_1^2-\frac{5}{3}\beta_1-\left(a_1-\frac{5}{3}\beta_0\right)^2\right],
\end{eqnarray}
we arrive at,
\begin{equation}
\label{eq:msco}
\alpha_{\overline{MS}}(\mu)=
\alpha_P+Y_1\alpha_P^2+Y_2\alpha_P^3,
\end{equation}
where
\begin{eqnarray}
\label{eq:msco2}
Y_1&=&\frac{2N}{3\pi}-\frac{n_f}{6\pi}\ln\left(1+\frac{C_0m^2}
{{\mu}^2}\right),\\
Y_2&=&Z_2-X_2.
\end{eqnarray}
For $m=0$ $Y_1$ is independent\footnote{
For $n_f$ degenerate massive flavours one could in principle maintain
the mass independence of this coefficient:
$Y_1\mapsto Y_1-11N/(12\pi)\ln(1+C_0m^2/\mu^2)$,
$\mu\mapsto \mu\times \sqrt{1+C_0m^2/\mu^2}$. 
In the most interesting case, QCD
with non-degenerate flavours, this is however not possible.}
of $n_f$. For $n_f=0$, $N=3$ we find
the numerical value, $Y_2\approx 0.9538$: the NNLO
correction is small.

In analogy to defining a coupling from the measured plaquette,
other couplings can be computed from force and potential.
We illustrate this procedure at NLO:
in a first step we can write,
\begin{eqnarray}
\frac{aV({\mathbf R}a)}{v_1({\mathbf R})}&=&\alpha_L
+\frac{v_2({\mathbf R})}{v_1({\mathbf R})}\alpha^2\nonumber\\\label{eq:avcalc}
&=&
\alpha_V(q^*)+\left[\frac{v_2({\mathbf R})}{v_1({\mathbf R})}
+B_1(q^*a)\right]\alpha^2,
\end{eqnarray}
where the function $B_1(x)$ is defined in Eq.~(\ref{eq:defB11}).
A similar expression can easily be written
down for the force. The respective $q^*a$ values
can be found in Tab.~\ref{tab:qstar}. Consequently, $\alpha_V(q^*)$ can
be obtained by solving the quadratic equation Eq.~(\ref{eq:avcalc})
and converted
into $\alpha_{\overline{MS}}(\mu)$ via Eqs.~(\ref{eq:msco}) and
(\ref{eq:msco2}), where we set $Y_2=0$ in our
${\mathcal O}(\alpha^2)$ calculation.

Finally, we can run $\alpha_{\overline{MS}}$ numerically to arbitrarily
high scales using the perturbative four-loop $\beta$
function~\cite{vanRitbergen:1997va,Tarasov:1980au} and
then determine the $\Lambda_{\overline{MS}}$-parameter, defined
as in Eq.~(\ref{eq:deflambda}).

\subsubsection{Pure gauge theory}

\begin{figure}[hbt]
\includegraphics[width=8cm]{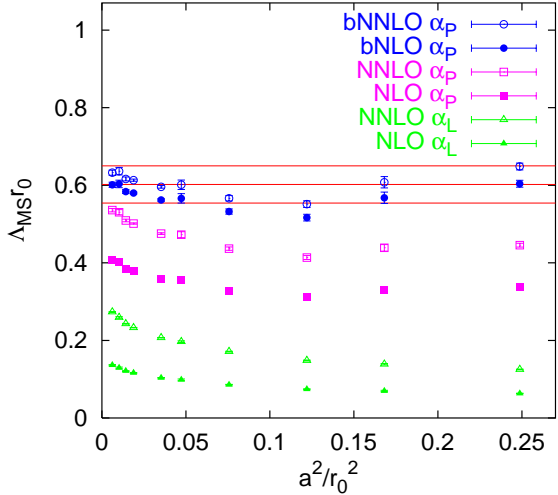}
\caption{
\label{fig:alphcomp}
Comparison of $\Lambda_{\overline{MS}}$-parameters for $SU(3)$ gauge theory.
The error band denotes the continuum limit result obtained by
the ALPHA Collaboration~\cite{Garden:2000fg}.
NNLO corresponds to $\mathcal O(\alpha^3)$,
NLO to ${\mathcal O}(\alpha^2)$, bNLO and bNNLO to ``boosted'' NLO and
NNLO, $\alpha_L$ to a conversion from the bare lattice coupling
and $\alpha_P$ to the $\alpha$-values,
``measured'' from the logarithm of
the average plaquette.}
\end{figure}

We display one- and two-loop results on $\alpha_{\overline{MS}}(\mu)$
as obtained from the logarithm of the average plaquette
following the boosted perturbation theory procedure detailed
above (as well as in Refs.~\cite{davi1,davi2}),
in Tab.~\ref{tab:plaqquench2}. This is then numerically converted into
estimates of the QCD $\Lambda_{\overline{MS}}$-parameter
which are displayed in the
last two columns of the table.
In Fig.~\ref{fig:alphcomp} we compare various methods of determining the
$\overline{MS}$ $\Lambda$-parameter as a function of the lattice spacing.
The horizontal error band corresponds to the continuum limit result as
obtained by the ALPHA Collaboration~\cite{Garden:2000fg} by use of
finite volume techniques.
$\alpha_L$ refers to a conversion from the bare
lattice coupling $\alpha_L=3/[2\pi\beta(a)]$ and $\alpha_P$ to a coupling,
``measured'' from the average plaquette. NLO and NNLO refer to
results from a na\"{\i}ve perturbative conversion of $\alpha_L$ into
$\alpha_{\overline{MS}}(a^{-1})$ at order $\alpha^2$ and $\alpha^3$
respectively
while the abbreviations bNLO and bNNLO
correspond to the boosted perturbation theory procedure detailed above.

\begin{table}[hbt]
\caption{\label{tab:plaqquench2}
$\mu=e^{-5/6}q^*$ in $SU(3)$ pure gauge theory
and one- (1-l) and two-loop (2-l) estimates of
$\alpha_{\overline{MS}}(\mu)$ and $\Lambda_{\overline{MS}}$
from the average plaquette. The errors of $\alpha$
do not reflect the uncertainty in the scale $\mu$.}
\begin{ruledtabular}
\begin{tabular}{cccccc}
$\beta$&$\mu r_0$&
$\alpha_{\overline{MS},\mbox{\tiny 1-l}}$&
$\alpha_{\overline{MS},\mbox{\tiny 2-l}}$&
$\Lambda_{\overline{MS},\mbox{\tiny 1-l}}r_0$&
$\Lambda_{\overline{MS},\mbox{\tiny 2-l}}r_0$\\\hline
5.5  & 2.97 (4)&0.26368(3)&0.27517(3)&0.604 (9)&0.648 (9)\\
5.6  & 3.61 (9)&0.22947(3)&0.23747(3)&0.568(15)&0.608(15)\\
5.7  & 4.23 (7)&0.20417(3)&0.21000(3)&0.517 (9)&0.551 (9)\\
5.8  & 5.38 (7)&0.18753(2)&0.19216(2)&0.532 (7)&0.566 (7)\\
5.9  & 6.80(13)&0.17582(2)&0.17970(2)&0.566(12)&0.601(12)\\
6.0  & 7.88 (4)&0.16667(1)&0.17000(1)&0.562 (3)&0.596 (3)\\
6.2  &10.78 (6)&0.15233(1)&0.15494(1)&0.580 (3)&0.613 (3)\\
6.3  &12.40(10)&0.14639(1)&0.14873(1)&0.584 (5)&0.616 (5)\\
6.4  &14.62(24)&0.14105(1)&0.14316(1)&0.603(10)&0.636(10)\\
6.6  &18.82(21)&0.13172(1)&0.13346(1)&0.601 (7)&0.632 (7)
\end{tabular}
\end{ruledtabular}
\end{table}

\begin{table}[hbt]
\caption{\label{tab:shortquench2}
Boosted NLO pure gauge estimates of $\alpha_{\overline{MS}}(\mu)$
from $aV(1)$, $aV(2)$, $F_{12}$ and $F_{23}$, respectively.
The respective scales $\mu r_0$ can easily be read off from
Tabs.~\ref{tab:qstar} and \ref{tab:plaqquench}.
The errors of $\alpha$
do not reflect the uncertainty in $\mu r_0$.}
\begin{ruledtabular}
\begin{tabular}{ccccc}
$\beta$&
$\alpha_{\overline{MS}}^{V_1}(\mu)$&
$\alpha_{\overline{MS}}^{V_2}(\mu)$&
$\alpha_{\overline{MS}}^{F_{12}}(\mu)$&
$\alpha_{\overline{MS}}^{F_{23}}(\mu)$\\\hline
5.5  &0.3414(17)&0.629(27) & ---      & --- \\
5.6  &0.2862 (7)&0.4409(32)&1.212(39) &1.594(89)\\
5.7  &0.2443(13)&0.3308(31)&0.682(19) &1.096(79)\\
5.8  &0.2173 (6)&0.2744(16)&0.4958(90)&0.721(30)\\
5.9  &0.1984 (5)&0.2426(10)&0.4104(54)&0.560(19)\\
6.0  &0.1856 (1)&0.2192(10)&0.3444 (5)&0.446(19)\\
6.2  &0.1669 (1)&0.1898 (5)&0.2736 (3)&0.332 (9)\\
6.4  &0.1527 (1)&0.1702 (2)&0.2341 (8)&0.271 (2)\\
6.6  &0.1411 (1)&0.1554 (1)&0.2073 (6)&0.226 (2)
\end{tabular}
\end{ruledtabular}
\end{table}

\begin{table}[hbt]
\caption{\label{tab:shortquench3}
One-loop pure gauge estimates of $\Lambda_{\overline{MS}}r_0$
from the average plaquette, $aV(1)$, $aV(2)$, $F_{12}$ and $F_{23}$,
respectively. In the last line we display our continuum limit estimates.
The errors are purely statistical.}
\begin{ruledtabular}
\begin{tabular}{cccccc}
$\beta$&
$\Lambda_{\overline{MS}}^{\Box}r_0$&
$\Lambda_{\overline{MS}}^{V_1}r_0$&
$\Lambda_{\overline{MS}}^{V_2}r_0$&
$\Lambda_{\overline{MS}}^{F_{12}}r_0$&
$\Lambda_{\overline{MS}}^{F_{23}}r_0$\\\hline
5.5  &0.604 (9)&0.736(15)&0.952(32)& ---     & --- \\
5.6  &0.568(15)&0.705(21)&0.935(29)& ---     & --- \\
5.7  &0.517 (9)&0.633(17)&0.819(23)& ---     & --- \\
5.8  &0.532 (7)&0.635(12)&0.797(18)&0.672(17)&0.708(19)\\
5.9  &0.566(12)&0.653(18)&0.813(23)&0.734(24)&0.806(32)\\
6.0  &0.562 (3)&0.639 (5)&0.768 (5)&0.708 (5)&0.805 (8)\\
6.2  &0.580 (3)&0.650 (4)&0.749 (4)&0.703 (4)&0.816 (7)\\
6.3  &0.584 (5)& ---     & ---     & ---     & --- \\
6.4  &0.603(10)&0.667(12)&0.755(15)&0.722(17)&0.829(25)\\
6.6  &0.601 (7)&0.654 (9)&0.736(10)&0.717(13)&0.764(24)\\\hline
$\infty$&0.609 (4)&0.666(5)&0.727(8)&0.735 (8)&0.770 (9)
\end{tabular}
\end{ruledtabular}
\end{table}

\begin{figure}[hbt]
\includegraphics[width=8cm]{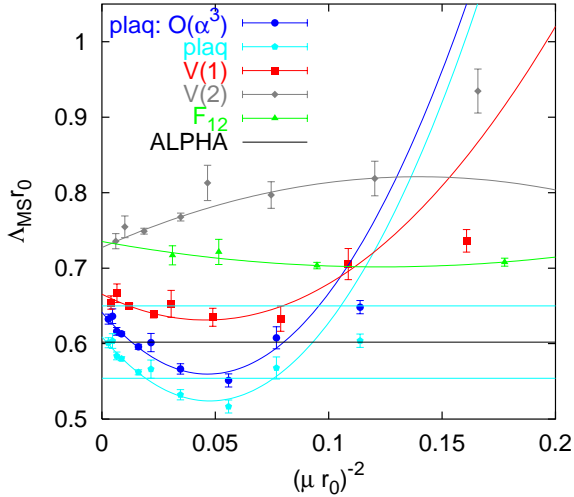}
\caption{
\label{fig:shortquench2}
Comparison of $\Lambda_{\overline{MS}}$-parameters for $SU(3)$ gauge theory
from different non-perturbative input quantities as a function
of the scale $\mu=e^{-5/6}q^*$. The curves correspond to
quadratic plus quartic fits in $1/\mu$ and
the error band is the continuum limit ALPHA Collaboration
result~\cite{Garden:2000fg}.}
\end{figure}

We find that the ratios
between NLO and NNLO results can be brought closer to unity, both
by ``measuring'' the coupling and by ``boosting'' the perturbation theory.
The combination of both methods indeed
brings the result in agreement with the known number
(error band).
Interestingly, almost all bNLO and bNNLO results are
within the expected range, even at rather coarse lattice spacings.

In Tab.~\ref{tab:shortquench2} we display values $\alpha_{\overline{MS}}(\mu)$
obtained from the short distance lattice potential
and force\footnote{
One could also imagine to repeat this procedure for a ``tadpole improved''
potential, $V(R)+\ln\Box/(2a)$ at the corresponding $q^*$s.
Since in this case the result is obviously correlated with that
which we obtained from the plaquette we leave this exercise
to the interested reader and remark that any such additive
corrections cancel in the force.}.
These are then converted into estimates of $\Lambda$-parameters
and compared to the corresponding bNLO estimates from the
average plaquette in Tab.~\ref{tab:shortquench2}.
We also display these numbers versus the inverse
momentum scale $1/\mu^2$
in Fig.~\ref{fig:shortquench2}. The curves represent results
from phenomenological\footnote{Since $\Lambda$ cannot directly be
obtained from position space Greens functions there is no theoretically
well founded reason to assume the leading order lattice corrections to
be quadratic in the lattice spacing. In fact we know that
perturbative corrections to
$\Lambda$ exist which should be of order $\Lambda/\ln[(a\Lambda)^{-2}]$.}
quadratic plus quartic fits in $1/(\mu r_0)$.
The $\Lambda_{\overline{MS}} r_0$ values resulting from these
continuum limit extrapolations are displayed
in the last row of the table. The results from the high $q^*$ quantities
$V(1)$ and $\Box$ turn out to be in reasonable agreement with
that from the ALPHA Collaboration,
$\Lambda_{\overline{MS}}r_0=0.602(48)$, however, this is not the case
for $F_{12}$ or $V(2)$.

Unfortunately,
there exists no {\em first principles} way to estimate systematic uncertainties
in any determination of $\Lambda_{\overline{MS}}$ that is partially based
on perturbation theory at momenta as low as a few GeV.
For instance the $F_{12}$ data exhibit a plateau
for $\beta\geq 5.9$ that is in no way inferior to that obtained from the
average plaquette. However, the two extrapolated values
differ by almost three standard deviations of the ALPHA Collaboration
result. This suggests that the good agreement obtained
for the plaquette might be partly accidental.
We are unaware of any convincing argument why an estimate
of $\alpha_{\overline{MS}}(\mu)$ extracted from $F_{12}$ at $\beta=6.6$,
i.e.\ $\mu r_0\approx 5.65$ should be less reliable than the value
obtained from the plaquette at say $\beta=5.8$ ($\mu r_0\approx 5.4$).
The same can be said about the combination $aV(2)$ at $\beta=6.6$,
$\mu r_0\approx 12.8$ and the plaquette at $\beta=6.3$, $\mu r_0\approx 12.4$
and yet the former two $\beta=6.6$ estimates
$\Lambda_{\overline{MS}}^{F_{12}}r_0=
0.717(13)$ and $\Lambda_{\overline{MS}}^{V_2}r_0=0.764(24)$ happen to lie
significantly higher than the known
continuum limit result.

We conclude that without the comparison
between predictions from different
non-perturbatively obtained observables we might easily have
underestimated the systematic uncertainties of the approach.
Having different observables as well as the ALPHA result
at hand it becomes obvious that the 4~\% difference
between bNLO and bNNLO estimates from the plaquette do not necessarily
reflect the whole truth, but that a conservative approach would
instead argue in favour of a 15~\% error on the $\Lambda$-parameter 
in an ${\mathcal O}(\alpha^2)$ determination like the above. 
Quantities with higher $q^\ast$ values are certainly likely to 
behave better, but incorporating a number of measurements to estimate
systematic errors is prudent when such a scale setting prescription
has been used.

The present article contains the NLO perturbation
theory that is necessary for the use
of additional short distance quantities such 
as $V(1)$ and the chair and parallelogram Wilson loops.
The observed scattering
between extrapolations from different input quantities will hopefully
shrink both as finer lattices are simulated,
and once higher order corrections are known for potential and
force. In the meantime we quote the value,
\begin{equation}
\Lambda_{\overline{MS}}^{(0)}r_0=0.609(4)(90),
\end{equation}
where the second error is systematic.

\subsubsection{Results with sea quarks}
We shall limit our discussion of the fermionic case
to ${\mathcal O}(\alpha^2)$ since only
the plaquette is known to NNLO and only for Wilson quarks at
some negative mass values~\cite{Alles:1998is}.
The fermionic
correction to $c_2$ within Eqs.~(\ref{eq:expand0})
-- (\ref{eq:expand}) can be calculated from Tab.~\ref{tab:fermiloop2}
with the help of Eqs.~(\ref{eq:wilex2}) and the definitions
Eqs.~(\ref{eq:wilex1}) -- (\ref{eq:wilex12}).
The corrections to the potential are available in
Tabs.~\ref{tab:potfermi1} -- \ref{tab:potfermi2}, with the conventions of
Eqs.~(\ref{eq:defv2}) and (\ref{eq:defvf}) and $v_i=(4\pi)^iV_i$.
We also include data
that has been obtained by the CP-PACS Collaboration~\cite{Yoshie:2000wd}
by combining SW fermions with the Iwasaki gluon action into the
analysis. The corresponding
perturbative expansion of the plaquette can be read off from
Tabs.~\ref{TableSmallLoops3} and \ref{tab:fermiloop}, with the definitions
Eqs.~(\ref{eq:wilex2}) -- (\ref{eq:wnlo}). In this case
$b_1\approx -1.2466+n_fK^{SW}_1(ma)$~\cite{Aoki:1998ar}
and $K_1$ is independent of the gauge action. We find $q^*a\approx 3.213$
for the plaquette.
We display CP-PACS simulation parameters and results on $r_0/a$ and the
plaquette in Tab.~\ref{tab:cppacs}. The quark masses are obtained via
Eq.~(\ref{eq:made}). We set $c_{SW}=1$ in the analysis
of SW-Wilson and SW-Iwasaki (SW-I) results which is consistent at NLO.

\begin{table}[htb]
\caption{\label{tab:cppacs}
$r_0/a$ and $\Box$ obtained on the CP-PACS ensemble~\cite{Yoshie:2000wd}.}
\begin{ruledtabular}
\begin{tabular}{ccccccc}
$\beta$&$\kappa$&$ma$&$r_0/a$&$\Box$\\\hline
1.80&0.1409&0.162&1.716(35)&0.49053(3)\\ 
1.80&0.1430&0.110&1.799(13)&0.49505(4)\\
1.80&0.1445&0.074&1.897(30)&0.49936(4)\\
1.80&0.1464&0.029&2.064(38)&0.50720(6)\\
1.95&0.1375&0.117&2.497(54)&0.55336(2)\\
1.95&0.1390&0.078&2.651(42)&0.55667(2)\\
1.95&0.1400&0.052&2.821(29)&0.55914(2)\\
1.95&0.1410&0.027&3.014(33)&0.56188(3)\\
2.10&0.1357&0.087&3.843(16)&0.59803(1)\\
2.10&0.1367&0.060&4.072(15)&0.59920(1)\\
2.10&0.1374&0.042&4.236(14)&0.60006(1)\\
2.10&0.1382&0.020&4.485(12)&0.60108(1)\\
2.20&0.1351&0.069&4.913(21)&0.62003(1)\\
2.20&0.1358&0.050&5.073(19)&0.62062(1)\\
2.20&0.1363&0.037&5.237(22)&0.62104(1)\\
2.20&0.1368&0.023&5.410(21)&0.62149(1)
\end{tabular}
\end{ruledtabular}
\end{table}

\begin{table}[hbt]
\caption{\label{tab:wilalpha}
Boosted NLO estimates of $\alpha_{\overline{MS}}(\mu)$ and
$\Lambda_{\overline{MS}}r_0$ from the average plaquette,
with two flavours of Wilson quarks. $\delta\Lambda_{m}^{\Box}$ and
$\delta\Lambda_{\Delta\! K_1}^{\Box}$ refer to the corrections,
included in $\Lambda_{\overline{MS}}$, that are due to the
finite quark masses and from the $ma$ dependent term
$\Delta K_1$ alone, respectively.}
\begin{ruledtabular}
\begin{tabular}{ccccccc}
$\beta$&$ma$&$\mu r_0$&$\alpha_{\overline{MS}}^{\Box}(\mu)$&
$\Lambda_{\overline{MS}}^{\Box}r_0$&$\delta\Lambda_{m}^{\Box}r_0$&
$\delta\Lambda_{\Delta\! K_1}^{\Box}r_0$\\\hline
5.5&0.060&5.95 (4)&0.20925(1)&0.549(3)&+0.049&+0.056\\
5.5&0.040&6.48 (4)&0.20475(2)&0.564(4)&+0.038&+0.043\\
5.5&0.028&6.91 (5)&0.20103(3)&0.578(4)&+0.030&+0.034\\
5.5&0.020&7.23 (4)&0.19988(5)&0.588(4)&+0.023&+0.026\\
5.6&0.050&7.54 (4)&0.19405(1)&0.564(3)&+0.045&+0.051\\
5.6&0.040&7.81 (8)&0.19247(1)&0.570(6)&+0.038&+0.043\\
5.6&0.030&8.09(11)&0.19076(1)&0.575(8)&+0.031&+0.035\\
5.6&0.020&8.71 (4)&0.18885(1)&0.601(3)&+0.023&+0.026\\
5.6&0.010&9.21 (9)&0.18702(1)&0.617(6)&+0.012&+0.013
\end{tabular}
\end{ruledtabular}
\end{table}

\begin{table}[hbt]
\caption{\label{tab:swalpha}
Estimates of $\alpha_{\overline{MS}}(\mu)$ and
$\Lambda_{\overline{MS}}r_0$ from the average plaquette,
with two flavours of SW quarks. $\delta\Lambda_{m}^{\Box}$
refers to the finite quark mass correction, included in
$\Lambda_{\overline{MS}}$.}
\begin{ruledtabular}
\begin{tabular}{cccccc}
$\beta$&$ma$&$\mu r_0$&$\alpha_{\overline{MS}}^{\Box}(\mu)$&
$\Lambda_{\overline{MS}}^{\Box}r_0$&$\delta\Lambda_{m}^{\Box}r_0$\\\hline
5.2&0.046&7.03 (6)&0.19109(1)&0.502(4) &+0.001\\
5.2&0.024&7.45 (6)&0.18903(1)&0.515(4) &+0.004\\
5.25&0.043&7.59 (7)&0.18568(1)&0.497(5)&+0.005\\
5.26&0.072&6.96 (8)&0.18647(3)&0.462(5)&+0.003\\
5.29&0.093&7.11 (7)&0.18407(5)&0.454(5)&+0.000\\
5.29&0.053&7.78(10)&0.18255(1)&0.483(7)&+0.005\\
5.29&0.035&8.31(13)&0.18143(1)&0.507(8)&+0.005
\end{tabular}
\end{ruledtabular}
\end{table}

\begin{table}[hbt]
\caption{\label{tab:ksalpha}
Estimates of $\alpha_{\overline{MS}}(\mu)$ and
$\Lambda_{\overline{MS}}r_0$ from the average plaquette,
with ``two'' flavours of KS quarks.}
\begin{ruledtabular}
\begin{tabular}{cccccc}
$\beta$&$ma$&$\mu r_0$&$\alpha_{\overline{MS}}^{\Box}(\mu)$&
$\Lambda_{\overline{MS}}^{\Box}r_0$&$\delta\Lambda_{m}^{\Box}r_0$\\\hline
5.3  &0.30  &2.44(4)&0.3362 (15)&0.587(15)&-0.052\\
5.35 &0.30  &2.65(2)&0.3069 (14)&0.552 (7)&-0.041\\
5.415&0.30  &2.91(8)&0.2739 (11)&0.496(16)&-0.032\\
5.3  &0.20  &2.59(8)&0.32891(55)&0.603(19)&-0.029\\
5.35 &0.20  &2.76(2)&0.29874(54)&0.551 (5)&-0.022\\
5.415&0.20  &3.18(2)&0.26530(39)&0.508 (4)&-0.018\\
5.3  &0.15  &2.63(2)&0.32810(55)&0.595 (5)&-0.019\\
5.35 &0.15  &2.87(2)&0.29190(26)&0.549 (4)&-0.015\\
5.415&0.15  &3.36(2)&0.25878(20)&0.510 (3)&-0.012\\
5.3  &0.10  &2.75(3)&0.31293 (1)&0.592 (6)&-0.010\\
5.35 &0.10  &3.02(2)&0.28296 (8)&0.546 (3)&-0.009\\
5.415&0.10  &3.62(2)&0.25067 (4)&0.515 (2)&-0.007\\
5.5  &0.10  &4.58(6)&0.22136 (7)&0.488 (7)&-0.006\\
5.3  &0.075 &2.82(3)&0.30792 (8)&0.592 (7)&-0.007\\
5.35 &0.075 &3.18(2)&0.27783 (4)&0.556 (3)&-0.006\\
5.3  &0.05  &2.90(3)&0.30438(15)&0.596 (7)&-0.005\\
5.35 &0.05  &3.47(4)&0.27364 (6)&0.590 (8)&-0.003\\
5.415&0.05  &4.02(2)&0.24217 (4)&0.530 (2)&-0.003\\
5.5  &0.05  &5.06(5)&0.21748 (1)&0.515 (5)&-0.002\\
5.3  &0.025 &3.00(2)&0.29847(15)&0.597 (4)&-0.002\\
5.35 &0.025 &3.49(3)&0.26552 (9)&0.559 (5)&-0.001\\
5.415&0.025 &4.36(2)&0.23761 (5)&0.551 (2)&-0.001\\
5.5  &0.025 &5.62(3)&0.21541 (2)&0.559 (3)&-0.002\\
5.6  &0.025 &7.10(2)&0.19960 (3)&0.575(12)&-0.001\\
5.415&0.0125&4.54(3)&0.23492(13)&0.558 (4)&-0.000\\
5.5  &0.0125&5.88(2)&0.21447 (2)&0.579 (2)&-0.000\\
5.6  &0.08  &6.03(2)&0.20033 (3)&0.494 (2)&-0.004\\
5.6  &0.04  &6.71(3)&0.19975 (1)&0.545 (3)&-0.002\\
5.6  &0.02  &7.15(2)&0.19929 (1)&0.577 (2)&-0.001\\
5.6  &0.01  &7.38(2)&0.19916 (1)&0.594 (2)&-0.000
\end{tabular}
\end{ruledtabular}
\end{table}

\begin{table}[hbt]
\caption{\label{tab:iwaalpha}
Estimates of $\alpha_{\overline{MS}}(\mu)$ and
$\Lambda_{\overline{MS}}r_0$ from the average plaquette,
with two flavours of SW quarks and Iwasaki glue.}
\begin{ruledtabular}
\begin{tabular}{ccccc}
$\beta$&$ma$&$\mu r_0$&$\alpha_{\overline{MS}}^{\Box}(\mu)$&
$\Lambda_{\overline{MS}}^{\Box}r_0$\\\hline
1.80&0.162&2.396(49)&0.28681(2)&0.445 (9)\\ 
1.80&0.110&2.512(18)&0.28378(3)&0.457 (4)\\
1.80&0.073&2.649(42)&0.28091(3)&0.473 (8)\\
1.80&0.029&2.882(53)&0.27573(4)&0.497 (9)\\
1.95&0.117&3.487(75)&0.24623(1)&0.477(10)\\
1.95&0.078&3.702(59)&0.24418(1)&0.497 (8)\\
1.95&0.052&3.939(40)&0.24265(1)&0.522 (6)\\
1.95&0.027&4.209(46)&0.24096(2)&0.549 (6)\\
2.10&0.087&5.366(22)&0.21910(1)&0.558 (3)\\
2.10&0.060&5.686(21)&0.21841(1)&0.586 (2)\\
2.10&0.042&5.915(20)&0.21790(1)&0.606 (2)\\
2.10&0.020&6.263(17)&0.21729(1)&0.637 (2)\\
2.20&0.069&6.860(29)&0.20616(1)&0.608 (3)\\
2.20&0.050&7.084(27)&0.20582(1)&0.625 (3)\\
2.20&0.037&7.313(31)&0.20558(1)&0.643 (3)\\
2.20&0.023&7.554(29)&0.20531(1)&0.662 (3)
\end{tabular}
\end{ruledtabular}
\end{table}

\begin{table}[hbt]
\caption{\label{tab:shortfullwil}
Boosted NLO estimates of $\Lambda_{\overline{MS}}^{(2)}r_0$ for Wilson quarks
from the average plaquette, $aV(1)$, $aV(2)$ and $F_{12}$,
respectively. The errors are purely statistical.}
\begin{ruledtabular}
\begin{tabular}{cccccc}
$\beta$&$ma$&
$\Lambda_{\overline{MS}}^{\Box}r_0$&
$\Lambda_{\overline{MS}}^{V_1}r_0$&
$\Lambda_{\overline{MS}}^{V_2}r_0$&
$\Lambda_{\overline{MS}}^{F_{12}}r_0$\\\hline
5.5&0.060&0.549(3)&0.616 (5)&0.786 (8)&0.687 (8)\\
5.5&0.040&0.564(4)&0.628 (5)&0.784 (8)&0.690 (8)\\
5.5&0.028&0.578(4)&0.641 (6)&0.790 (8)&0.697 (8)\\
5.5&0.020&0.588(4)&0.650 (6)&0.788 (8)&0.693 (8)\\
5.6&0.050&0.564(3)&0.621 (5)&0.759 (7)&0.691 (7)\\
5.6&0.040&0.570(6)&0.627 (8)&0.758(11)&0.686(11)\\
5.6&0.030&0.575(8)&0.634(10)&0.755(14)&0.678(15)\\
5.6&0.020&0.601(3)&0.659 (4)&0.784 (6)&0.709 (7)\\
5.6&0.010&0.617(6)&0.675 (8)&0.805(13)&0.734(15)
\end{tabular}
\end{ruledtabular}
\end{table}

\begin{table}[hbt]
\caption{\label{tab:shortfullsw}
Estimates of $\Lambda_{\overline{MS}}^{(2)}r_0$ for SW quarks.}
\begin{ruledtabular}
\begin{tabular}{cccccc}
$\beta$&$ma$&
$\Lambda_{\overline{MS}}^{\Box}r_0$&
$\Lambda_{\overline{MS}}^{V_1}r_0$&
$\Lambda_{\overline{MS}}^{V_2}r_0$&
$\Lambda_{\overline{MS}}^{F_{12}}r_0$\\\hline
5.2&0.046&0.502(4)&0.537(5)&0.653 (8)&0.672 (9)\\
5.2&0.024&0.515(4)&0.546(5)&0.655 (7)&0.670(10)\\
5.2&0.018&---     &0.559(6)&0.666 (8)&0.678(10)\\
5.26&0.072&0.462(5)&0.500(7)&0.611(10)&0.645(12)\\
5.29&0.093&0.454(5)&0.496(6)&0.605 (8)&0.648(11)
\end{tabular}
\end{ruledtabular}
\end{table}

\begin{table}[hbt]
\caption{\label{tab:shortfullks}
Estimates of $\Lambda_{\overline{MS}}^{(2)}r_0$ for KS quarks.}
\begin{ruledtabular}
\begin{tabular}{cccccc}
$\beta$&$ma$&
$\Lambda_{\overline{MS}}^{\Box}r_0$&
$\Lambda_{\overline{MS}}^{V_1}r_0$&
$\Lambda_{\overline{MS}}^{V_2}r_0$&
$\Lambda_{\overline{MS}}^{F_{12}}r_0$\\\hline
5.3  &0.300&0.587(15)&0.684(19)&---      &---\\
5.3  &0.200&0.603(19)&0.699(27)&---      &---\\
5.3  &0.150&0.595 (5)&0.694 (8)&---      &---\\
5.3  &0.100&0.592 (6)&0.683(13)&---      &---\\
5.3  &0.075&0.592 (7)&0.699 (8)&---      &---\\
5.3  &0.050&0.596 (7)&0.696(11)&---      &---\\
5.3  &0.025&0.697 (4)&0.692 (6)&---      &---\\
5.35 &0.300&0.552 (7)&0.670 (7)&---      &---\\
5.35 &0.200&0.551 (5)&0.661 (7)&---      &---\\
5.35 &0.100&0.546 (3)&0.650 (6)&0.840(31)&---\\
5.35 &0.075&0.556 (3)&0.664 (6)&0.900(33)&---\\
5.35 &0.050&0.590 (8)&0.702(12)&0.959(34)&0.434(71)\\
5.35 &0.025&0.559 (5)&0.660 (8)&0.904(25)&0.526 (9)\\
5.415&0.300&0.496(16)&0.618(19)&0.879(76)&---\\
5.415&0.200&0.508 (4)&0.626 (7)&0.846(26)&0.477(13)\\
5.415&0.150&0.510 (3)&0.627 (5)&0.854(25)&0.517(12)\\
5.415&0.100&0.515 (2)&0.620 (5)&0.881(18)&0.561 (7)\\
5.415&0.050&0.530 (2)&0.625 (4)&0.836(11)&0.596 (7)\\
5.415&0.025&0.551 (2)&0.634 (4)&0.853(11)&0.630 (7)\\
5.415&0.0125&0.558 (4)&0.634 (6)&0.840(13)&0.635(10)\\
5.5  &0.100&0.488 (7)&0.580 (9)&0.787(14)&0.623(11)\\
5.5  &0.050&0.515 (5)&0.592 (6)&0.772 (8)&0.631 (7)\\
5.5  &0.025&0.559 (3)&0.629 (4)&0.800 (7)&0.661 (7)\\
5.5  &0.0125&0.579 (2)&0.645 (3)&0.814 (5)&0.676 (5)\\
5.6  &0.080&0.494 (2)&0.564 (2)&0.717 (4)&0.627 (4)\\
5.6  &0.040&0.545 (3)&0.607 (3)&0.756 (5)&0.662 (5)\\
5.6  &0.020&0.577 (2)&0.634 (2)&0.776 (4)&0.676 (4)\\
5.6  &0.010&0.594 (2)&0.650 (2)&0.783 (4)&0.676 (4)
\end{tabular}
\end{ruledtabular}
\end{table}

\begin{table}[hbt]
\caption{\label{tab:chiextra}
Chirally extrapolated $a/r_0$ values and extrapolations of
$\Lambda_{\overline{MS}}^{(2)}r_0$, according to Eq.~(\ref{eq:chiextra}).}
\begin{ruledtabular}
\begin{tabular}{cccccc}
Action&$\beta$&$a/r_0$&$c$&$[\Lambda_{\overline{MS}}^{(2)}r_0](0)$&$\chi/{N_{DF}}$\\\hline
Wilson&5.5&0.178(4)&-0.271(17)&0.613(3)&0.51/2\\
Wilson&5.6&0.148(3)&-0.335(47)&0.639(7)&1.67/2\\
SW&5.2&0.185&-0.136&0.532&$N_{DF}=0$\\
SW&5.29&0.162(4)&-0.198(23)&0.542(9)&0.48/1\\
KS&5.415&0.312(2)&-0.311(30)&0.573(3)&0.56/1\\
KS&5.5&0.236(6)&-0.508(38)&0.605(3)&1.06/1\\
KS&5.6&0.194(1)&-0.371 (8)&0.613(1)&0.16/1\\
SW-I&1.8&0.460(11)&-0.251(34)&0.508(7)&0.78/2\\
SW-I&1.95&0.312 (5)&-0.358(34)&0.575(6)&0.83/2\\
SW-I&2.1&0.213 (1)&-0.327(12)&0.666(3)&2.08/2\\
SW-I&2.2&0.177 (1)&-0.254(18)&0.692(4)&2.32/2
\end{tabular}
\end{ruledtabular}
\end{table}

\begin{figure}[hbt]
\includegraphics[width=8cm]{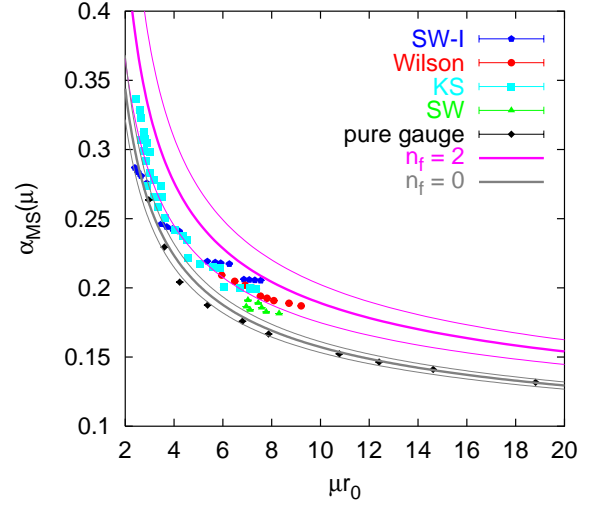}
\caption{
\label{fig:alpharuncom}
Comparison between bNLO $\alpha_{\overline{MS}}$ estimates from
pure gauge theory simulations and simulations with
$n_f = 2$ Wilson, KS and SW quarks, the latter with Wilson and Iwasaki (SW-I)
gluon actions. The $n_f = 0$ error band
corresponds to the ALPHA result $\Lambda_{\overline{MS}}=0.602(48)r_0^{-1}$
while the $n_f = 2$ band corresponds to the four-loop running
with $\Lambda_{\overline{MS}}^{(2)}=0.69(15)$.}
\end{figure}

\begin{figure}[hbt]
\includegraphics[width=8cm]{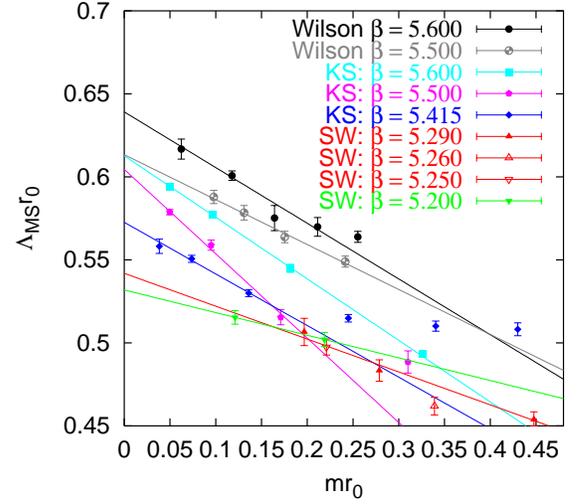}
\caption{Chiral extrapolations of $\Lambda_{\overline{MS}}^{(2)}r_0$ at finite
lattice spacings for the Wilson gluonic action combined with
the three different fermionic actions.
\label{fig:chiextra}}
\end{figure}

\begin{figure}[hbt]
\includegraphics[width=8cm]{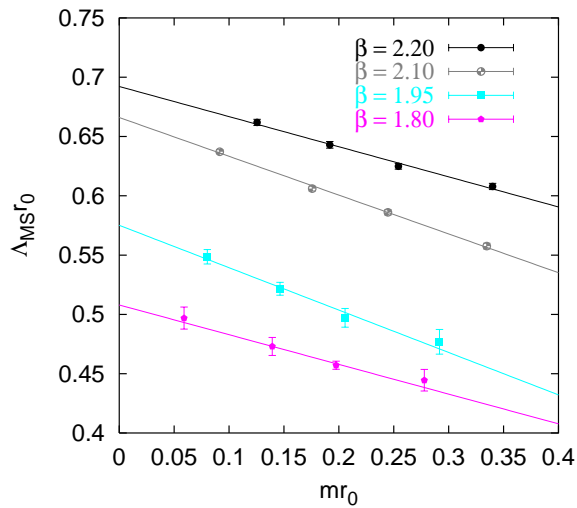}
\caption{Chiral extrapolations of $\Lambda_{\overline{MS}}^{(2)}r_0$ at finite
lattice spacings for the SW-I action.
\label{fig:chiiwa}}
\end{figure}

\begin{figure}[hbt]
\includegraphics[width=8cm]{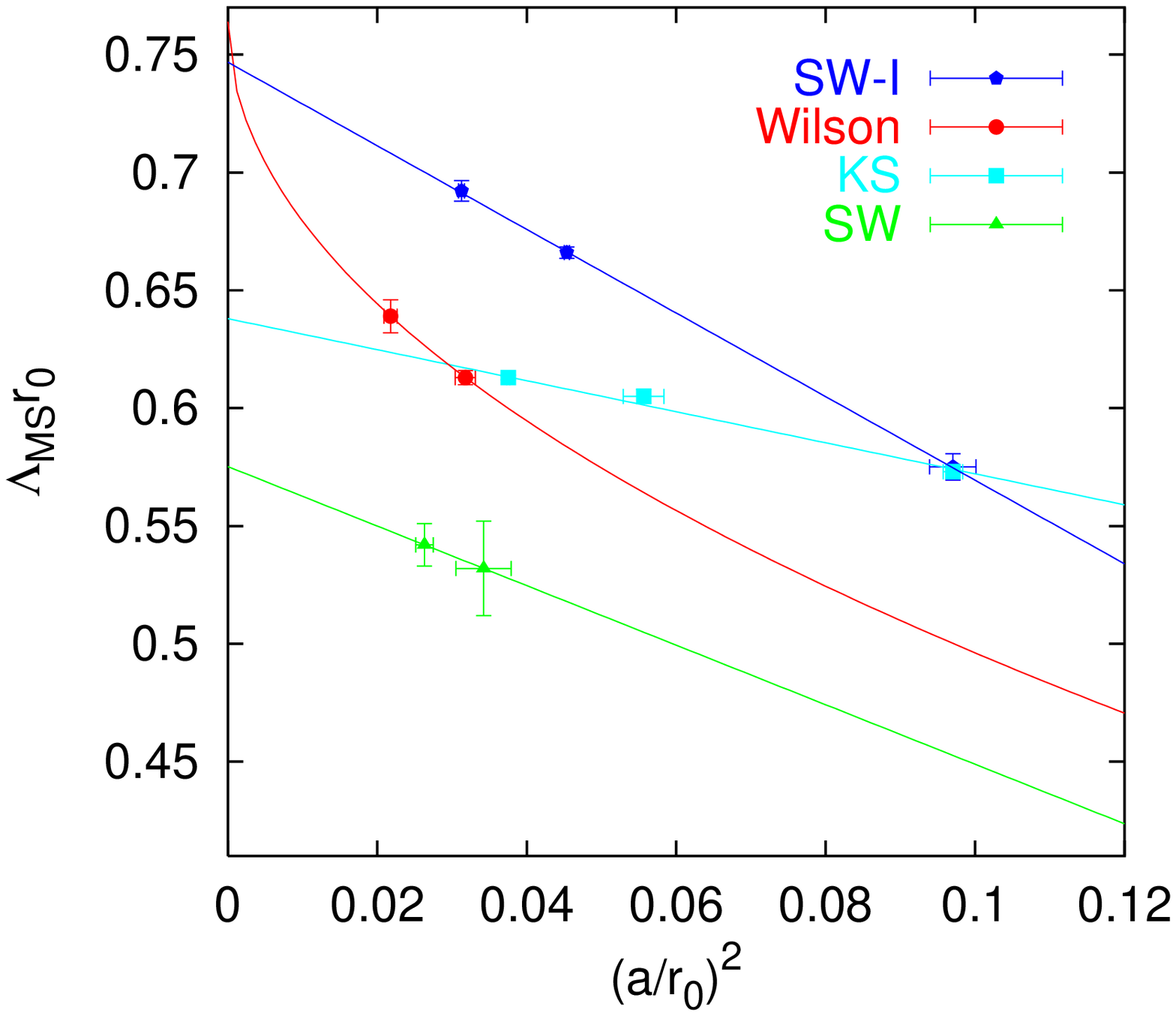}
\caption{Attempts of a continuum limit extrapolation (linear in
$a/r_0$ for Wilson and quadratic for KS and SW quarks) of
$\Lambda_{\overline{MS}}^{(2)}r_0$ from the available
``world lattice data''.
\label{fig:contextra}}
\end{figure}

We display our results on $\alpha_{\overline{MS}}(\mu)$ as well
as on $\Lambda_{\overline{MS}}r_0$ in Tabs.~\ref{tab:wilalpha}
-- \ref{tab:iwaalpha}. $\delta\Lambda_{m}^{\Box}$ refers to
the contribution to the $\Lambda_{\overline{MS}}$-parameters
due to the mass dependence of the coefficients $c_2$, $b_1$ (through
$\Delta K_1$) and is included into the $\Lambda_{\overline{MS}}$ estimates.
This mass dependence has not been considered
prior to the present study.
For SW and KS fermions the effect turns
out to be
of the order of the statistical uncertainty as long as $ma<0.1$.
However, for Wilson fermions this is a substantial effect, ranging from
10~\% at $ma\approx 0.06$ to 2~\% at $ma\approx 0.01$. In this case
the main effect is due to the
term $\Delta K_1(ma)$ in the matching
between lattice and $\overline{MS}$ scheme and we hence conclude
that un-improved Wilson fermions are not particularly well suited
for unambiguous
determinations of the running coupling, unless data at rather small
quark masses are available.
We have neglected the mass dependence in our analysis of the SW-I
data (Tab.~\ref{tab:iwaalpha}), which is equally small
as in the SW-Wilson case.

We use our perturbative result on the
potential to estimate the ratio $\Lambda_{\overline{MS}}r_0$
from the potential $aV(1)$, $aV(2)$ and force $F_{12}$ for the three
$n_f=2$ data sets with Wilson glue.
The resulting numbers are compared to the results from the plaquette
in Tabs.~\ref{tab:shortfullwil} -- \ref{tab:shortfullks}. Like in the quenched
case the estimates from $V(2)$ consistently turn out to be bigger by
about 15~\%, in comparison to those obtained from $V(1)$ or the plaquette,
however, in the case of KS quarks the estimates from $F_{12}$ are somewhat
more in line
with those from the plaquette. At present we do not have data at
sufficiently many
different lattice spacings for a detailed comparison like the
one presented
in Fig.~\ref{fig:shortquench2} for the $n_f=0$ case. 
Hence, we will only use
the most ultra violet quantity, the average plaquette, to extract the
$n_f=2$ $\Lambda$-parameter and assume a systematic uncertainty of 15~\%
due to higher order perturbative corrections,
based on our study of the pure gauge case as well as on the numbers in
Tabs.~\ref{tab:shortfullwil} -- \ref{tab:shortfullks}.

In Fig.~\ref{fig:alpharuncom} we compare our $n_f=2$ estimates on
$\alpha_{\overline{MS}}(\mu)$ from the plaquette to the $n_f=0$ results.
Due to the variation in the sea quark masses, action
and lattice spacings used, the $n_f=2$ points scatter around quite a bit but
yield consistently larger values of $\alpha$, compared to
the quenched case as already observed in lattice simulations of
the static potential, for instance in Ref.~\cite{Bali:2000vr} as well
as in Sec.~\ref{sec_data} below.
It is also clear from the figure that at present the $\mu$-window
covered by $n_f=2$ results is much smaller than the one available in
pure gauge studies.

Since chiral symmetry is explicitly broken for the fermionic actions used
we should first extrapolate our $\Lambda_{\overline{MS}}^{(2)}$ estimates
obtained at similar quark masses in physical units $mr_0$ to the continuum
limit, before attempting a chiral extrapolation:
$m\rightarrow m_u,m_d\approx 0$.
Due to the small range of lattice spacings covered by the
presently available data,
however, we are forced to interchange the ordering of these two
limits: after discarding lattices with $2.5\,a>r_0$,
i.e.\ the KS data with $\beta<5.415$ or $ma>0.05$,
we extrapolate in $\Lambda_{\overline{MS}}^{(2)}r_0$
at fixed $\beta$-values to small quark masses
and subsequently send the lattice spacing to zero.

The chiral extrapolations are illustrated in Fig.~\ref{fig:chiextra}
for Wilson glue and in Fig.~\ref{fig:chiiwa} for the SW-I action.
We include the $\beta=1.8$ data points into the latter figure but
we will discard these coarse lattice results from our continuum limit
extrapolation.
The last column of Tab.~\ref{tab:shortfullwil} reveals that the slopes
of $\Lambda_{\overline{MS}}^{(2)}r_0$ as a function of $mr_0$ would
have been much steeper for Wilson fermions
had we neglected the $ma$ dependence of the
matching between lattice and $\overline{MS}$ schemes.
In the course of each extrapolation the lattice spacing in physical
units $a/r_0$ changes. However, $\Lambda$ is no spectral quantity and
there is no theoretical handle on the functional form of such an extrapolation
anyway. We find linear fits,
\begin{equation}
\label{eq:chiextra}
\left[\Lambda_{\overline{MS}}^{(2)}r_0\right](mr_0) =
\left[\Lambda_{\overline{MS}}^{(2)}r_0\right](0)+c\,mr_0,
\end{equation}
with parameters $[\Lambda_{\overline{MS}}^{(2)}r_0](0)$ and $c$,
to be consistent with the $mr_0<0.25$ ($mr_0<0.35$ for
the SW-I action) data.
The resulting parameter values as well as $a/r_0(m=0)$
can be found in Tab.~\ref{tab:chiextra}. For the SW data
at $\beta=5.20$ no fit is possible since
$N_{DF}=0$ and at $\beta=5.29$ we included a rather heavy
mass, $mr_0\approx 0.45$.

Finally we attempt a continuum limit extrapolation in
Fig.~\ref{fig:contextra}, where we have generously estimated
error bars for the SW quarks at $\beta=5.20$.
The lines represent extrapolations, linear in $a/r_0$ for Wilson fermions
and quadratic in $a/r_0$ for KS and SW quarks. The extrapolated values are
$\Lambda_{\overline{MS}}^{(2)}r_0=0.638(3)$ with $\chi^2/N_{DF}=0.75/1$
for KS quarks, $\Lambda_{\overline{MS}}^{(2)}r_0\approx 0.764$
and $\Lambda_{\overline{MS}}^{(2)}r_0\approx 0.575$
for Wilson and SW fermions, respectively, in which cases $N_{DF}=0$.
For the SW-I action we obtain
$\Lambda_{\overline{MS}}^{(2)}r_0=0.747(2)$ with $\chi^2/N_{DF}=0.05/1$.
As a final result we quote the (unweighted) average of the KS
and SW-I data sets,
\begin{equation}
\Lambda_{\overline{MS}}^{(2)}r_0=0.69\pm 0.15,
\end{equation}
where the error estimate reflects
the uncertainties due to higher order perturbative corrections
as well as in the extrapolations, both chiral and to the
continuum limit.

The above number should be related to the recent
more optimistic estimate
$\Lambda_{\overline{MS}}^{(2)}r_0=0.553(34)$ by the QCDSF and UKQCD
collaborations~\cite{Booth:2001qp} which is
based on the UKQCD SW data alone. Note that their result agrees
with our above value, $\Lambda_{\overline{MS}}^{(2)}r_0\approx 0.575$,
obtained from the same SW data set by use of somewhat different methods
in the conversion between schemes and extrapolations. For instance
in Ref.~\cite{Booth:2001qp} $\alpha_{\overline{MS}}(\mu)$ was extracted
at NNLO, estimating
the as yet unknown NNLO fermionic contribution
to the plaquette, while we
consistently  worked at NLO.

Fig.~\ref{fig:contextra} paints a superficially disappointing
picture for universality of the continuum limit. Universality
absolutely must hold for lattice field theory to be a worthwhile exercise,
so one might be concerned about the large difference
in particular between the two different results obtained with SW fermions.
While we try to estimate systematics by
incorporating this difference into our error, 
it is interesting to consider how one could expect the figure to 
mature as improved data from future simulations become available.

Firstly, we note that the KS and SW-I continuum limit extrapolations are
much better
controlled than those for Wilson or SW quarks with Wilson glue,
due to the availability
of an additional data point. 
There are sizable ${\mathcal O}(a^2)$
corrections to the current KS and SW data. 
The Wilson continuum limit result has
been obtained by drawing a straight line through two data points but
there is no reason to believe ${\mathcal O}(a^2)$ corrections to
be any smaller (nor indeed any larger) in this case than in the
SW and KS cases. Likewise, it is not clear whether part of the large slope
of the SW-I data set (which is only perturbatively
improved such that ${\mathcal O}(\alpha_L a)$ corrections
are still present) might be due to a residual linear contribution.
One might hope that the scatter of the results will
be reduced by future data at different lattice spacings that
would allow us to discriminate between
linear and quadratic terms.

Secondly, we note that perturbative coefficients for the effects
of the clover term in the SW action are particularly large throughout the
literature, and refer the reader to Eq.~(\ref{eq:a16}) for
an example of an ${\mathcal O}(\alpha^2)$ coefficient and
Eqs.~(\ref{eq:k2222}) and (\ref{eq:k3333}) for ${\mathcal O}(\alpha^3)$. 
There is no automatism that guarantees ${\mathcal O}(\alpha^3)$
corrections to the plaquette in
determinations of $\alpha_{\overline{MS}}$ to be small, just because
this happened to be the case in the quenched theory with Wilson glue.
Given the experience
of the large $c_{SW}^{\nu}$ corrections we find 
it at least plausible that perturbative corrections to the SW 
determinations are larger than for the two other formulations. A calculation
of these might result in a better agreement between the different
continuum extrapolations too.

We remain confident that a combination of more lattice data and
higher order perturbative calculations
will greatly reduce the systematic spread shown in Fig.~\ref{fig:contextra}.
Of course eventually the ordering of chiral and continuum limit extrapolations
should be interchanged, unless data with
chirally symmetric lattice fermions become available.

We finally note that efforts are on the
way of calculating the $n_f=2$ $\Lambda$-parameter with the help
of finite size
scaling techniques in the Schr\"odinger functional scheme. At present
the error from this approach~\cite{DellaMorte:2002vm} is
of ${\mathcal O}(50 \mbox{\%})$ and the
scale has still to be related to a physical quantity like $r_0$ or
$f_{\pi}$, however, some improvement can be expected in the near future.

\subsubsection{Implications for the ``real'' world}
Running the estimate $\Lambda_{\overline{MS}}^{(2)}r_0=0.69(15)$
to the mass of the $Z$ vector boson $m_Z\approx 91.2$~GeV results in the
value,
\begin{equation}
\alpha_{\overline{MS}}^{(2)}(m_Z)=0.0945(30)(7),
\end{equation}
where the second error that is smaller than 1~\%
is due to the 5~\% scale ambiguity~\cite{Bali:1997am}
in the phenomenological
value $r_0\approx 0.5$~fm. This subdominant uncertainty can
be reduced by using a quantity directly accessible in experiment
like $f_{\pi}$ or $m_N$ to set the scale. Nature of course does not
contain two but five quark flavours lighter than the
$Z$ boson and we have to address this problem if we intend to
produce numbers that have phenomenological relevance.

\begin{figure}[hbt]
\includegraphics[width=8cm]{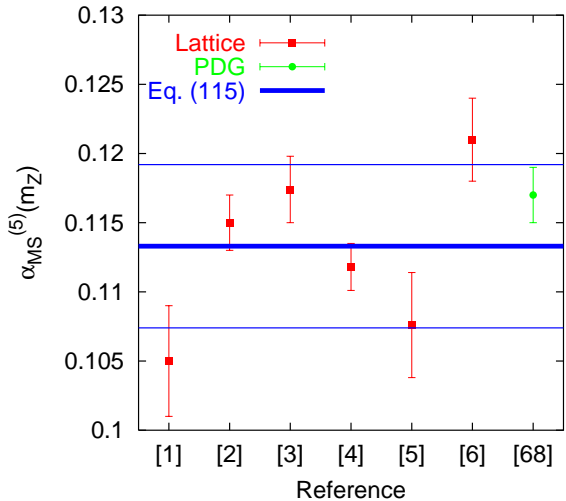}
\caption{Our estimate of $\alpha_{\overline{MS}}^{(5)}$ (error band),
in comparison with numbers quoted in previous lattice studies as well
as with the average from the ``Review of
Particle Physics'' (PDG).
\label{fig:alpha5}}
\end{figure}

The main difficulty of a QCD prediction is the effect
of the strange quark. 
Given that we cannot resolve the shift
between $\Lambda^{(0)}_{\overline{MS}}r_0=0.60(5)$ and
$\Lambda^{(2)}_{\overline{MS}}r_0=0.69(15)$
within the errors, for the moment being we assume that
the strange quark which is heavier than the up and the down
quarks anyway
will not have a large impact on this ratio either
and guess $\Lambda^{(3)}_{\overline{MS}}r_0
\approx \Lambda^{(2)}_{\overline{MS}}r_0$.
However, we try to incorporate the uncertainty of this assumption
(as well as the scale error of $r_0$) by somewhat inflating the
systematic error:
\begin{equation}
\Lambda^{(3)}_{\overline{MS}}=270(70)\,\,\mbox{MeV}.
\end{equation}

Our result of Sec.~\ref{sec:beta} indicates that a perturbative
matching of the $n_f=3$ and $n_f=4$ couplings around the charm threshold
should be reliable. Further, the difference between a three and a
four-loop $\overline{MS}$
running from a scale of 1.1~GeV upwards corresponds to a relative
shift in $\alpha_{\overline{MS}}$ at the $Z$ resonance
of only 0.26~\%. 
Hence, we perturbatively match the coupling at the charm and
bottom thresholds.
We allow pole masses $1.1$~GeV$ <m_c<1.5$~GeV and
4.4~GeV $<m_b<5.0$~GeV and arrive at
the estimate,
\begin{equation}
\label{eq:preal}
\alpha_{\overline{MS}}^{(5)}(m_Z)=0.1133(59),
\end{equation}
using the four-loop $\beta$-function.
The effect of varying the respective quark masses $m_c$ and $m_b$
as well as differences between a three and four-loop running of the coupling
have been added linearly in the above error.

The result is compatible with the average quoted in the ``Review of
Particle Physics''~\cite{Hagiwara:pw}:
$\alpha_{\overline{MS}}^{(5)}(m_Z)=0.1172(20)$ but lower than
results from D0/CDF or LEP experiments, a tendency that is consistent
with previous lattice
estimates~\cite{El-Khadra:1992vn,davi1,davi2,Spitz:1999tu,Booth:2001qp}.
In Fig.~\ref{fig:alpha5} we relate various
lattice studies in chronological order to ours. We find that,
although all lattice results on this
quantity~\cite{El-Khadra:1992vn,davi1,davi2,Spitz:1999tu,Booth:2001qp,davi3}
(as well as the PDG world average~\cite{Hagiwara:pw})
lie within the error bars that we obtain,
many of the previous numbers turn out to be incompatible with each other,
within their quoted uncertainties. For instance we obtain
a $\chi^2/N_{DF}=17.2/5$ when averaging the six lattice
results, and this
assuming that the quoted errors, that are dominated by systematics,
only represent one standard deviation!

The least well controlled step in our prediction Eq.~(\ref{eq:preal})
is certainly
the extrapolation $n_f=0,2\longrightarrow n_f=3$.
A first lattice study with ``$n_f=3$'' exists~\cite{davi3}, however,
it is too early to investigate quark mass effects and to
attempt a continuum
limit extrapolation in this case.
Once lattice results with 2+1 quark flavours 
of a similar quality as those
analysed here become available
a determination of $\alpha_{\overline{MS}}(m_Z)$ with about
4~\% accuracy using
the methods explored above should become realistic,
and by both, a full fledged
${\mathcal O}(\alpha^3)$ calculation in lattice perturbation theory
and more simulation points that would allow for a better controlled
continuum limit extrapolation
this error should be reducible to 1 -- 2~\% within the next few years.

We experienced that systematic errors are easily underestimated in this
kind of study, in particular since neither the size of higher order
perturbative corrections nor
the functional form of the chiral and continuum limit extrapolations
are fully controlled.
However, using a set of different lattice observables and actions
certainly helps in arriving at realistic error estimates.
Perturbation theory is naturally most reliable when the scales involved are
high and it is very expensive to increase the energy $a^{-1}$ in
lattice simulations, at least if one keeps the simulation volume
fixed in physical units. This makes the average plaquette a
particularly valuable quantity in studies like the present one. Future
predictions might benefit if lattice data on the plaquette, not only in
the fundamental but also in the
adjoint and higher representations,
were available as well as data on other short distance quantities
like the ``chair'', the ``parallelogram'' and the ``rectangle''.

\subsection{Comparison to the non-perturbative static potential\label{sec_data}}We shall compare our perturbation theory results with
data from simulations of QCD without and with sea quarks.
First we explain what we mean by boosted and tadpole improved boosted
perturbation theory in the present context. Next we investigate different
possibilities to parametrize
the pure gauge static potential  at short distances, based on our results
of Sec.~\ref{sec_potl}. We conclude with an attempt to resolve
the different running of the coupling when including sea quarks
{}from non-perturbative data.

\subsubsection{Parametrizations of the potential}
In what follows we will refer to the parametrization
\begin{equation}
aV({\mathbf R}a)=g^2V_1({\mathbf R})
\end{equation}
with $V_1$ defined in Eq.~(\ref{eq:v1per}) [and Eq.~(\ref{eq:defv1})]
as leading order (LO). Apart from a different value of
$g^2=6/\beta$ this result does not depend on $n_f$.
The NLO result can be found in Eqs.~(\ref{eq:edcba}) --
(\ref{eq:defvf}). The LO result above
corresponds to
\begin{equation}
aV(Ra)-aV_S(a)\longrightarrow -C_F\frac{g^2}{4\pi R}\quad(R\rightarrow\infty),
\end{equation}
where $aV_S=C_Fg^2\times 0.252731\ldots$ while the parametrizations
of the NLO and NNLO curves in the $R\gg 1$ limit
with and without massive sea quarks
can be found in Appendix~\ref{sec:larger}. In the pure gauge case
we will also compare against NNLO expectations for which
the small $R$ lattice effects have not yet been calculated.
When including fermions
this is not possible since even in the massless case
the contribution to the self energy
is only known to NLO.

In addition to expansions in terms of the bare coupling
we shall also compare against ``boosted'' perturbation theory expectations.
In this case we use the $\Lambda_{\overline{MS}}$ values of
Tabs.~\ref{tab:shortquench3} and \ref{tab:wilalpha}, calculated from
the respective plaquette at NLO as input.
Since we shall compute the coupling $\alpha_V(q)$ at NNLO this is somewhat
inconsistent. However, we also aim at a comparison with the fermionic case
where such a determination of the $\Lambda$-parameter is at present
only possible at NLO and prefer to compare like with like. As we have
seen above, the numerical difference
between NLO and NNLO turns out to be small anyway for this quantity.
Using $\Lambda_{\overline{MS}}r_0$ as input
we numerically run $\alpha_{\overline{MS}}(\mu)$ to the
required scales $\mu=e^{-5/6}q^*$, according
to the four-loop $\beta$-function. Some of the
respective $aq^*({\mathbf R})$
values
can be read off Tab.~\ref{tab:qstar} and for large $R$ we
employ the parametrization Eq.~(\ref{eq:qstarp}).
Slightly deviating from Sec.~\ref{sec_alpha_s}, the resulting
$\alpha_{\overline{MS}}(\mu)$ is then
always converted into $\alpha_V(q^*)$ at the
two loop level [${\mathcal O}(g^6)$], at bLO as well as at bNLO
and bNNLO, via
Eqs.~(\ref{eq:alphavex}) -- (\ref{eq:a2shift})
and Eq.~(\ref{eq:a2lll}). This means that
what we call bLO includes some NLO and NNLO
contributions and bNLO includes a conversion from the
$\overline{MS}$ to the $V$ scheme at NNLO. However, the conversion from
the momentum space continuum to
the position space lattice potential is done at the order indicated.

Finally, we introduce ``tadpole improved'' boosted perturbation theory.
We start from the observation that each link that appears within
any lattice observable contains classes of tadpole diagrams
that are lattice specific. We try to cancel these ultra violet contributions
in part by a non-perturbative prescription~\cite{Parisi:1980pe,lepage}:
the second of the two Wilson loops that appear within
the difference $aV(a{\mathbf R})=
\lim_{T\rightarrow\infty}\{\ln[W({\mathbf R},T)]-\ln[W({\mathbf R},T+1)]\}$
contains two more links than the first one.
We hence prefer to perturbatively expand
the combination $aV(a{\mathbf R})+\ln\Box/2$ instead, within which
the net number of links is zero. We have already made use of this
trick in Eq.~(\ref{eq:selft}). Finally, we reexpress the result in
terms of $\alpha_V(q^*_N)$, boosting our perturbative series, where
the new $q^*_N({\mathbf R})$ values can easily be calculated from those
of the potential $q^*({\mathbf R})$ and the plaquette $q^*_{\Box}$: 
\begin{equation}
\ln(aq^*_N({\mathbf R}))=\frac{\ln[aq^*({\mathbf R})]
V_1({\mathbf R})-\ln(aq^*_{\Box})c_1/2}{V_1({\mathbf R})-c_1/2}.
\end{equation}
The $\alpha_V(q^*_N)$ are then obtained as described above.
We shall refer to this third method as tbLO, tbNLO and tbNNLO, depending
on the order of the conversion from $\alpha_V[q^*_N({\mathbf R})]$
to $aV({\mathbf R})$.

\subsubsection{Pure gauge theory}
\begin{figure}[hbt]
\includegraphics[width=8cm]{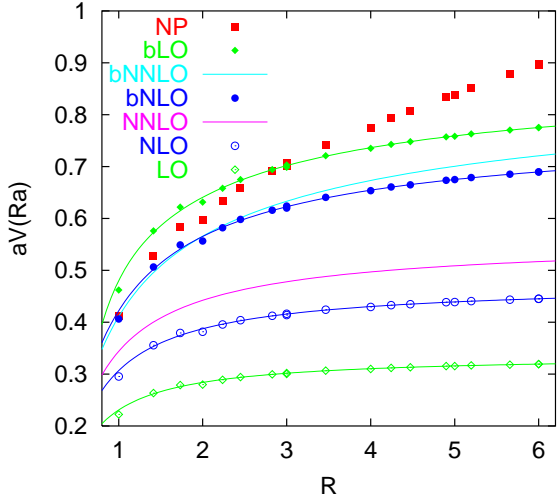}
\caption{
\label{fig:quenched} Comparison between quenched lattice data at $\beta=6.0$
and boosted (b) and un-boosted LO, NLO and NNLO
perturbative potentials. The curves are the large $R$ expectations
from continuum perturbation theory.}
\end{figure}

In Fig.~\ref{fig:quenched} we compare LO, NLO, NNLO and bLO, bNLO
and bNNLO expectations against the static potential at $\beta=6.0$.
At this lattice spacing
we use $\Lambda_{\overline{MS}}r_0=0.562(3)$ and there is
no free parameter.
The un-boosted results come closer to the non-perturbative data as
the order of the calculation increases but even at NNLO only about half of the
potential can be explained, even at distances as short as $R=1$.
The boosted results are more in line with the data but still only on a very
qualitative level at best. It is also obvious that if one tuned the
$\Lambda$-parameter such that $V(1)$ is reproduced, $V(2)$ would be undershot,
in agreement with our experience from Sec.~\ref{sec_alpha_s} above.
We can define an effective Coulomb coupling,
\begin{equation}
\label{eq:erun}
e(R)=-R[aV(R)-aV_S],
\end{equation}
and find $e\approx 0.106$ at LO while $e(3)\approx 0.19$
and $e(3)\approx 0.26$ at NLO and NNLO, respectively.
Finally, bNLO results in $e(3)\approx 0.49$.

\begin{figure}[hbt]
\includegraphics[width=8cm]{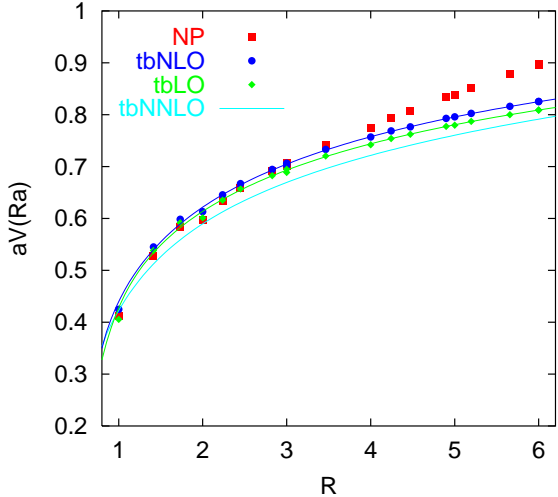}
\caption{
\label{fig:quenched2} The non-perturbative potential at $\beta=6.0$,
in comparison with boosted and tadpole improved perturbation.}
\end{figure}

\begin{figure}[hbt]
\includegraphics[width=8cm]{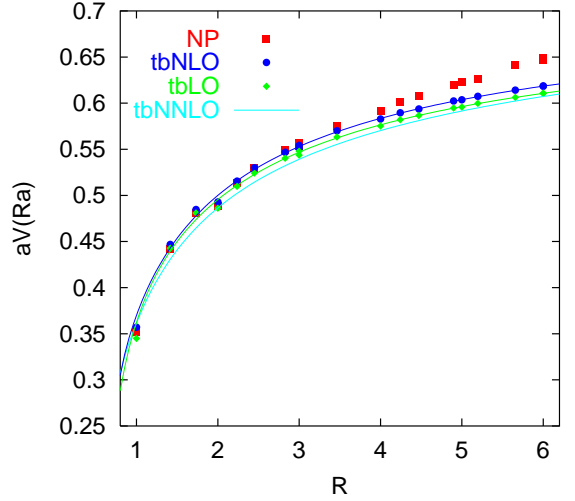}
\caption{
\label{fig:quenched3} The same as Fig.~\ref{fig:quenched2} at $\beta=6.4$.}
\end{figure}

Next in Fig.~\ref{fig:quenched2} we try out the tbLO, tbNLO and tbNNLO
parametrizations and indeed
tadpole improvement vastly reduces the difference with the NP potential.
To exclude that this
is just accidental we also display a comparison with the potential at 
$\beta=6.4$ in Fig.~\ref{fig:quenched3}
where the plaquette yields $\Lambda_{\overline{MS}}r_0=0.603(10)$,
and with the smaller gauge coupling the situation becomes even more convincing.
It turns out to be absolutely essential to compare
like with like.
Contrary to previous
claims~\cite{Bali:1999ai,Necco:2001gh}
not only the force but also the short range potential can be understood in
terms of perturbation theory, provided that the self energy contribution
is dealt with in a consistent way:
when comparing with NP lattice results the self energy
should not be subtracted from
the perturbative expansion while when matching
with the $\overline{MS}$ scheme in which $V_S^{\overline{MS}}=0$
by definition
it has to be subtracted at the same scales $\mu(r)$ at which
the interaction energy $V_{\mbox{\scriptsize int}}(r)=V(\mu;r)-V_S(\mu)$
is calculated,
to avoid renormalon ambiguities.
A similar observation
has been made in a recent comparison with continuum perturbation
theory in a renormalon based approach~\cite{Pineda:2002se}.
In Ref.~\cite{Bali:1999ai}
a comparison was made between the
NP lattice potential at short distances and the perturbative
$\overline{MS}$ scheme prediction. In this study
one and the same perturbative
self energy had been used at all distances, which was obtained from
a fit to the data, and consequently, no agreement was found.

\begin{figure}[hbt]
\includegraphics[width=8cm]{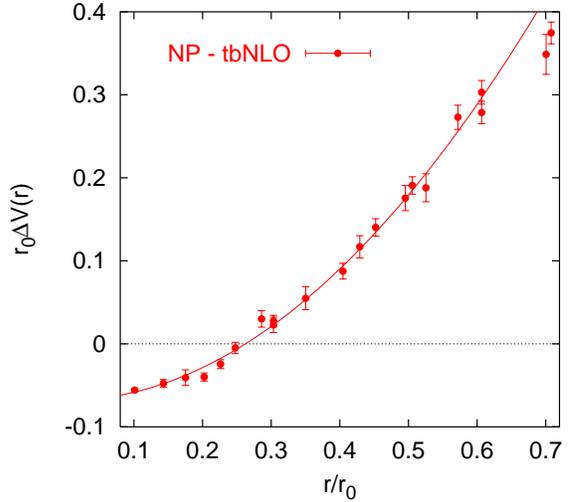}
\caption{Non-Perturbative minus perturbative potentials
at $\beta=6.4$, together with a quadratic fit, Eq.~(\ref{eq:quad}).
The distance $r=0.4\,r_0\approx 1\,\mbox{GeV}^{-1}$ corresponds
to about 4 lattice spacings.
\label{fig:quenched4} }
\end{figure}

In Fig.~\ref{fig:quenched4}
we display the difference between non-perturbative and
perturbative potentials, $\Delta V(r)=V(r)-V^{\mbox{\scriptsize tbNLO}}(r)$ for the
example of $\beta=6.4$ as a function of $R=r/a$ in physical units
$r_0/a=9.89(16)$. Within the statistical errors of the non-perturbative
potential all points lie on a smooth curve indicating
that the lattice effects are
accounted for by the perturbative result.
However, there are deviations between prediction and lattice data:
at short distances we overestimate the NP potential by 30 MeV and
around 0.3~fm we underestimate the result by 110~MeV.
We fit the difference for $r<0.61\,r_0\approx 0.3$~fm by
the phenomenological ansatz,
\begin{equation}
\label{eq:quad}
\Delta V(r)=b+cr^2
\end{equation}
with $b=-0.068(3)\,r_0^{-1}$, $c=0.99(6)\,r_0^{-3}$ and $\chi^2/N_{DF}=
18.2/17$. Allowing for an additional linear term does not improve the quality
of the fit and the term turns out to be in agreement with zero:
unlike in Ref.~\cite{Bali:1999ai}
no linear term is required at short distances to understand the data and this
without any free parameter in the computation of $\Delta V(r)$!
We notice that the coefficient
$c\approx (1.6\Lambda_{\overline{MS}})^3$ is of order one in units
of the QCD $\Lambda$-parameter.

\begin{figure}[hbt]
\includegraphics[width=8cm]{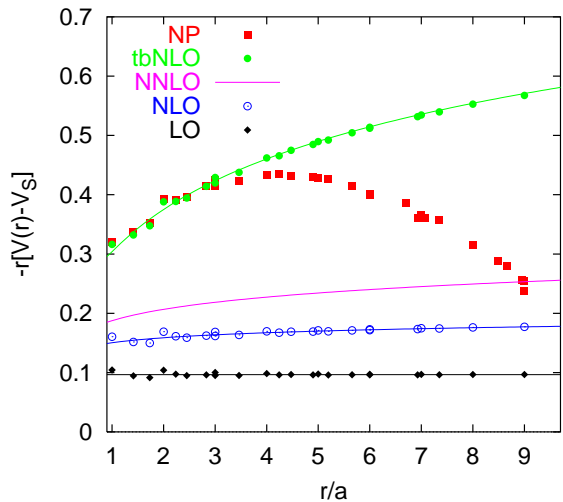}
\caption{
\label{fig:quenched5} The running of the Coulomb coupling at $\beta=6.6$,
where $5.1\,a\approx 0.4\,r_0\approx 0.2$~fm~$\approx 1$~GeV${}^{-1}$.
We have subtracted the value $aV_S^{\mbox{\scriptsize tbNLO}}\approx
0.65$ from the non-perturbative
potential.}
\end{figure}

\begin{figure}[hbt]
\includegraphics[width=8cm]{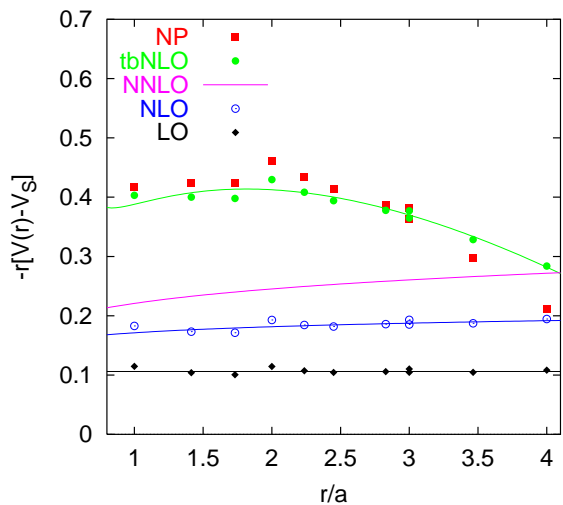}
\caption{
 \label{fig:quenched6} The same as Fig.~\ref{fig:quenched5} at $\beta=6.0$,
where $2.1\, a\approx 0.4\,r_0$ and we
subtract the estimate $aV_S^{\mbox{\scriptsize NP}}\approx
0.83$.}
\end{figure}

In view of our plan to resolve the differences in the running of the
coupling between quenched and un-quenched simulations,
in Fig.~\ref{fig:quenched5} we display the dimensionless combination
$e(r/a)$
of Eq.~(\ref{eq:erun}) for $\beta=6.6$ where $5.1\,a\approx 0.4\,r_0\approx
1\,\mbox{GeV}^{-1}$. At this $\beta$ value
the average plaquette yields $\Lambda_{\overline{MS}}r_0=0.601(7)$.
We have also included bare LO, NLO and NNLO results for comparison.
The self energies have been subtracted from all perturbative potentials,
LO, NLO, NNLO as well as tbNLO at  the respective orders
in perturbation theory.
As what we are doing is a linear transformation
we do not wish to subtract different $V_S$ values
from the NP and tbNLO potentials which would spoil the qualitative
agreement between
the results found above. Ideally
one would subtract a non-perturbatively determined $V_S$ which then
guarantees scaling between different NP data sets. The difference between
NP and tbNLO however will then result in the tbNLO coupling to diverge
linearly at large $R$ where perturbation theory will become unreliable anyway.
In the absence of a non-perturbative determination of $V_S$ we
will take the $\beta=6.6$ tbNLO value $aV^{\mbox{\scriptsize tbNLO}}_S\approx 0.650$ as
an estimate which is indeed close at least to the tbNNLO result
$aV^{\mbox{\scriptsize tbNNLO}}_S\approx 0.670$.
Doing this means that the tbNLO
Coulomb coupling will increase logarithmically with $R$ and
no linear term is present.

In the figure we see a gap opening up between NP and tbNLO potentials
for $Ra>4a\approx 0.15$~fm. The NP data starts diverging towards
negative values, which is expected as $e(r)\rightarrow -\sigma r^2$
as $r\rightarrow\infty$ where $\sigma$ denotes the string tension.
In addition, our estimate of an NP self energy might be
wrong by a few per cent which will result in an unwanted
linear contribution to $e$.
We find relatively large values of the Coulomb coupling, $e\approx 0.4$,
that naturally disagree with $e\approx 0.3$ as obtained
from phenomenological fits to a Cornell type
potential~\cite{Bali:2000vr,Bali:2000gf}, $aV(R)=V_0-e/R+a^2\sigma\,R$,
in which the linear confining term contributes at all distances:
the parameter $e$ within this parametrization can only be interpreted
as a Coulomb coupling at distances $r\ll\sqrt{\sigma}$.
We observe that the running of the coupling is quite steep at short distances
which provides us with hope that differences between the quenched
and un-quenched $\beta$-functions can be resolved from simulation
data. 

In Fig.~\ref{fig:quenched6} we display the same
comparison for a similar $R$ range in physical units at
$\beta=6.0$ where
$2.1\,a\approx 0.4\,r_0\approx
1\,\mbox{GeV}^{-1}$. In this case the difference between
tbNLO and tbNNLO self energies is 15~\% such that we
cannot regard these values as reliable anymore.
Instead we compute $a^{6.0}V_{S}^{6.0}
=a^{6.0}V^{6.0}(r_0)-(R_{0}^{6.6}/R_{0}^{6.0})a^{6.6}
[V^{6.6}(r_0)-V_{S}^{6.6}]$, where the superscripts refer to
the respective $\beta$ values,
and add this non-perturbative difference to the $\beta=6.6$
$aV_S$ estimate above.
In doing this we arrive at $aV_S^{\mbox{\scriptsize NP}}\approx 0.83$ which we subtract from both
NP and tbNLO data sets.
Now at large $R$ the tbNLO coupling will diverge linearly towards negative
values as our $V_S^{\mbox{\scriptsize NP}}$ estimate differs from $V_S^{\mbox{\scriptsize tbNLO}}$ by as much
as $0.35\,a^{-1}$.
It is obvious when comparing for instance the dimensionless
NP couplings at $R\approx 3.5$ with $R\approx 8.5$
that both NP data sets scale nicely.
We see however that at short distances
the agreement with the perturbative expectation at $\beta=6.0$
is somewhat worse
than that at $\beta=6.6$ above.
The gap between NP and tbNLO curves opens up in the same region
in lattice units, around $R\approx 4$, which corresponds
to a larger distance in physical units. Since the perturbative expectation
decreases at large $r$ only because of the difference between
$aV_S^{\mbox{\scriptsize NP}}$ and $aV_S^{\mbox{\scriptsize tbNLO}}$
we should not expect
agreement at $R>\sqrt{5}$ anymore and
interpret this behaviour as accidental. This view is confirmed by our
$\beta=6.2$ and $\beta=6.4$ data sets: the improved agreement at small
$r$ goes along with the gap between perturbative and NP potentials
opening up earlier.

\subsubsection{Comparison with $n_f=2$}\label{sec:nf2c}
We compare the quenched potential with a potential obtained with sea quarks
for the example of two flavours of Wilson fermions at $\beta=5.6$
and $\kappa=0.1565$. These parameter values have been chosen to correspond
to a quenched lattice coupling of $\beta=6.0$:
in the $n_f=2$ case one finds~\cite{Bali:2000vr}
$r_0=5.28(5)a$ while a quenched $\beta=6.0$ yields
$r_0=5.33(3)a$.
The effective string tensions, determined by three parameter
Cornell fits agree within errors too:
$\sigma a^2=0.0466(14)$ vs.\ $\sigma a^2=0.0479(7)$, such that
the linear term vanishes when subtracting the two potentials
from each other. The $\kappa$ value corresponds to a quark mass
$ma\approx 0.040$ and we will use this value to obtain the perturbative
expectations.

\begin{figure}[hbt]
\includegraphics[width=8cm]{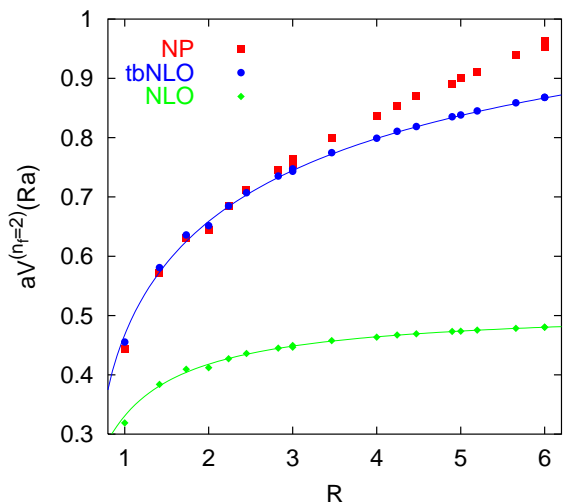}
\caption{The non-perturbative potential for Wilson fermions
at $\beta=5.6$, $\kappa=0.1565$, i.e.\ $ma\approx 0.04$,
in comparison with NLO tadpole improved boosted and bare perturbation
theory.
\label{fig:ferm}}
\end{figure}

In Fig.~\ref{fig:ferm} we compare the NP $n_f=2$ potential with
NLO and tbNLO perturbation theory. In the latter case we used the
value $\Lambda_{\overline{MS}}r_0=0.570(6)$, as obtained from the
plaquette $\Box=0.57073$. The situation is very similar to
that of the quenched potential at $\beta=6.0$ depicted in
Figs.~\ref{fig:quenched} and \ref{fig:quenched2} which is why
we refrain from including LO, bLO, bNLO and tbLO curves. The respective
NNLO results are not known at present, even at large $R$.

\begin{figure}[hbt]
\includegraphics[width=8cm]{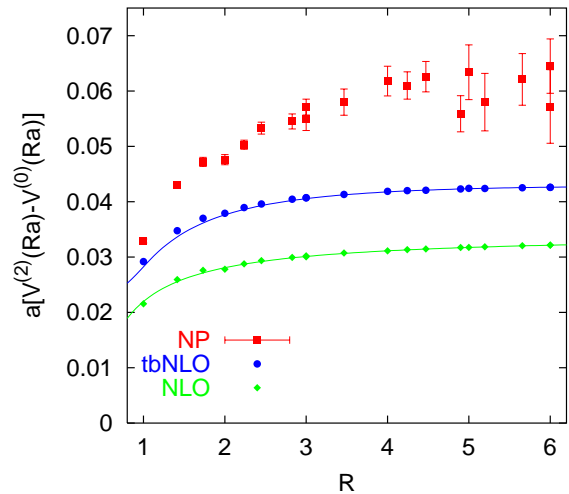}
\caption{
\label{fig:differm} The difference between the $n_f=2$ lattice potential
with Wilson fermions at $\beta=5.6$, $\kappa=0.1565$
and the quenched potential at $\beta=6.0$: non-perturbatively vs.\
perturbatively.}
\end{figure}

In a next step we calculate the difference between the $n_f=2$ and
the $n_f=0$ potentials. This is depicted in Fig.~\ref{fig:differm},
together with the respective differences calculated
in tbNLO and NLO.
The NP linear confining term cancels in this combination,
such that there is some reason to believe that
this difference should be more perturbative than the potentials themselves,
at least as long as $R<12$, the distance at which the QCD string
will eventually ``break'' in
the $n_f=2$ case.
In the NP as well as perturbative data sets lattice artefacts are clearly
visible that can be ascribed to the effect of the sea quarks.
In the case of NLO we have depicted the $n_f=0$ expectation using
the slightly smaller one loop
matched value $\beta=5.969$ rather than $\beta=6.0$ for internal
consistency. 

As $R\rightarrow\infty$ all three curves approach constant values as
the massive fermions start to decouple from the running. Although the shape of
all curves is similar, even the tbNLO prediction underestimates the NP
data by almost 30~\%. This is not surprising as the vertical scale 
has been vastly inflated after subtracting two quantities of similar size
from each other. For instance we can almost completely
eliminate the gap between
the NP and tbNLO results by increasing the
$\Lambda_{\overline{MS}}^{(2)}r_0$ value (or decreasing the
$\Lambda_{\overline{MS}}^{(0)}r_0$ value) by only 3~\%.
We find it quite encouraging that at least
on a qualitative level we are able to
resolve sea quark effects that only appear
{}from ${\mathcal O}(\alpha^2)$ onwards in perturbation theory.
Unsurprisingly, a quantitative understanding is not possible
without knowledge of the difference to second non-trivial
order, i.e.\ ${\mathcal O}(\alpha^3)$ in this case.

\begin{figure}[hbt]
\includegraphics[width=8cm]{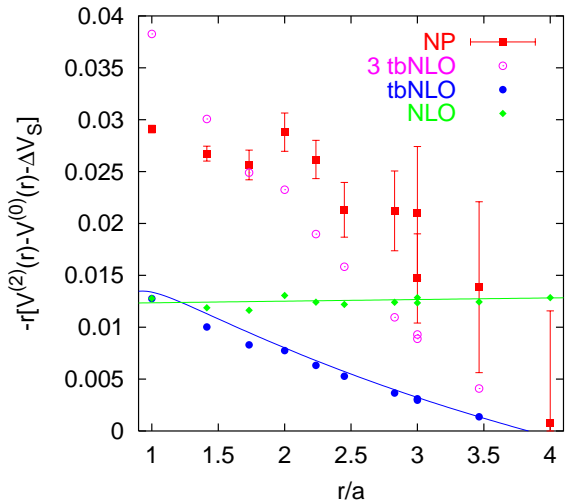}
\caption{
\label{fig:differm2}
The same as Fig.~\protect\ref{fig:differm} but with the
difference of the respective self energy contributions subtracted,
and multiplied by $-r$. We estimate $a\Delta V_S^{\mbox{\scriptsize NP}}\approx 0.062$
and find $a\Delta V_S^{\mbox{\scriptsize tbNLO}}\approx 0.042$
and $a\Delta V_S^{\mbox{\scriptsize NLO}}\approx 0.034$. ``3 tbNLO'' refers to the
tbNLO result, multiplied by three.}
\end{figure}

We now turn to the difference in the running of the two respective
Coulomb couplings, between $n_f=2$ and $n_f=0$ and depict the quantity
$\Delta e(R)=-Ra[V^{(2)}(Ra)-V^{(0)}(Ra)-\Delta V_S]$
in Fig.~\ref{fig:differm2}. This combination further inflates scale
and errors, in particular at large $R$. 
In the tbNLO case
we can now identify deviations between the curve that constitutes the
large $R$ limit and the points calculated in lattice perturbation
theory at short distances. This bias, which also appeared
in Fig.~\ref{fig:massive} and which turns out to be
an ${\mathcal O}(ma)$ lattice artefact,
has been discussed in Sec.~\ref{sec:contu} above.
Since at this lattice spacing, and within the scale of the figure,
perturbative predictions and NP data cannot be expected to
agree with each other anymore (cf.~Fig.~\ref{fig:differm})
we subtract different $\Delta V_S$ values from the three data
sets. While we compute $a\Delta V_S^{\mbox{\scriptsize NLO}}$ and $a\Delta V_S^{\mbox{\scriptsize tbNLO}}$
in perturbation theory, we fit $a\Delta V_S^{\mbox{\scriptsize NP}}\approx 0.062$
from the NP data depicted in Fig.~\ref{fig:differm}.

The NP data indicates an enhancement of the Coulomb coupling in the $n_f=2$
case at the given mass by almost 0.03 over the quenched coupling
at short distances. This enhancement then
reduces to zero at large distances where the two data sets are matched.
Contrary to this we find the bare NLO expectation to be rather flat
as a function of
$R$: the running of the coupling with $R$ is faster when fermions
are included, however, this effect is compensated for by a larger
value of $g^2$, in comparison with the quenched case. This
larger bare coupling can
be translated into a deflation of
the distance scale which in turn slows down the running again.
In the end one would expect an increase of $e$ by 0.012 -- 0.013
to NLO.
The tbNLO prediction includes some slope but falls short of the
NP data by a factor of about 3. After multiplying the points by this
factor (open circles) the agreement with the NP $\Delta e(R)$ is
at best qualitative. We should not forget however that in the figure
we have vastly inflated tiny differences.
Unfortunately, $n_f=2$ lattice simulations at finer lattice spacings
are not yet available such that we cannot test to what extent
the quality of the parametrization improves as the continuum limit
is approached.

\section{Summary}
\label{sec:conclude}
We have calculated Wilson loops, the static potential and the static
source self energy to ${\mathcal O}(\alpha^2)$ in lattice perturbation
theory with the Wilson gluonic action as well as with
massive Wilson-Sheikholoslami-Wohlert and Kogut-Susskind fermions
for arbitrary representation of the external sources.
Some numbers for ``improved''  gluonic actions are provided too.
The results are useful for understanding violations of rotational symmetry
and the effect of including sea quarks on
the lattice. Lattice results can be related to other
regularization schemes like the perturbatively defined $\overline{MS}$
scheme, for instance to allow for an extraction of the QCD coupling
at high energies from low energy hadronic phenomenology.

We find that at presently employed quark masses
lattice spacing effects on the short distance potential
are not smaller in absolute terms for the two non-perturbatively
${\mathcal O}(a)$ improved fermionic actions, relative to the
Wilson action. However, improvement results in a reduced mass dependence
around $ma=0$
of quantities like the static source self energy, that is required for
determinations of the $b$ quark mass in an effective field theory framework,
small Wilson loops and in the matching between lattice and
$\overline{MS}$ schemes at finite $a$.

We perturbatively determined $\beta=2N/g^2$ shifts between
the pure gauge theory and QCD with massive sea quarks.
The results appear to be reliable within 10~\% for quark masses
down to 20~MeV for Wilson type fermions (which translates
into less than 1~\% uncertainty
in the value of the predicted matched $\beta$) and within 25~\% for
``$n_f=2$'' KS fermions within a window of lattice spacings
that correspond to quenched $\beta$ values $5.8<\beta<6.15$.
This method might turn out to be useful in predicting the impact
of a shift in the sea quark mass onto correlation lengths like $r_0/a$
in future simulations. In particular we confirm that for SW fermions
the lines of constant $r_0/a$ and $m_{\pi}a$ are significantly tilted
with respect to those of constant $\beta$ and $ma$, respectively,
a fact that complicates continuum limit extrapolations.

The success of the perturbative matching of the running coupling
at flavour thresholds via an intermediate mass-dependent scheme,
in our case the potential scheme, in lattice simulations (which
allow us to compare with the correct non-perturbative result)
means that the standard procedures employed in perturbative
calculations~\cite{Bernreuther:1983zp,Brodsky:1998mf} should be
quite reliable even at the charm quark mass.

We investigated so-called boosted perturbation theory
methods~\cite{Parisi:1980pe,lepage} and found them to work well in many cases.
However, such methods have to be digested with some caution: as
boosted ${\mathcal O}(\alpha^2)$ perturbation theory is tuned
to reduce higher order and in particular ${\mathcal O}(\alpha^3)$
corrections, differences between results calculated at these two orders
are not necessarily indicative of the systematic uncertainties involved.
In the absence of even higher order calculations, the perturbative
uncertainty in  for instance predictions of the running coupling
can be estimated by varying both, the quantity from which the coupling is
determined and the lattice spacing $a$.

In analysing quenched lattice data on the static potential and
plaquette we obtain the value $\Lambda_{\overline{MS}}^{(0)}r_0=0.61(9)$
which is in agreement with $\Lambda_{\overline{MS}}^{(0)}r_0=0.60(5)$
obtained by use of finite size techniques by the ALPHA
Collaboration~\cite{Bode:1999hd}. The error which is dominated by
the uncertainty of higher order perturbative corrections
can be systematically reduced by an ${\mathcal O}(\alpha^3)$
calculation of the static potential.
Using Wilson, KS and SW $n_f=2$ lattice data, the latter with
Wilson as well as with Iwasaki glue, we estimate
$\Lambda_{\overline{MS}}^{(2)}r_0=0.69(15)$ in the continuum and
chiral limit, where the error is again
dominated by systematics. We find
this value to depend strongly on both, the quark mass and 
the lattice spacing.
Chiral and in particular continuum limit extrapolations
are not fully under control but the use of different input quantities
and lattice actions allows us to estimate the systematic uncertainties.

Based on the $n_f=0$ and $n_f=2$ estimates we guess
$\Lambda_{\overline{MS}}^{(3)}=270(70)$~MeV. This value corresponds to
$\alpha_{\overline{MS}}^{(5)}(m_Z)=0.1133(59)$.
In spite of the fact that more lattice data entered our analysis
than has been used in any of the previous attempts to
determine the $n_f=2$
running coupling, our error estimate is not smaller but larger
than those that have been quoted by various lattice groups in the
past~\cite{El-Khadra:1992vn,davi1,davi2,Spitz:1999tu,Marcantonio:2001fc,Booth:2001qp,davi3}.
The main reason is that the use of different quark actions and observables
for the first time enabled us to realistically assess the uncertainties
connected to higher order perturbative corrections and chiral
and continuum extrapolations.

We estimate that
the error on the QCD coupling at the $Z$ boson mass can be reduced to
about 4~\%, once a set of $n_f=2+1$ lattice simulations,
similar to those that are already available
for $n_f=2$, becomes available. By improving the
perturbative calculation to NNLO as well as by
incorporating a wider range
of lattice spacings into the numerical simulations, errors of
1 -- 2 \% appear to be realistic within the next few years.
It might be possible to further improve on this quality
by a dedicated lattice simulation of running quark masses and
coupling~\cite{DellaMorte:2002vm}.

We found that the lattice potential at distances smaller than about
1~GeV${}^{-1}$ is quite well described by NLO boosted tadpole improved
perturbation theory without any free parameter, as long as no attempt
is made to subtract the power divergent static self energy.
This observation is somewhat in contrast to previous
beliefs~\cite{Bali:1999ai,Necco:2001gh} and
supports the view that perturbation theory for this quantity 
(and not only for the force)
might be quite reliable in the continuum too.
This is so as long as renormalons (in the ${\overline{MS}}$ scheme)
and power divergences (in the lattice scheme)
are properly dealt with when
rearranging the perturbative series~\cite{Pineda:2002se,sumino}.
At distances larger than 1~GeV${}^{-1}$ the linear confining
term sets in rather rapidly and it is doubtful that
this behaviour can be emulated completely by higher order
perturbative corrections. At distances shorter than 1~GeV${}^{-1}$
the difference between non-perturbative data and perturbative expectation
is dominated by a quadratic term with a coefficient of ${\mathcal O}(1)$
in units of $\Lambda_{\overline{MS}}$. The dependence of this coefficient
on the way in which the perturbative series is organized, on the
$\Lambda_{\overline{MS}}r_0$ value that is used and on the lattice spacing
deserves further study.

It turned out to be possible to resolve the difference in the logarithmic
running of the coupling between two flavour QCD and the quenched
approximation from
non-perturbative data on the static potential. This difference
is in rough qualitative agreement with the NLO expectation, however,
about three times larger than expected. Unfortunately,
at present we are limited to inverse lattice spacings of 2~GeV
such that it is not possible to judge whether the discrepancy
reduces as the scale is increased. As the magnitude of the effect should
also increase as the quark mass is reduced, next generation lattice
data at smaller quark masses
should turn out helpful too.

\begin{acknowledgments}
This work has been supported by the EU network
HPRN-CT-2000-00145.
During this work G.B.\ has benefited from
a Heisenberg Fellowship (DFG grant Ba~1564/4-1)
and a PPARC Advanced Fellowship
(grant PPA/A/S/2000/00271). He
has also received funding from PPARC grant
PPA/G/O/1998/00559. P.B.\ acknowledges support
{}from PPARC grant PPA/J/S/1998/00756.
We thank Christine Davies, Antonio Pineda and
Rainer Sommer for helpful comments and
Sonali Tamhankar, Urs Heller and Alan Irving
for making available to us MILC, HEMCGC and UKQCD results, respectively.
We express our gratitude to the SESAM/T$\chi$L Collaboration
for the permission to include some unpublished results on
plaquette values and short distance potentials.
\end{acknowledgments}
\appendix
\section{Perturbative relations between lattice and continuum
schemes\label{AppendixMSlat}}
We review the known perturbative relations between
the lattice $\beta$-function and the 
modified minimal subtraction $(\overline{MS})$ scheme for various
lattice actions in $SU(N)$ gauge theory with $n_f$ fermion flavours.
We also display continuum perturbation theory results
for the (mass dependent) static potential in position space, which
constitute the limit against which our calculations will converge
for $r\gg a$.
\subsection{Conventions}
We define the QCD $\beta$-function as,
\begin{equation}
\label{eq:beta}
\beta(\alpha)=\frac{d\alpha}{d\ln\mu^2}=-\beta_0\alpha^2-\beta_1\alpha^3-
\beta_2\alpha^4-\cdots.
\end{equation}
While in mass-independent schemes like lattice
regularization\footnote{Strictly speaking, in lattice regularization,
$\beta_i$ only become
scale independent in the continuum limit $a\rightarrow 0$, i.e.\
when scales $\mu\ll a^{-1}$ and quark masses
$m_i\ll a^{-1}$ are considered. In general
$\beta_i$ will obtain additional non-universal contributions
that are functions of
$am_i$ and $a\mu$. While contributions of the latter type are
genuinely non-perturbative in character, quark mass
dependent contributions
arise in perturbation theory too.} or minimal
subtraction
the first two coefficients,
\begin{eqnarray}
\label{eq:beta0}
\beta_0&=&\left(\frac{11}{3}N-\frac{2}{3}n_f\right)\frac{1}{4\pi},\\
\label{eq:beta1}
\beta_1&=&\left[\frac{34}{3}N^2-\left(\frac{13}{3}N-\frac{1}{N}\right)n_f
\right]\frac{1}{(4\pi)^2},
\end{eqnarray}
are universal,
$\beta_2$ will in general depend on the renormalization scheme.
In the $\overline{MS}$ scheme $\beta_2$ reads~\cite{Tarasov:1980au},
\begin{eqnarray}
\beta_2^{\overline{MS}}&=&\left[\frac{2857}{54}N^3-
\left(\frac{1709}{54}N^2-\frac{187}{36}-\frac{1}{4N^2}\right)n_f\right.
\nonumber\\
&&+\left.
\left(\frac{56}{27}N-\frac{11}{18N}\right)n_f^2\right]\frac{1}{(4\pi)^3}.
\end{eqnarray}

The QCD $\Lambda$-parameter in a given scheme is defined as,
\begin{equation}
\label{eq:deflambda}
\Lambda = \lim_{\mu\rightarrow\infty} \mu\, \exp\left(
-\frac{1}{2\beta_0\alpha(\mu)}\right)
\left[\beta_0\alpha(\mu)\right]^{-\frac{\beta_1}{2\beta_0^2}}.
\end{equation}
We discuss the conversion between two different renormalization schemes:
let,
\begin{equation}
\label{eq:defpr}
\alpha'(\mu)=\alpha(\mu)+c_1\alpha^2(\mu)+c_2\alpha^3(\mu)+\cdots.
\end{equation}
{}From the definition of the $\beta$-function, Eq.~(\ref{eq:beta}),
one infers,
\begin{eqnarray}
\beta_0'&=&\beta_0,\quad\beta_1'=\beta_1,\\
\beta_2'&=&\beta_2-c_1\beta_1+(c_2-c_1^2)\beta_0,\label{eq:beta2}\\
\Lambda'&=&\Lambda\, e^{c_1/(2\beta_0)}\label{eq:lambda}.
\end{eqnarray}
If not only the scheme but also the matching scale $\mu$ is shifted,
one easily obtains the relation,
\begin{equation}
\label{eq:conv1}
\alpha'(\mu')=\alpha(\mu)+c_1'\alpha^2(\mu)+c_2'\alpha^3(\mu)+\cdots,
\end{equation}
with,
\begin{eqnarray}
\label{eq:scalec1}
c_1'&=&c_1-2\beta_0\ln\left(\frac{\mu'}{\mu}\right),\\
c_2'&=&c_2-\left(2\beta_1+4c_1\beta_0\right)\ln\left(\frac{\mu'}{\mu}\right)
\nonumber\\\label{eq:scalec2}
&&+4\beta_0^2\ln^2\left(\frac{\mu'}{\mu}\right)\\\nonumber
&=&c_2-2\beta_1\ln\left(\frac{\mu'}{\mu}\right)+{c_1'}^2-c_1^2.
\label{eq:conv3}
\end{eqnarray}

\subsection{Conversion between the $\overline{MS}$ and lattice schemes}
The $\overline{MS}$ coupling is related to the lattice coupling
$\alpha_L=g^2/(4\pi)$ via,
\begin{equation}
\label{eq:cmsbarl}
\alpha_{\overline{MS}}(a^{-1})=\alpha_L+b_1\alpha_L^2+b_2\alpha^3+\cdots,
\end{equation}
where for the Wilson and Sheikholeslami-Wohlert (SW) quark actions one
obtains~\cite{Hasenfratz:1980kn,Weisz:1981pu,Luscher:1995np,Alles:1998is,Christou:1998ws},
\begin{equation}
\label{eq:b1}
b_1=-\frac{\pi}{2N}+k_1N+K_1(ma)n_f,
\end{equation}
where $K_1(ma)=K_1(0)+\Delta K_1(ma)$,
with the numerical values,
\begin{eqnarray}
k_1&=&2.135730074078457(2),\\\label{eq:K1}
K_1(0)&=&-0.08414443(6)\nonumber\\&&
+0.0634188788(1)c_{SW}\label{eq:a16}\\\nonumber
&&-0.3750240693(1)c_{SW}^2,
\end{eqnarray}
for the Wilson parameter $r=1$. The tree-level value of the
SW coefficient, $c_{SW}=1+3.3414(9)\alpha_L+\cdots
=c_{SW}^{(0)}+c_{SW}^{(1)}g^2+\cdots$,
is $c_{SW}^{(0)}=1$.
In the case of Kogut-Susskind (KS) fermions $K_1(0)$
reads~\cite{Weisz:1981pu,Luscher:1995zz},
\begin{equation}
\label{eq:K12}
K^{KS}_1(0)=-0.03298341916(1).
\end{equation}
Note that in our notation $n_f$ KS quark flavours correspond to
$n_f$ (not $4\,n_f$)
flavours of Dirac quarks in the continuum limit. Therefore, $n_f$ has to
be a multiple of four.
The value of $k_1$ applies to the Wilson gluonic action only.
However, the contribution $K_1(ma)$ (whose quark mass dependence has been
calculated for the first time in this publication and is displayed in
Tab.~\ref{tab:shift}) remains the same for Symanzik type
improvements of the Wilson gluonic
action.

At vanishing quark mass or lattice spacing,
$m\ll a^{-1}$ we obtain
the massless limit, i.e.\ $\Delta K_1(0)=0$.
The mass dependent term results in a change of
the QCD $\beta$-function at finite $a$:
\begin{equation}
\beta_0^L=\beta_0^{\overline{MS}}-
\sum_i\frac{a}{2}\frac{d\Delta K_1(m_ia)}{da}.
\end{equation}
From this we can infer, $\lim_{ma\rightarrow 0}d\Delta K_1(ma)/d\ln a
=\lim_{ma\rightarrow 0}d\Delta K_1(ma)/d\ln(am)=0$: the dependence
is regular in $ma$ near the origin. Since the $\beta$-function is no
on-shell quantity, we cannot exclude terms proportional to
$ma$, even for improved quark actions.
By demanding that
in the limit $ma\rightarrow\infty$ the quenched $\beta$-function is
re-obtained, we find,
\begin{equation}
\label{eq:k1limit}
\lim_{x\rightarrow\infty}\frac{d\Delta K_1(x)}{d\ln x}=-\frac{1}{3\pi},
\end{equation}
i.e.\ $-\Delta K_1$ will grow logarithmically
for large $ma$. This is expected since $ma\rightarrow\infty$ means
that the coupling on the left hand side of Eq.~(\ref{eq:cmsbarl})
runs with $n_f$ active flavours while the running
on the lattice side of this equation
corresponds to that of pure Yang-Mills theory. The
difference has to be compensated for by $\Delta K_1(ma)$.
We will determine the integration constant in
the region $ma\rightarrow\infty$ 
in Eq.~(\ref{eq:k1la}) below from matching the un-quenched and quenched
static potentials in the infra red.

The coefficient $b_2$ is at present only known for massless Wilson-SW quarks
and Wilson gluonic action:
\begin{equation}
\label{eq:b2}
b_2=b_1^2+\frac{3\pi^2}{8N^2}+k_2+k_3N^2+\left(\frac{K_2}{N}+K_3N\right)n_f
\end{equation}
with~\cite{Christou:1998ws,Bode:2001uz}
\begin{eqnarray}
k_2&=&-2.8626215972(6),\\
k_3&=&1.24911585(3),\\
K_2&=&0.1890(2)\label{eq:k2222}\\\nonumber
&&-0.02492(5)c_{SW}-0.83585(3)c_{SW}^2\\\nonumber
&&-0.79942(5)c_{SW}^3-0.012947(2)c_{SW}^4,\\
K_3&=&-0.1579(3)\label{eq:k3333}\label{eq:K3}\\\nonumber
&&-0.00540(6)c_{SW}+0.7684(1)c_{SW}^2\\\nonumber
&&+0.033842(5)c_{SW}^3+0.006920(2)c_{SW}^4.
\end{eqnarray}

From Eqs.~(\ref{eq:beta2}) and (\ref{eq:lambda})
we infer,
\begin{eqnarray}
\beta_2^L&=&\beta_2^{\overline{MS}}+b_1\beta_1-(b_2-b_1^2)\beta_0,\\
\Lambda_{\overline{MS}}&=&\Lambda_L\exp\left(\frac{b_1}{2\beta_0}\right).
\end{eqnarray}
This results for instance in the numerical values,
\begin{eqnarray}
\Lambda^{(0)}_{\overline{MS}}&\approx& 28.809\,\Lambda_L^{(0)},\\
\Lambda^{(2)}_{\overline{MS}}&\approx& 41.053\,\Lambda_L^{W,(2)}
\approx 27.379\,\Lambda_L^{SW,(2)},
\end{eqnarray}
for $N=3$ colours.

\subsection{Conversion between the $\overline{MS}$ scheme and
schemes based on the static potential\label{sec:conve}}
We define the coupling $\alpha_V$ through the static QCD potential in
momentum space,
\begin{equation}
\tilde{V}(q)=-4\pi C_F\frac{\alpha_V(q)}{q^2},
\end{equation}
where $q=|{\mathbf q}|$ and $C_F=(N^2-1)/(2N)$.
The momentum space potential is related to that in position space
via a Fourier transformation,
\begin{equation}
\tilde{V}(q)=\int\frac{d^3x}{(2\pi)^3}[V(r)-V_S]e^{i{\mathbf q}
{\mathbf x}},
\end{equation}
where $r=|{\mathbf x}|$.
$V_S$ denotes the self energy term,
that vanishes in dimensional regularization
but will in general be present, and the position space potential,
\begin{equation}
\label{eq:vrdef}
V(r)=-C_F\frac{\alpha_R(r^{-1})}{r}+V_S,
\end{equation}
defines the coupling $\alpha_R$.
By Fourier transformation one obtains the relation between the two schemes,
\begin{equation}
\label{eq:RV}
\alpha_R(\mu)=\alpha_V\left(e^{-\gamma}\mu\right)
\left(1+\frac{\pi^2\beta_0^2}{3}\alpha^2\right),
\end{equation}
which holds for fermion masses $m_i\ll\mu$.
$\gamma=0.57721566\ldots$ is the Euler constant.

We write,
\begin{equation}
\label{eq:alphavex}
\alpha_V(\mu)=\alpha_{\overline{MS}}(\mu)+a_1\alpha_{\overline{MS}}^2(\mu)
+a_2\alpha^3+\cdots.\label{eq:defa1}
\end{equation}
The coefficients $a_1$ and $a_2$ are known\footnote{
In fact the leading log contribution to $a_3$~\cite{Brambilla:1999qa}
and, more recently, the complete NNLO result~\cite{Pineda:2000gz}
have been calculated too.}. For $n_f$ massless
quark flavours they read~\cite{Schroder:1999vy}:
\begin{eqnarray}
a_1&=&\left(\frac{31}{9}N-\frac{10}{9}n_f\right)\frac{1}{4\pi},\label{eq:a1}\\
a_2&=&\left\{\left(\frac{4343}{162}+4\pi^2-\frac{\pi^4}{4}+\frac{22}{3}
\zeta_3\right)N^2\right.
\nonumber\\\nonumber
&-&\left[\left(\frac{1798}{81}+\frac{56}{3}\zeta_3\right)\frac{N}{2}
+\left(\frac{55}{3}-16\zeta_3\right)\frac{C_F}{2}\right]n_f\\
&+&
\left.\frac{100}{81}n_f^2\right\}\frac{1}{16\pi^2}.
\end{eqnarray}
$\zeta_3$ denotes the value of the
$\zeta$-function
$\zeta(3)=1.20205690\ldots$.
The corresponding result for massive quark flavours with
masses $m_i$ has been obtained in Ref.~\cite{Melles:2000dq}:
\begin{widetext}
\begin{eqnarray}
\label{eq:a1run}
a_1(\{m_i\})&=&a_1+\frac{1}{6\pi}\sum_{i=1}^{n_f}\ln
\left(1+\frac{C_0m_i^2}{\mu^2}\right),\\
a_2(\{m_i\})&=&a_2+\frac{19}{48\pi^2}\sum_{i=1}^{n_f}\left\{
\ln\left(1+\frac{C_3}{C_2}\frac{m_i^2}{\mu^2}+\frac{1}{C_2}
\frac{m_i^4}{\mu^4}\right)\right.\nonumber\\
&&+\left.\frac{C_3-2C_1}{\sqrt{C_3^2-4C_2}}
\left\{2\ln\left(1+\frac{C_3+\sqrt{C_3^2-4C_2}}{2C_2}\frac{m_i^2}{\mu^2}\right)
-\ln\left[1+\left(2+\frac{C_3}{C_2}\right)\frac{m_i^2}{\mu^2}\right]\right\}
\right\}\label{eq:a2run}\\\nonumber
&&+\frac{1}{54\pi^2}\sum_{i=1}^{n_f}\left(\frac{31}{2}N-5n_f\right)
\ln\left(1+\frac{C_0m_i^2}{\mu^2}\right)
+\frac{1}{36\pi^2}\sum_{i,j=1}^{n_f}
\ln\left(1+\frac{C_0m_i^2}{\mu^2}\right)
\ln\left(1+\frac{C_0m_j^2}{\mu^2}\right),
\end{eqnarray}
\end{widetext}
with the numerical values~\cite{Brodsky:1999fr},
\begin{eqnarray}
C_0&=&5.19(3),\label{eq:c0def}\\
C_1&=&-0.571(34),\\
C_2&=&0.221(15),\\
C_3&=&1.33(12).
\end{eqnarray}
A change in the argument on the left hand side of
Eq.~(\ref{eq:alphavex}), $\alpha_V(\mu)
\mapsto\alpha_V(\mu')$, results in the substitutions,
\begin{eqnarray}
\label{eq:a1shift}
a_1(\mu',\{m_i\})&=&a_1\left(\left\{m_i\frac{\mu}{\mu'}\right\}\right)
-2\beta_0\ln\left(\frac{\mu'}{\mu}\right),\\
a_2(\mu',\{m_i\})&=&a_2\left(\left\{m_i\frac{\mu}{\mu'}\right\}\right)
+4\beta_0^2\ln^2\left(\frac{\mu'}{\mu}\right)\label{eq:a2shift}\\\nonumber
&&-\left[2\beta_1+4\beta_0a_1\left(\left\{m_i\frac{\mu}{\mu'}\right\}\right)
\right]\ln\left(\frac{\mu'}{\mu}\right),
\end{eqnarray}
in analogy to Eqs.\ (\ref{eq:scalec1}) -- (\ref{eq:scalec2}).

The masses $m_i$ in Eqs.\ (\ref{eq:a1run}) -- (\ref{eq:a2run})
denote pole masses,
\begin{equation}
m=\left[1+\frac{C_F}{4\pi}\left(4-3\ln\frac{m^2}{\mu^2}\right)\alpha\right]
m_{\overline{MS}}(\mu).
\end{equation}
Re-expressing Eq.~(\ref{eq:a2run}) in terms of $\overline{MS}$
quark masses $m_i$
at a scale $\mu$, therefore, results in the replacement,
\begin{equation}
a_2^{\overline{MS}}(\{m_i\})
=a_2(\{m_i\})
+\frac{C_F}{12\pi^2}\sum_{i=1}^{n_f}\frac{4-3\ln\frac{m_i^2}{\mu^2}}{
1+\frac{\mu^2}{C_0m_i^2}},
\end{equation}
while Eq.\ (\ref{eq:a1run}) remains unaffected.
For Wilson fermions one obtains~\cite{Capitani:2001xi},
\begin{equation}
m_{\overline{MS}}(a^{-1})=\left(1+16.95241 \frac{C_F}{4\pi}\alpha\right)m_0.
\end{equation}
Hence, in terms of lattice quark masses $m_i(a)$, $a_2$ reads:
\begin{equation}
\label{eq:a2lll}
a_2^L(\{m_i\})=a_2(\{m_i\})
+\frac{C_F}{12\pi^2}\sum_{i=1}^{n_f}\frac{16.95241-3\ln m_i^2a^2}{
1+\left(C_0m_i^2a^2\right)^{-1}}.
\end{equation}

We now go to position space and write,
\begin{equation}
\label{eq:alpharms}
\alpha_R(\mu)=\alpha_{\overline{MS}}(\mu)
+a_1^R\alpha_{\overline{MS}}^2(\mu)
+a_2^R\alpha^3+\cdots.
\end{equation}
In following Eqs.~(\ref{eq:conv1}) -- (\ref{eq:conv3})
we obtain from
Eqs.~(\ref{eq:RV}) and (\ref{eq:alphavex}),
\begin{eqnarray}
a_1^R&=&a_1+2\gamma\beta_0,\\
a_2^R&=&a_2+4\gamma\beta_0a_1+2\gamma\beta_1+\left(4\gamma^2+
\frac{\pi^2}{3}\right)\beta_0^2.
\end{eqnarray}
The one loop result for massive quark flavours reads,
\begin{equation}
\label{eq:a1rm}
a_1^R(\{m_i\})=a_1^R+\frac{1}{3\pi}
\sum_{i=1}^{n_f}\mbox{Ein}\left(\frac{\sqrt{C_0}m_i}{\mu}
\right)
\end{equation}
with
\begin{equation}
\label{eq:defa1r}
\mbox{Ein}(x)=\gamma+
\ln(x)+E_1(x)=-\sum_{\nu=1}^{\infty}\frac{(-x)^{\nu}}{\nu\,\nu!}.
\end{equation}
For the corresponding two-loop mass dependence of the
position space potential we refer to Ref.~\cite{Melles:2000dq}.
The exponential integral $E_1(x)$
is defined by,
\begin{equation}
E_1(x)=\int_x^{\infty}dt\frac{e^{-t}}{t},
\end{equation}
i.e.
\begin{equation}
\mbox{Ein}(x)=\int_0^xdt\frac{1-e^{-t}}{t}.
\end{equation}

{}From Eqs.~(\ref{eq:defpr}), (\ref{eq:lambda}),
(\ref{eq:defa1}) and (\ref{eq:a1}) one obtains the
numerical values,
\begin{equation}
\label{eq:cvms}
\Lambda_V^{(0)}\approx 1.5995\,\Lambda_{\overline{MS}}^{(0)},\quad
\Lambda_V^{(2)}\approx 1.5213\,\Lambda_{\overline{MS}}^{(2)},
\end{equation}
while Eq.~(\ref{eq:RV}) results in,
\begin{equation}
\Lambda_R^{(n_f)}=e^\gamma\Lambda_V^{(n_f)}\approx 1.7811\,\Lambda_V^{(n_f)}.
\end{equation}
Note that for massive quarks the coefficients
of the $V$- and $R$-scheme $\beta$-functions
become explicitly scale dependent.
However, this does not affect the respective QCD $\Lambda$-parameters,
that are defined in the limit $\mu\rightarrow\infty$
[Eq.~(\ref{eq:deflambda})].

We obtain:
\begin{widetext}
\begin{eqnarray}
\frac{d\alpha_V}{d\ln\mu^2}=-\left[\beta_0^{(n_f)}
+\frac{1}{6\pi}\sum_{i=1}^{n_f}\left(1+\frac{\mu^2}{C_0m_i^2}\right)^{-1}\right]
\alpha_V^2(\mu)+\cdots\\
\frac{d\alpha_R}{d\ln\mu^2}=-\left\{\beta_0^{(n_f)}
+\frac{1}{6\pi}\sum_{i=1}^{n_f}
\left[1-\exp\left(-\frac{\sqrt{C_0}m_i}{\mu}\right)\right]\right\}
\alpha_R^2(\mu)+\cdots.
\end{eqnarray}
\end{widetext}
In the situation, $m_1=m_2=\cdots=m_{n_f-1}=0, m_{n_f}>0$,
the above coefficient functions smoothly
interpolate between $\beta_0^{(n_f-1)}$ (for $\mu/m_{n_f}\rightarrow 0$)
and $\beta_0^{(n_f)}$ (for $\mu/m_{n_f}\rightarrow\infty$) as one would
expect.

\subsection{The perturbative potential at large $R$.\label{sec:larger}}
We denote a dimensionless
distance measured in lattice units by $R$
while $r=Ra$ refers to the distance in physical units.
At separations $R\gg 1$, i.e.\ $r\gg a$,
rotational invariance should be restored and
the static potential obtained in lattice perturbation theory should
coincide with the one computed above by use of dimensional regularization.
In the large $R$ limit 
our results should therefore approach
the continuum expressions given in
Appendix~\ref{sec:conve} above.

We define,
\begin{eqnarray}
\label{eq:LR}
aV({\mathbf R}a)&=&v_1({\mathbf R})\alpha_L+v_2({\mathbf R})
\alpha_L^2+\cdots,\\
aV_S&=&v_1(\infty)\alpha_L+v_2(\infty)\alpha_L^2+\cdots,
\end{eqnarray}
where the self-energy
$V_S=2\delta m_{\mbox{\scriptsize stat}}$ is twice the lattice pole
mass of a static colour source.
The coefficient functions $v_i({\mathbf R})$
have been calculated in the present article.
At large separations we expect $v_i({\mathbf R})$ only to
depend on the modulus $R=|{\mathbf R}|$ and can write,
\begin{equation}
v_{c,i}(R)=v_i(R)-v_i(\infty).
\end{equation}
Eq.~(\ref{eq:vrdef}) now implies,
\begin{equation}
\label{eq:LRc}
a\left[V({\mathbf R}a)
-V_S\right]\longrightarrow -C_F\frac{\alpha_R\left[(Ra)^{-1}\right]}{R}\quad
(R\rightarrow\infty).
\end{equation}
It follows,
\begin{equation}
\label{eq:LRc1}
v_1({\mathbf R})-
v_1(\infty)\longrightarrow v_{c,1}(R)=-\frac{C_F}{R}\quad(R\rightarrow\infty),
\end{equation}
where
\begin{equation}
v_1(\infty)=C_F\times 3.1759115\ldots,
\end{equation}
for the Wilson gluonic action.
$v_2(\infty)$ is also known. Here we state the result for
massless Wilson-SW quark flavours and $N=3$:
\begin{eqnarray}
v_2(\infty)&=&C_F\times\{16.714(4)+[-0.42333(6)\\\nonumber
&&+0.0516(2)c_{SW}-0.5870(2)c_{SW}^2]n_f\}.
\end{eqnarray}
In the case of massless KS quarks the numerical value of the
$n_f$ coefficient is $-C_F\times 0.36846(6)$.
Results on the mass dependence of the fermionic coefficients
can be found in Tabs.~\ref{tab:mself} and \ref{tab:const}.
The $n_f=0$ result has first been derived in
Refs.~\cite{Weisz:1984bn,Heller:1985hx}.
The precision has subsequently
been increased by Martinelli and Sachrajda~\cite{Martinelli:1999vt}.
We do, however, slightly disagree with this reference and
obtain $\delta m_{\mbox{\scriptsize stat},2}=C_Fv_2(\infty)/2=11.143(3)$,
rather than 11.152~\cite{Martinelli:1999vt}.
The latter reference
includes the Wilson and SW contributions for massless fermions as well,
which we were able to reproduce with increased precision.
The mass dependence of $v_2(\infty)$
as well as the KS result are new.
The $n_f=0$ value for $v_3(\infty)$ has recently been obtained by
means of stochastic perturbation theory~\cite{DiRenzo:2001nd}:
\begin{equation}
\label{eq:v33l}
v_3(\infty)=C_F\times 129(2).
\end{equation}
This result agrees with the one of Ref.~\cite{Trottier:2001vj}
$v_3(\infty)=C_F\times
130(1)$, obtained from a fit to high $\beta$ Monte Carlo
data.

{}From Eqs.~(\ref{eq:LR}) -- (\ref{eq:LRc1}) we obtain,
\begin{equation}
\label{eq:relxxx}
\alpha_R\left[(Ra)^{-1}\right]=\alpha_L+e_1(R)\alpha_L^2+
e_2(R)\alpha_L^3+\cdots
\end{equation}
with
\begin{equation}
e_i(R)=\frac{v_{c,i+1}(R)}{v_{c,1}(R)}=-\frac{R}{C_F}v_{c,i+1}(R).
\end{equation}
In converting the
scheme and scale by use of Eqs.~(\ref{eq:alpharms}),
(\ref{eq:cmsbarl}) and (\ref{eq:conv1}) -- (\ref{eq:scalec2}),
we obtain the coefficients
$e_1(R)$ and $e_2(R)$:
\begin{widetext}
\begin{eqnarray}
\alpha_R\left[(Ra)^{-1}\right]&=&\alpha_{\overline{MS}}[(Ra)^{-1}]+
a_1^R(\{Rm_i\})\alpha_{\overline{MS}}^2[(Ra)^{-1}]+
a_2^R(\{Rm_i\})\alpha^3\nonumber\\
&=&\alpha_L
+\left[
b_1+a_1^R(\{Rm_i\})+2\beta_0\ln R\right]\alpha_L^2
+\left\{b_2+a_2^R(\{Rm_i\})+2b_1a_1^R(\{Rm_i\})\right.\nonumber\\
&+&\left.
[2\beta_1+4(b_1+a_1^R(\{Rm_i\}))\beta_0]\ln R+
4\beta_0^2\ln^2 R\right\}\alpha^3.
\end{eqnarray}
Therefore,
\begin{eqnarray}
\label{eq:b1fit}
v_{c,2}(R)&=&-\frac{C_F}{R}\left[b_1+a_1^R(\{Rm_i\})+
2\beta_0\ln R\right],\\
v_{c,3}(R)&=&-\frac{C_F}{R}\left\{b_2+a_2^R(\{Rm_i\})+2b_1a_1^R(\{Rm_i\})
+[2\beta_1+4(b_1+a_1^R(\{Rm_i\}))\beta_0]\ln R+
4\beta_0^2\ln^2 R\right\}.\label{eq:vc3}
\end{eqnarray}
This amounts to
\begin{equation}\label{eq:resvc2}
v_{c,2}(R)=-\frac{C_F}{R}\left\{A_0+A_1\ln R+\sum_{i=1}^{n_f}\left[
\Delta K_1(m_ia)+\frac{1}{3\pi}
\mbox{Ein}(\sqrt{C_0}m_iaR)\right]\right\}
\end{equation}
\end{widetext}
with\footnote{We have factorized out the Casimir constant
$C_F$ such that the above results also apply to the potential between
charges in representation $D$, replacing $C_F$ by $C_D$.
For $N\neq 3$ the constant $C_0$ has not yet been
calculated.}\footnote{Rather than
comparing the perturbative lattice potential
at large distances with known results, one can also use
Eq.~(\ref{eq:b1fit}) to determine the parameters $b_i$
from the calculated $v_{i+1}(R)-v_{i+1}(\infty)$ at large
$R$. While in precision
this procedure can in no way compete with the 
values displayed in Eqs.~(\ref{eq:K1}) or (\ref{eq:K12}),
calculating big Wilson loops and extracting
the lattice potential by means of stochastic
perturbation theory~\cite{Parisi:1981ys,DiRenzo:2001nd}
might turn out to be
a feasible alternative to diagrammatic techniques
for more involved quark and gluonic
actions.}\footnote{$v_{c,i}$ are subject to $(am)^{\nu}$ and
$(a/r)^{\nu}$ lattice corrections, where $\nu=1$ in the case of
Wilson fermions and $\nu=2$ for the KS and SW actions. While the $r$ dependent
corrections vanish at large $r$ where we match to the continuum scheme,
the $am$ corrections will in general
be present in this limit. For $0<ma<\infty$ $\Delta K_1$
can be defined by matching the lattice scheme to a
mass-dependent scheme like the $R$-scheme. We do this by
imposing that $\Delta K_1$ contains all $am$ corrections to $v_{c,2}$.
One can use different quantities to define $\Delta K_1$, for instance
the potential in momentum space. Since in general
the $(ma)^{\nu}$ corrections for different observables will differ,
$\Delta K_1$ can only be fixed up to an $(ma)^{\nu}$ ambiguity.
This is a reflection of the loss of universality of the
$\beta$-function,
due to the mass dependence of $\beta_0$ (and $\beta_1$).
In the continuum limit, $a\rightarrow 0$ universality will be restored.}
the numerical values for $n_f$ flavours of
Wilson-SW fermions in $SU(3)$ QCD,
\begin{eqnarray}
\label{eq:respvc21}
A_0&\approx&7.7164259753\\\nonumber
&-&[0.233808-0.063419(1)c_{SW}+0.37524(1)c_{SW}^2]n_f,\\
A_1&\approx&1.75070437401-0.10610329540\,n_f.\label{eq:respvc2}
\end{eqnarray}
For KS fermions the fermionic contribution to
$A_0$ differs:
\begin{equation}\label{eq:respvc22}
A_0^{KS}\approx 7.7164259753-0.1826473162\,n_f.
\end{equation}
The function $\mbox{Ein}(x)$ is defined in Eq.~(\ref{eq:defa1r})
and can be approximated for small arguments by,
\begin{equation}
\label{eq:a1rm2}
\frac{1}{R}\mbox{Ein}\left(\sqrt{C_0}m_iaR\right)=
\sqrt{C_0}m_ia+{\mathcal O}\left[(m_iaR)^2\right]:
\end{equation}
to leading order the coefficient $v_{c,2}(R)$ is parallel shifted
in proportion to $ma$ at short distances.

At infra red scales, $Ra\gg m_i^{-1}$, $E_1(x)\propto e^{-x}/x$,
such that the massive flavours decouple from the
logarithmic running of the potential.
For $n_f$ massive degenerate flavours
with masses $m_i=m$, we find:
\begin{eqnarray}
\label{eq:mat}
v_{c,2}^{(n_f)}(R)&&\stackrel{R\rightarrow\infty}{\longrightarrow}
v_{c,2}^{(0)}(R)\\\nonumber
&&-\frac{n_fC_F}{R}\left\{K_1(ma)+\frac{1}{3\pi}
\left[\ln\left(\sqrt{C_0}ma\right)-\frac{5}{6}\right]\right\}.
\end{eqnarray}

For massless fermions we
can parametrize the next order term in the expansion
of the lattice potential $v_{c,3}$ by [Eq.~(\ref{eq:vc3})],
\begin{equation}
\label{eq:v3c}
v_{c,3}=-\frac{C_F}{R}\left(B_0+B_1\ln R+B_2\ln^2R\right).
\end{equation}
The fermionic part of
$b_2$, which is required to calculate 
the contribution to $B_0$ that is proportional to $n_f$,
is only known for massless Wilson-SW fermions.
Restricting ourselves to massless Wilson quarks and $SU(3)$ we obtain the
numerical values,
\begin{eqnarray}
B_0\approx&&73.8166085803\\\nonumber
&&-3.667(1)\,n_f+0.0530952\,n_f^2,\\
B_1\approx&&28.3102065046\\\nonumber
&&-2.295710\,n_f+0.0496157\,n_f^2,\\
B_2\approx&&3.06496580518\label{eq:v3cc}\\\nonumber
&&-0.3715110067\,n_f+0.01125790929\,n_f^2.
\end{eqnarray}
$B_2$ is known exactly while the accuracy of the
term within $B_0$ that is
proportional to $n_f$ is limited by the precision
of the constants $K_2$ and $K_3$ [Eqs.~(\ref{eq:k2222}) -- (\ref{eq:k3333})].

\section{The perturbative $\beta$-shift\label{sec:match}}
\subsection{Matching quenched and un-quenched}
In Eq.~(\ref{eq:mat}) we have seen
that at distances $r\gg m^{-1}$ the running of the coupling is not
affected by the presence of sea quarks anymore: 
at large distances, at least in
perturbation theory, the effect of massive
quarks can be integrated out into a shift of the coupling constant of
the quenched theory. 
The matching can be done within an intermediate mass-dependent scheme,
such as the $V$- or $R$- schemes. For simplicity we assume
$n_f$ mass-degenerate flavours. The discussion below can easily be generalized
to the non-degenerate case.
For $Ra\gg m^{-1}$ we write,
\begin{eqnarray}
\frac{a[V(Ra)-V_S]}{v_{c,1}(R)}&=&\alpha_L^{(n_f)}+e_1^{(n_f)}(R)
\alpha_L^2+\cdots\nonumber\\
&=&\alpha_L^{(0)}+e_1^{(0)}(R)
\alpha_L^2+\cdots,
\end{eqnarray}
where $e_i^{(n_f)}(R)$ are the expansion coefficients of
Eq.~(\ref{eq:relxxx}) for $n_f$ massive flavours with lattice quark mass
$m=(\kappa^{-1}-\kappa_c^{-1})/(2a)$.
The matched coupling that results in the same physics at
$R\gg (ma)^{-1}$ is,
\begin{equation}
\alpha_L^{(0)}=Z(m)\alpha_L^{(n_f)}
\end{equation}
with
\begin{equation}
\label{eq:matchqu}
Z(m)=1+z_1(m)\alpha_L^{(n_f)}+z_2(m)\alpha_L^2+\cdots,
\end{equation}
\begin{eqnarray}
z_1(m)=&&\lim_{R\rightarrow\infty}
\left[e_1^{(n_f)}(R)-e_1^{(0)}(R)\right],\\
z_2(m)=&&\lim_{R\rightarrow\infty}
\left\{e_2^{(n_f)}(R)-e_2^{(0)}(R)\right.\nonumber\\&&\left.-2
e_1^{(0)}(R)\left[e_1^{(n_f)}(R)-e_1^{(0)}(R)\right]\right\}.
\end{eqnarray}
In fact it turns out to be easier to determine
$Z(m)$ in momentum space
in the limit $q\rightarrow 0$. Since one and the same
physical potential is matched, momentum and
position space results have to agree [modulo the $(am)^{\nu}$ ambiguity
within $\Delta K_1$] and
we explicitly checked this in our calculation.

Using the $\overline{MS}$ scheme in momentum space we
demand for $q\rightarrow 0$:
\begin{eqnarray}
\alpha_V^{(n_f)}(q)&=&\alpha_{\overline{MS}}^{(n_f)}\left(a^{-1}\right)+
a_1^{(n_f)}(q,\{m\})\alpha^2+\cdots\nonumber\\
&=&
\alpha_{\overline{MS}}^{(0)}\left(a^{-1}\right)+
a_1^{(0)}(q)\alpha^2
+\cdots\\\nonumber
&=&\alpha_V^{(0)}(q),
\end{eqnarray}
where the $a_i(q,\{m\})$ are defined in Eqs.~(\ref{eq:a1shift})
-- (\ref{eq:a2shift}).
This amounts to,
\begin{equation}
\label{eq:conver1}
\alpha_{\overline{MS}}^{(0)}\left(a^{-1}\right)
=W(m)\alpha_{\overline{MS}}^{(n_f)}\left(a^{-1}\right)
\end{equation}
with
\begin{equation}
W(m)=1+w_1(m)\alpha_{\overline{MS}}^{(n_f)}\left(a^{-1}\right)
+w_2(m)\alpha_{\overline{MS}}^2\left(a^{-1}\right)+\cdots,
\end{equation}
\begin{eqnarray}
\label{eq:conver3}
w_1(m)&=&\lim_{q\rightarrow 0}\left[
a_1^{(n_f)}(q,\{m\})-a_1^{(0)}(q)\right]\\\nonumber
&=&a_1^{(n_f)}-a_1^{(0)}+\frac{n_f}{3\pi}\ln\left(\sqrt{C_0}ma\right),\\
w_2(m)&=&\lim_{q\rightarrow 0}
\left[a_2^{(n_f)}(q,\{m\})-a_2^{(0)}(q)\right.\\\nonumber
&&\left.-2a_1^{(0)}(q)w_1(m)\right].
\end{eqnarray}
By use of Eq.~(\ref{eq:cmsbarl}) one can then relate the above
$\overline{MS}$ scheme coefficients to the lattice coefficients
of Eq.~(\ref{eq:matchqu}):
\begin{equation}
z_1=w_1+b_1^{(n_f)}-b_1^{(0)}=w_1+K_1n_f,
\end{equation}
\begin{eqnarray}
z_2&=&w_2+b_2^{(n_f)}-b_2^{(0)}+2\left(b_1^{(n_f)}-b_1^{(0)}\right)
\left(w_1-b_1^{(0)}\right)
\nonumber\\
&=&w_2+\left(\frac{K_2}{N}+K_3N+2K_1w_1+K_1^2n_f\right)n_f.
\end{eqnarray}

\subsection{Results}
In following the above considerations
we obtain the one loop result,
\begin{equation}
z_1(m)=\left[K_1(ma)-\frac{5}{18\pi}+
\frac{1}{3\pi}\ln\left(\sqrt{C_0}ma\right)\right]n_f:
\end{equation}
the infra red behaviour of the theory with massive
fermions is the same as that in the absence of massive flavours,
at a gauge coupling that decreases with the quark mass:
\begin{eqnarray}
\label{eq:bshift}
\alpha^{(0)}_L&=&\alpha^{(n_f)}_L\\\nonumber
&+&\frac{n_f}{3\pi}\left[\ln(Dma)
+3\pi\Delta K_1(ma)\right]\alpha^2,\\
D&=&\sqrt{C_0}\exp[3\pi K_1(0)-5/6].
\end{eqnarray}
The constant $D$ has the numerical values,
\begin{eqnarray}
D^W&=&0.448(2),\\
D^{SW}&=&0.0238(1),\\
D^{KS}&=&0.726(2),
\end{eqnarray}
for Wilson, SW and KS
fermions, respectively.
From the values of $D$ we see that
the coupling shift is much more pronounced for SW fermions
than it is for Wilson or KS flavours.
{}From Eq.~(\ref{eq:k1limit}) we infer that for large $ma$
$\Delta K_1(ma)=-\ln(D'ma)/(3\pi)$.
Note that Eq.~(\ref{eq:bshift}) fixes the coefficient $D'=D$ and therefore the
leading order behaviour of $\Delta K_1(ma)$:
\begin{eqnarray}
\label{eq:k1la}
\Delta K_1(ma)&\stackrel{ma\rightarrow\infty}{\longrightarrow}&
-\frac{1}{3\pi}\ln(Dma)\\\nonumber
&=&0.00105(30)-K_1(0)-\frac{1}{3\pi}\ln ma.
\end{eqnarray}
In the above limit as well as in the limit $am=0$
$\Delta K_1(ma)$ is uniquely determined. This is different in the
intermediate $ma$ range where $\Delta K_1(ma)$ will depend on
the quantity used in the matching and is therefore subject to
${\mathcal O}[(am)^{\nu}]$ ambiguities. This in turn will also affect the
predicted $\beta$-shift.

The coefficient of the $\alpha^2$ term
diverges for $m\ll a^{-1}$:
for small values of $m$, the physics at
distance scales $r\gg m^{-1}$, at which the matching
to the quenched theory has to be performed, will 
be dominated by non-perturbative effects. Eventually, at very small
quark masses
the quenched and un-quenched theories
cannot be matched to each other anymore at any distance scale
with relevance to hadronic physics and the un-quenched theory
becomes genuinely different from the quenched one.

Finally we state the result for $z_2(m)$:
\begin{widetext}
\begin{eqnarray}
z_2(m)&=&\left\{\left[-\left(\frac{62}{9}+\frac{56}{3}\zeta_3\right)\frac{N}{2}
-\left(\frac{55}{3}-16\zeta_3\right)\frac{C_F}{2}
+\frac{19}{3}\frac{C_3-2C_1}{\sqrt{C_3^2-4C_2}}\ln
\left(\frac{C_3+\sqrt{C_3^2-4C_2}}{C_3-\sqrt{C_3^2-4C_2}}\right)
\right.\right.\nonumber\\&+&
\left.\left.\frac{38}{3}\ln\left(\frac{m^2a^2}{\sqrt{C_2}}\right)\right]
\frac{1}{16\pi^2}
+\left(16.95241-3\ln m^2a^2\right)\frac{C_F}{12\pi^2}+
\frac{K_2}{N}+K_3N\right\}n_f\label{eq:nextorder}\\\nonumber
&+&\left\{\left[\frac{100}{81}+
\frac{4}{9}\ln^2\left(C_0m^2a^2\right)\right]\frac{1}{16\pi^2}+
\left[-\frac{20K_1}{9}+
\left(\frac{4K_1}{3}-\frac{95}{81\pi}\right)\ln\left(C_0m^2a^2\right)\right]
\frac{1}{4\pi}+K_1^2\right\}n_f^2,
\end{eqnarray}
\end{widetext}
where $m$ denotes the lattice mass
$(\kappa^{-1}-\kappa_c^{-1})/(2a)$, rather than the pole mass
(hence the correction term
containing the numerical factor 16.95241).
Unfortunately, we cannot yet apply this result to lattice data.
For even in the case of Wilson-SW fermions where $K_2(0)$ and
$K_3(0)$ are known, the quark mass dependence of these coefficients
still has to be computed\footnote{In the case of Wilson fermions
one might be tempted to argue that ${\mathcal O}(ma)$ terms are
lattice artefacts anyway.
However, in the determination of the
$\beta$-shift it is exactly this $ma$ dependence of $r_0/a$
that we are looking for and this indeed depends on the lattice action.
Hence the variation of $K_2$ and
$K_3$ with $ma$ cannot be neglected.}.

\end{document}